%% file: manuscript.tex
\documentclass[aip,jcp,reprint,noshowkeys,superscriptaddress]{revtex4-2}
\usepackage{graphicx,dcolumn,bm,xcolor,microtype,multirow,amscd,amsmath,amssymb,amsfonts,physics,wrapfig,bbold,siunitx,xspace,braket,txfonts}
\usepackage[version=4]{mhchem}

\usepackage[utf8]{inputenc}
\usepackage[T1]{fontenc}
\usepackage{feynmp-auto}
\usepackage{longtable}
\usepackage{threeparttable}
\usepackage{multirow}

\usepackage{hyperref}
\hypersetup{
    colorlinks,
    linkcolor={red!50!black},
    citecolor={red!70!black},
    urlcolor={red!80!black}
}

\usepackage{listings}
\definecolor{codegreen}{rgb}{0.58,0.4,0.2}
\definecolor{codegray}{rgb}{0.5,0.5,0.5}
\definecolor{codepurple}{rgb}{0.25,0.35,0.55}
\definecolor{codeblue}{rgb}{0.30,0.60,0.8}
\definecolor{backcolour}{rgb}{0.98,0.98,0.98}
\definecolor{mygray}{rgb}{0.5,0.5,0.5}

\definecolor{sqred}{rgb}{0.85,0.1,0.1}
\definecolor{sqgreen}{rgb}{0.25,0.65,0.15}
\definecolor{sqorange}{rgb}{0.90,0.50,0.15}
\definecolor{sqblue}{rgb}{0.10,0.3,0.60}

\lstdefinestyle{mystyle}{
    backgroundcolor=\color{backcolour},
    commentstyle=\color{codegreen},
    keywordstyle=\color{codeblue},
    numberstyle=\tiny\color{codegray},
    stringstyle=\color{codepurple},
    basicstyle=\ttfamily\footnotesize,
    breakatwhitespace=false,
    breaklines=true,
    captionpos=b,
    keepspaces=true,
    numbers=left,
    numbersep=5pt,
    numberstyle=\ttfamily\tiny\color{mygray},
    showspaces=false,
    showstringspaces=false,
    showtabs=false,
    tabsize=2
  }

  \newcolumntype{d}{D{.}{.}{-1}}

\newcommand{\cre}[1]{a_{#1}^\dagger}
\newcommand{\ani}[1]{a_{#1}}

\newcommand{\ERI}[2]{\braket{#1|#2}}
\newcommand{\aERI}[2]{\bra{#1}\ket{#2}}
\newcommand{\sERI}[2]{(#1|#2)}

\newcommand{\fnt}{\footnotetext}
\newcommand{\fnm}{\footnotemark}

\newcommand{\eqrref}[1]{Eq.~\eqref{#1}}

\newcommand{\HF}{\text{HF}}

\newcommand{\GOWO}{\text{$G_0W_0$}}
\newcommand{\CCD}{\text{CCD}}
\newcommand{\drCCD}{\text{drCCD}}
\newcommand{\RPA}{\text{RPA}}
\newcommand{\IP}{\text{IP}}
\newcommand{\EA}{\text{EA}}
\newcommand{\IPEA}{\text{IP/EA}}

\newcommand{\bx}{\boldsymbol{x}}
\newcommand{\bA}{\boldsymbol{A}}
\newcommand{\bB}{\boldsymbol{B}}
\newcommand{\bX}{\boldsymbol{X}}
\newcommand{\bY}{\boldsymbol{Y}}
\newcommand{\bT}{\boldsymbol{T}}
\newcommand{\bt}{\boldsymbol{t}}
\newcommand{\bR}{\boldsymbol{R}}
\newcommand{\bE}{\boldsymbol{E}}
\newcommand{\bM}{\boldsymbol{M}}
\newcommand{\bH}{\boldsymbol{H}}
\newcommand{\bO}{\boldsymbol{0}}
\newcommand{\bI}{\boldsymbol{1}}

\newcommand{\blam}{\boldsymbol{\lambda}}
\newcommand{\bOm}{\boldsymbol{\Omega}}
\newcommand{\weq}{\ensuremath{\overset{!}{=}}}

\newcommand{\hT}{\Hat{T}}

\newcommand{\hF}{\Hat{F}}
\newcommand{\hL}{\Hat{L}}
\newcommand{\hR}{\Hat{R}}
\newcommand{\hH}{\Hat{H}}

\newcommand{\hV}{\Hat{V}}
\newcommand{\hZ}{\Hat{Z}}
\newcommand{\hLam}{\Hat{\Lambda}}
\newcommand{\hXi}{\Hat{\Xi}}

\newcommand{\cL}{\mathcal{L}}
\newcommand{\wtH}{\widetilde{H}}
\newcommand{\wtV}{\widetilde{V}}

\newcommand{\T}[1]{#1^{\intercal}}
\newcommand{\Ec}{E_\text{c}}
\newcommand{\Eh}{\ensuremath{\ E_\mathrm{h}}}

\newcommand{\ldr}{\text{$\lambda$-dr}}
\newcommand{\dr}{\text{dr}}

\newcommand{\SupInf}{\textcolor{blue}{Supplementary Material}\xspace}

\newcommand{\LCPQ}{Laboratoire de Chimie et Physique Quantiques (UMR 5626), Universit\'e de Toulouse, CNRS, Toulouse, France}
\newcommand{\UnHam}{Department of Chemistry, University of Hamburg, 22761 Hamburg, Germany; The Hamburg Centre for Ultrafast Imaging (CUI), Hamburg 22761, Germany}

\begin{document}	

\title{Analytic $G_0W_0$ gradients based on a double-similarity transformation equation-of-motion coupled-cluster treatment}

\author{Marios-Petros \surname{Kitsaras}}
	\email{kitsaras@irsamc.ups-tlse.fr}
	\affiliation{\LCPQ}

\author{Johannes T\"olle}
        \email{johannes.toelle@uni-hamburg.de}
        \affiliation{\UnHam}
	
\author{Pierre-Fran\c{c}ois \surname{Loos}}
	\email{loos@irsamc.ups-tlse.fr}
	\affiliation{\LCPQ}
	
\begin{abstract}
The accurate prediction of ionization potentials (IPs) is central to understanding molecular reactivity, redox behavior, and spectroscopic properties. 
While vertical IPs can be accessed directly from electronic excitations at fixed nuclear geometries, the computation of adiabatic IPs requires nuclear gradients of the ionized states, posing a major theoretical and computational challenge, especially within correlated frameworks. 
Among the most promising approaches for IP calculations is the many-body Green's function $GW$ method, which provides a balanced compromise between accuracy and computational efficiency. 
Furthermore, it is applicable to both finite and extended systems.
Recent work has established formal connections between $GW$ and coupled-cluster doubles (CCD) theory, leading to the first derivation of analytic $GW$ nuclear gradients via a unitary CCD framework. 
In this work, we present an alternative, fully analytic formulation of $GW$ nuclear gradients based on a modified version of the traditional equation-of-motion CCD formalism, enabling the inclusion of missing correlation effects in the traditional CCD methods.

\end{abstract}

\maketitle

\section{Introduction}

The ionization potential (IP) of a molecule, defined as the minimum energy required to remove an electron from a neutral species in its ground state, is a fundamental property that governs its reactivity, redox behavior, and spectroscopic signatures. 
Two main types of IPs are commonly distinguished: the vertical ionization potential (VIP) and the adiabatic ionization potential (AIP). 
These differ in whether or not nuclear relaxation of the ionic state is considered.

The VIP refers to the energy required to remove an electron without allowing the nuclei to move. 
In other words, the ionized state is constrained to the equilibrium geometry of the neutral molecule. 
This vertical transition approximates the Franck–Condon principle and is representative of ultrafast ionization events, where the nuclei remain effectively frozen during the electronic transition. 
Experimentally, VIPs manifest as sharp, intense peaks in photoelectron spectroscopy (PES), particularly when probed using ultraviolet or X-ray photons for valence and core levels, respectively.

In contrast, the AIP accounts for full nuclear relaxation in both the neutral and ionized states. 
It represents the true thermodynamic threshold for ionization, corresponding to the energy difference between the neutral ground state and the fully relaxed ionic ground state. 
As such, AIPs are directly linked to gas-phase thermochemistry and redox potentials. 
In PES experiments, AIPs appear as the onset of a given ionization band, while the corresponding VIP aligns with the maximum of the corresponding band. 
High-resolution PES, and in particular threshold photoelectron spectroscopy (TPES), are capable of resolving this onset with sufficient precision to extract both AIP and VIP from a single measurement. \cite{Baer_2017} 

The energy difference between VIP and AIP, known as the relaxation energy, provides insight into the structural reorganization of the molecule upon ionization. 
A large relaxation energy indicates a significant geometric change, whereas a small value suggests that the neutral and ionic states share similar geometries. 
The simultaneous analysis of AIP and VIP in high-resolution PES thus enables a detailed understanding of the interplay between electronic structure and nuclear dynamics during ionization.

From a theoretical standpoint, the evaluation of VIPs is relatively straightforward. 
It requires knowledge of the ground-state equilibrium geometry and the ability to compute total energies of the neutral and cationic species at this fixed geometry. 
This can be achieved using standard electronic structure methods or, more efficiently, by computing excitation energies corresponding to electron removal (i.e., ionization) from the reference neutral ground state.

In contrast, the calculation of AIPs is more demanding, as it involves geometry optimization of the ionized species. 
This requires access to nuclear gradients of the ionized state, i.e., the first derivatives of the cationic energy with respect to the nuclear coordinates. 
Unlike ground-state gradients, which are routinely available in most quantum chemistry packages, \cite{Handy_1984,Schaefer_1986,Pulay_2014,Park_2020} analytic gradients for charged excited states are significantly more challenging to derive and implement, \cite{Lischka_2018} especially within correlated or many-body frameworks. \cite{Stalring_2001,Celani_2003,Szalay_2012,Iino_2023}
Nonetheless, these gradients are essential to perform geometry optimization, typically using a Newton-Raphson procedure, until a stationary point corresponding to the relaxed ionic minimum is reached.

While various computational approaches exist for accessing charged excitations, ranging from state-specific (e.g., $\Delta$SCF) to equation-of-motion/linear-response formalisms, \cite{Rowe_1968a,Sneskov_2012} one particularly promising method is the $GW$ approximation \cite{Hedin_1965,Aryasetiawan_1998,Reining_2017,Golze_2019,Marie_2024a} from many-body Green's function theory. \cite{Onida_2002,Martin_2016}
The $GW$ formalism provides a robust framework to compute IPs \cite{vanSetten_2015,Caruso_2016,Krause_2017,Lewis_2019,Bruneval_2021,Monino_2023,Marie_2024b,vanSetten_2018,Golze_2018,Golze_2020,Mejia-Rodriguez_2021,Li_2022,Mukatayev_2023,Panades-Barrueta_2023} and electron affinities (EAs)\cite{vanSetten_2015,Caruso_2016,Krause_2017,Gallandi_2016,Richard_2016,Knight_2016,Dolgounitcheva_2016} by approximating Hedin’s equations. \cite{Hedin_1965,Marie_2024c}
It gained popularity due to its favorable balance between accuracy and computational cost. 
Its results often rival those of high-level wavefunction-based approaches, while remaining applicable to much larger molecular systems \cite{Neuhauser_2013,Neuhauser_2014,Kaltak_2014,Govoni_2015,Vlcek_2017,Wilhelm_2018,Golze_2018,Duchemin_2019,DelBen_2019,Forster_2020,Duchemin_2020,Kaltak_2020,Forster_2021,Duchemin_2021,Wilhelm_2021,Forster_2022,Yu_2022,Panades-Barrueta_2023,Tolle_2024} and, importantly, periodic materials. \cite{Onida_2002,Martin_2016}
Although traditionally limited to finite systems, equation-of-motion coupled-cluster (EOM-CC) has very recently been extended to periodic systems as well, using both Gaussian and plane-wave basis sets.\cite{Vo_2024,Moerman_2025}

Recently, based on the established theoretical connections \cite{Lange_2018,Quintero_2022,Tolle_2022} between $GW$ and coupled-cluster doubles (CCD) theory, \cite{Cizek_1966,Paldus_1972,Crawford_2000,Piecuch_2002,Bartlett_2007,Shavitt_2009} T\"olle derived and implemented the first fully analytic $GW$ nuclear gradients \cite{Tolle_2025} (see also Refs.~\onlinecite{Lazzeri_2008,Faber_2011b,Faber_2015,Montserrat_2016,Li_2019a,Li_2024}).
His work has subsequently been extended to the first fully analytic Bethe-Salpeter equation (BSE) \cite{Salpeter_1951,Strinati_1988} nuclear gradients, \cite{Ismail-Beigi_2003,Kaczmarski_2010,Caylak_2021,Knysh_2022,Villalobos-Castro_2023,Knysh_2023a,Knysh_2023b,Tolle_2025b} effectively generalizing the present formalism to neutral (optical) excited states. \cite{Blase_2018,Blase_2020}
These pioneering results exploit theoretical connections between $GW$ and the unitary CCD (UCCD) \cite{Kutzelnigg_1982,Kutzelnigg_1983,Kutzelnigg_1984,Bartlett_1989,Szalay_1995,Taube_2006,Kutzelnigg_2010} framework. \cite{Tolle_2022}
However, this approach necessitates the numerical evaluation of an infinite series of nested (anti)commutators, which is a non-standard numerical technique in traditional CCD implementations.

It would be, therefore, desirable to reformulate the $GW$ nuclear gradients to leverage the well-established traditional EOM-CCD framework, as well as its extension to analytic properties. 
Unfortunately, a direct application of the traditional EOM-CCD formalism to $GW$ is hindered by the missing correlation effects, as analyzed in detail in Refs.~\onlinecite{Berkelbach_2018,Tolle_2022}. 
In this work, we demonstrate how this obstacle can be overcome by introducing a modified version of the traditional EOM-CCD formalism, which we refer to as IP/EA-EOM-$\lambda$-direct-ring CCD. 
This approach allows (i) for the inclusion of missing correlation effects, and (ii) the derivation of analytic $GW$ nuclear gradients within a more standard CC Lagrangian framework.

\section{Theory}

\subsection{The $GW$ approximation}
\label{subsec:GW}

Similar to other Green's function approaches, the $GW$ one-body Green's function $G$ is expressed using the recursive Dyson equation
\begin{equation} \label{eq:Dyson}
    G = G_0 + G_0 \Sigma G. 
\end{equation}
The mean-field reference one-body Green's function $G_0$, which for the current study corresponds to the Hartree-Fock (HF) approximation, is given by
\begin{equation} \label{eq:G0}
    G_0 (\bx_1\bx_2;\omega) =\sum_i \frac{\phi_i^{*} (\bx_1) \phi_i (\bx_2)}{\omega-\epsilon_i^\HF}+\sum_a \frac{\phi_a^{*} (\bx_1) \phi_a (\bx_2)}{\omega-\epsilon_a^\HF}, 
\end{equation}
where $\phi_p(\bx)$ and $\epsilon_p^\HF$ are the canonical HF orbitals and their corresponding energies. 
Likewise, the exact one-body Green's function is expressed as
\begin{equation} \label{eq:G}
    G (\bx_1\bx_2;\omega) =\sum_I \frac{\psi_I^{*} (\bx_1) \psi_I (\bx_2)}{\omega-\epsilon_I}+\sum_A \frac{\psi_A^{*} (\bx_1) \psi_A (\bx_2)}{\omega-\epsilon_A}, 
\end{equation}
where $\psi_I(\bx)$ and $\psi_A(\bx)$ are the so-called Dyson orbitals. 
The exact Green's function has poles at the energy differences associated with the electron-detached (hole states) and electron-attached (particle states) processes, namely, $\epsilon_I =  E^N_0 - E^{N-1}_I$ and $\epsilon_A = E^{N+1}_A - E^N_0,$ where $E^N_0$ is the ground-state energy of the neutral $N$-electron system, and $E^{N-1}_I$ and $E^{N+1}_A$ are the energies of the corresponding $(N-1)$- and $(N+1)$-electron states.
Within the quasiparticle approximation, where only ionized and electron-attached states with a dominant single-particle character are retained, the Green’s function reduces to the quasiparticle form
\begin{equation} \label{eq:GQP}
    G (\bx_1\bx_2;\omega) =\sum_{i} \frac{\psi_i^{*} (\bx_1) \psi_i (\bx_2)}{\omega-\epsilon_i}+\sum_{a} \frac{\psi_a^{*} (\bx_1) \psi_a (\bx_2)}{\omega-\epsilon_a},
\end{equation}
where $\epsilon_i$ and $\epsilon_a$ denote the hole and particle quasiparticle energies, respectively.
In the following, the standard convention is used for the orbital indices, with $i,j,k,\dots$ denoting hole states, $a,b,c,\dots$ denoting particle states, and $p,q,r,\dots$ indicating either occupied or virtual orbitals. 

The dynamical self-energy $\Sigma$ describes correlation effects beyond the mean-field approximation.
In $GW$, the self-energy is constructed using the dynamically-screened Coulomb interaction computed at the random-phase approximation (RPA) level of theory \cite{Bohm_1951,Pines_1952,Bohm_1953,Nozieres_1958,SchuckBook,Chen_2017,Ren_2015} and is given by
\begin{equation} \label{eq:SigmaGW}
    \Sigma_{pq} (\omega) = \sum_{i \mu} \frac{\sERI{pi}{\mu} \sERI{\mu}{qi}}{\omega - \epsilon_i + \Omega_\mu} + \sum_{a \mu} \frac{\sERI{ap}{\mu} \sERI{\mu}{aq}}{\omega - \epsilon_a - \Omega_\mu}. 
\end{equation}
The quasiparticle energies and the effective two-electron integrals 
\begin{subequations} \label{eq:scr_2int}
\begin{align} 
    \sERI{pq}{\mu} & = \sum_{ia} \qty[ \ERI{pa}{qi} x^a_{i,\mu} + \ERI{pi}{qa}  y^i_{a,\mu} ],
    \\
    \sERI{\mu}{pq} & = \sum_{ia} \qty[ \ERI{qi}{pa} x^{a*}_{i,\mu} + \ERI{qa}{pi}  y^{i*}_{a,\mu} ],
\end{align}
\end{subequations}
are required to construct the elements of the self-energy. 
The index $\mu$ enumerates the solutions to the RPA problem where $x^a_{i,\mu}$ and $y^i_{a,\mu}$ are the elements of the eigenvector solutions corresponding to the excitation energy $\Omega_\mu$ (see Sec.~\ref{subsubsec:RPA}). 
The two-electron integrals $\ERI{pq}{rs}$ are given in Dirac notation i.e., $\ERI{12}{12}$.

By substituting Eqs.~\eqref{eq:G0} and \eqref{eq:SigmaGW} into Eq.~\eqref{eq:Dyson}, one finds that the quasiparticle energies $\epsilon_p$ and corresponding Dyson orbitals $\psi_p(\bx)$ are obtained as the eigenvalues and eigenvectors of the following non-linear, frequency-dependent Fock-like operator:
\begin{equation} \label{eq:workingGW}
    \qty[ \hF + \Sigma(\omega = \epsilon_p) ] \psi_p(\bx) = \epsilon_p \psi_p(\bx).
\end{equation}
$\hF$ is the usual Fock operator including Hartree and exchange contributions, and the self-energy accounts for correlation effects. 
Although the so-called quasiparticle equation defined in Eq.~\eqref{eq:workingGW} acts as the working equation for $GW$, its non-linear and dynamical nature makes it difficult to solve. 
In practice, it is solved iteratively due to the dependence of the self-energy on the quasiparticle energies [see Eq.~\eqref{eq:SigmaGW}].
A hierarchy of approximations has been developed to address this issue, each defined by how the self-energy is constructed and updated at each iteration, leading to various levels of self-consistency (e.g., ev$GW$, \cite{Hybertsen_1986,Shishkin_2007a,Blase_2011a,Faber_2011,Rangel_2016} qs$GW$, \cite{Gui_2018,Faleev_2004,vanSchilfgaarde_2006,Kotani_2007,Ke_2011,Kaplan_2016,Forster_2021,Marie_2023} etc.). 
Further details can be found in Ref.~\onlinecite{Marie_2024a}.
The widely used $G_0W_0$ approximation corresponds to a single-shot calculation in which both the Green's function and the screened interaction are fixed. \cite{Strinati_1980,Hybertsen_1985a,Godby_1988,Linden_1988,Northrup_1991,Blase_1994,Rohlfing_1995}
When combined with the diagonal approximation, the $G_0W_0$ quasiparticle energies are commonly obtained by solving the following one-dimensional, nonlinear equation for each state $p$:
\begin{equation} \label{eq:GW_std}
	\epsilon_p = \epsilon^\HF_p + \Sigma_{pp} (\omega=\epsilon_p).
\end{equation}

Green's function approaches are designed to describe properties of the many-body electronic state.
In the case of $G_0W_0$, the quasiparticle energies, associated with IPs in the hole space and EAs in the particle space, are improved relative to the mean-field reference.
However, the total energy of the reference state is not uniquely defined. Various functionals have been proposed to calculate the correlation energy, including the Klein functional, \cite{Klein_1961} which is based on $G$, the Luttinger-Ward functional, \cite{Luttinger_1960,Baym_1961,Potthof_2003} which is based on $\Sigma$, and the Galitskii-Migdal functional.\cite{Galitskii_1958,Holm_2000} 
These expressions yield identical results only when a fully self-consistent Green's function (beyond the quasiparticle approximation) is used. \cite{Almbladh_1999,Dahlen_2004,Dahlen_2004a,Dahlen_2005,Dahlen_2005a,Dahlen_2006,Stan_2006,Stan_2009,Ismail-Beigi_2010}
In other words, for non-self-consistent schemes such as $\GOWO$, they generally differ.
Consequently, the total energy for the electron-detached or electron-attached state is also not uniquely defined within the $\GOWO$ approximation. 
In the present work, the correlation energy of the ground state is evaluated at the RPA level (see Sec.~\ref{subsec:RPACC}), which can be shown to be equivalent to the Klein functional.

By analogy with the extended Koopmans' theorem (EKT), in which the total energy of a charged state is approximated as
\begin{equation}
	E^\text{EKT}_{p} = E^\HF_0 \mp \epsilon_p^\HF,
\end{equation}
where $E^\HF_0$ is the HF energy of the reference state, we define the $G_0W_0$ total energy as
\begin{equation} \label{eq:G0W0_totene}
	E^\GOWO_{p} = E^\GOWO_0 \mp \epsilon_p^\GOWO,
\end{equation}
where $E^\GOWO_0$ consists of the mean-field HF energy and the RPA correlation energy, that is,
\begin{equation}
	E^\GOWO_0 = E^\HF_0 + \Ec^\RPA.
\end{equation}

Furthermore, Eq.~\eqref{eq:GW_std} is not well-suited for deriving analytic derivatives.
Unlike in standard Lagrangian formulations for non-variational ans\"atze, the identification of wavefunction parameters is obscured by the frequency dependence of the self-energy.
In this work, we derive a Lagrangian formulation that correctly reproduces the electronic energy of the charged states at the $G_0W_0$ level [see \eqrref{eq:G0W0_totene}].
This formulation exploits the connection between the RPA and a particular approximation to the CC expansion, as discussed in Sec.~\ref{subsec:RPACC}.
Through this connection, an exact block-diagonalization of the RPA matrix is achieved.
In Sec.~\ref{subsec:GWEOM}, this block-diagonalization is used to reformulate the nonlinear, frequency-dependent $G_0W_0$ problem into an EOM-CC problem.
The resulting formalism not only provides a clear definition of the electron-detached (or -attached) energies, but also enables a standard Lagrangian construction for computing analytic derivatives. 

\subsection{Connection between RPA and drCCD}
\label{subsec:RPACC}
\subsubsection{RPA}
\label{subsubsec:RPA}

The RPA equations can be cast as a linear eigenvalue problem, $\bH_\RPA \cdot \bR = \bR \cdot \bE$, that takes the form of Casida's equations \cite{Casida_1995}
\begin{equation} \label{eq:casida}
	\begin{pmatrix}
        \bA & \bB 
        \\
        -\bB^*& -\bA^*
    \end{pmatrix}
    \cdot
    \begin{pmatrix}
        \bX &\bY^* 
        \\
        \bY & \bX^*
    \end{pmatrix} = 
    \begin{pmatrix}
        \bX &\bY^* 
        \\
        \bY & \bX^*
    \end{pmatrix}
    \cdot
    \begin{pmatrix}
        \bOm & \bO 
        \\
        \bO& -\bOm
    \end{pmatrix},
\end{equation}
where the elements of matrices $\bA$ and $\bB$ are given by 
\begin{subequations} \label{eq:general_AB}
\begin{align}
    A_{ia,jb} &= \bra{\Phi^a_i} \hH \ket{\Phi^b_j},
    \\
    B_{ia,jb} &= \bra{\Phi^{ab}_{ij}} \hH \ket{\Phi_0}.
\end{align}
\end{subequations}
The determinants $\ket{\Phi_0}$, $\ket{\Phi^{a}_{i}}$, $\ket{\Phi^{ab}_{ij}}$ are the reference, singly- and doubly-excited determinants constructed within the mean-field orbital space.
The matrices $\bX$ and $\bY$ are associated with single excitations and deexcitations, respectively, and gather the elements of the eigenvectors, $x^a_{i,\mu}$ and $y^i_{a,\mu}$ previously defined in Sec.~\ref{subsec:GW}. 
The diagonal matrix $\bOm$ contains the positive excitation energy eigenvalues $\Omega_\mu$. 

For the purposes of the $GW$ approximation, the so-called \textit{direct} version of RPA is considered. This approximation only affects the two-electron part of the normal-ordered Hamiltonian $\hH_\text{N} = \hF_\text{N} + \hV_\text{N}$, with $\hF_\text{N} = \sum_{pq} f_{pq} \{ \cre{p}\ani{q} \}$ the normal-ordered Fock operator. 
Specifically, the exchange contribution in the fluctuation potential,
\begin{equation} \label{eq:fluctuation_operator}
	\begin{split}
	\hV_\text{N}
	& = \frac{1}{4}\sum_{pqrs} \aERI{pq}{rs}  \{ \cre{p}\cre{q}\ani{s}\ani{r} \} 
	\\
	& = \frac{1}{4}\sum_{pqrs} \qty(\ERI{pq}{rs}  - \ERI{pq}{sr} ) \{ \cre{p}\cre{q}\ani{s}\ani{r} \},
	\end{split}
\end{equation}
is neglected.
In this form,  only the direct (Coulomb) component of the electron-electron interaction is retained, while the exchange term is omitted, consistent with the assumptions underlying the $GW$ and direct RPA formalisms.
Operators $\cre{p}$ and $\ani{p}$ are the usual fermionic creators and annihilators, respectively, and the curly braces denote normal ordering with respect to the Fermi vacuum.    
Under this approximation, the matrix elements defined in Eq.~\eqref{eq:general_AB} reduce to
\begin{subequations}
\begin{align}
    A_{ia,jb} & = \qty( \epsilon_a^\HF - \epsilon_i^\HF ) \delta_{ij} \delta_{ab} + \ERI{aj}{ib},
    \\
    B_{ia,jb} & = \ERI{ab}{ij},
\end{align}
\end{subequations}
and the direct RPA correlation energy is given by the plasmon (or trace) formula: \cite{Sawada_1957a,Sawada_1957b,Rowe_1968b,SchuckBook}
\begin{equation} \label{eq:RPA_corr}
    \Ec^\RPA = \frac{1}{2} \qty[ \sum_\mu \Omega_\mu - \Tr(\bA) ]. 
\end{equation}
In the remainder of the manuscript, we will drop the term ``direct'' when referring to RPA, as it is always implied.

\subsubsection{drCCD} 
\label{subsubsec:rCCD}

It has been previously shown that RPA is equivalent to the direct-ring (dr) CCD (drCCD) truncation.\cite{Freeman_1977,Scuseria_2008,Jansen_2010,Scuseria_2013,Peng_2013,Berkelbach_2018,Rishi_2020} 
This approach constitutes an approximation to standard CCD, where the cluster operator includes all double excitations
\begin{equation} \label{eq:T2}
    \hT_2 
    = \frac{1}{4} \sum_{ijab} t^{ab}_{ij} \cre{a} \cre{b} a_j a_i . 
\end{equation}
Within the \textit{``ring''} approximation, only those terms of the similarity-transformed Hamiltonian that correspond to loop (ring) diagrams are retained.
It is well known that the ring approximation comes at the cost of breaking the fundamental antisymmetry of the electronic wavefunction.\cite{Scuseria_2008}
A bosonic character is introduced to a fundamentally fermionic system, and certain contractions among individual fermionic particle operators are ignored. As a result, the wavefunction fails to remain fully antisymmetric under electron exchange. 
In addition, only the direct Coulomb contribution of the fluctuation potential contributes, that is, excluding exchange.
A modified set of rules for computing the drCCD Hamiltonian
\begin{equation} \label{eq:wtHCCD}
	\wtH_\text{CCD} = e^{-\hT_2} \hH_\text{N} e^{\hT_2} \approx \wtH_\dr
\end{equation}
using diagrammatic techniques is presented in Appendix~\ref{app:rules}.

The drCCD amplitudes equations are given by projections of the drCCD Hamiltonian onto doubly-excited determinants 
\begin{equation} \label{eq:rCCD_ampl}
\begin{split}
    \bra{\Phi_{ij}^{ab}} \wtH_\dr \ket{\Phi_0} & \weq 0 \\
    \bB +\bA \cdot \bt + \bt \cdot \bA^* + \bt \cdot \bB^* \cdot \bt & = \bO. 
\end{split}
\end{equation}
The matrix $\bt$ collects the drCCD amplitudes $\bt_{ia,jb} = t^{ab}_{ij}$.
Within a standard CCD treatment, the correlation energy is given by
\begin{equation}\label{eq:ECC_ener_ver}
	\begin{aligned}
	\Ec^\CCD &= \bra{\Phi_0} \wtH_\CCD \ket{\Phi_0} \\
	&=  \frac{1}{4} \sum_{ijab} t^{ab}_{ij} \aERI{ij}{ab} \\
	& = \frac{1}{2} \sum_{ijab} t^{ab}_{ij} \ERI{ij}{ab} ,
	\end{aligned}
\end{equation}
where the antisymmetric form of the CCD amplitudes ($t^{ab}_{ij} = - t^{ab}_{ji} = t^{ba}_{ji}$) has been exploited. 
Applying the direct treatment within drCCD (where the amplitudes are not antisymmetric) is equivalent to using the last line of \eqrref{eq:ECC_ener_ver} for the correlation energy expression \cite{Scuseria_2008} 
\begin{equation}
	\begin{aligned}
		\Ec^\drCCD
		 &= \bra{\Phi_0} \wtH_\dr \ket{\Phi_0}  \\
		 &=\frac{1}{2} \sum_{ijab} t^{ab}_{ij} \ERI{ij}{ab} \\
		 & = \frac{1}{2} \Tr(\bB^* \cdot \bt ).
	\end{aligned}
\end{equation}
Thus, an effective factor of $2$ arises in the calculation of the expectation values of the drCCD Hamiltonian. 

It can be further shown \cite{Scuseria_2008} that the drCCD correlation energy coincides with the RPA correlation energy
\begin{align}
    \Ec^\drCCD = \Ec^\RPA.
\end{align}
The latter becomes apparent when the term $\sum_\mu \Omega_\mu$ in Eq.~\eqref{eq:RPA_corr} is rewritten as the trace of a block-diagonal RPA matrix, as seen below in \eqrref{eq:casida_simple_blockdiag}.

In order to obtain analytic derivatives of the RPA correlation energy, which is part of the total $G_0W_0$ electronic energy [see \eqrref{eq:G0W0_totene}], the method of Lagrange multipliers is employed. \cite{Gauss_2000,Helgakerbook} 
Using the established connection to the drCCD approximation,\cite{Rekkedal_2013} the corresponding Lagrangian is defined as
\begin{equation}
    \cL^\drCCD_\text{c} = \bra{\Phi_0} \qty( 1 + \hLam_2 ) \wtH_\dr \ket{\Phi_0}, \label{eq:RPA_lang}
\end{equation} 
with the $\lambda$-amplitude equations obtained by enforcing the following stationarity conditions
\begin{equation} \label{eq:RPA_lambdaamps}
    \pdv{\cL^\drCCD_\text{c}}{t^{ab}_{ij}} =
    \bra{\Phi_0} \qty(1 + \hLam_2 ) \pdv{\wtH_\dr}{t^{ab}_{ij}} \ket{\Phi_0} \weq 0. 
\end{equation}
The deexcitation operator $\hLam_2$ is given by 
\begin{equation}
    \hLam_2 
    = \frac{1}{4} \sum_{abij} \lambda^{ij}_{ab} \cre{i} \cre{j} \ani{b} \ani{a}.
\end{equation}
The $\lambda$ Lagrange multipliers can also be viewed as wavefunction parameters of the CC bra state within the bivariational formulation of CC.\cite{Stanton_1993a,Levchenko_2005,Kvaal_2013} 
Eq.~\eqref{eq:RPA_lambdaamps} can then be recast in matrix form as:
\begin{equation} \label{eq:drCCD_lambda}
     \bB^*  + \blam \cdot \qty(\bA + \bA^* + \bB^* \cdot \bt + \bt \cdot \bB^*  )  = \bO. 
\end{equation}
The $\lambda$ amplitude equations are linear in $\blam$, which collects the elements of the $\lambda$ amplitudes as $\blam_{ia,jb}=\lambda^{ij}_{ab}$. 
A diagrammatic derivation accompanied by working equations can be found in Appendix \ref{app:rCCD}.

\subsubsection{Block diagonalization of the RPA Hamiltonian}
\label{subsec:blockdiag}

Previously, it has been shown that the exact correspondence between $G_0W_0$ and EOM-CC requires the block diagonalization of the RPA matrix \cite{Tolle_2022}
\begin{align}
	\boldsymbol{U}^\dagger \cdot \boldsymbol{H}_\text{RPA} \cdot \boldsymbol{U}  = 
	\begin{pmatrix}
		\tilde{\boldsymbol{A}} & \boldsymbol{0} 
		\\
		\boldsymbol{0} & - \tilde{\boldsymbol{A}}^*
	\end{pmatrix}.
\end{align}
This can be achieved either through a Bogoliubov transformation or through an alternative unitary transformation, e.g., a unitary drCCD (drUCCD) treatment. \cite{Tolle_2022}
In contrast, the \textit{similarity-transformed} drCCD Hamiltonian is not equivalent to a \textit{unitary transformation} even if it manages to reproduce the correlation energy $\Ec^\RPA$.
The differences stem from the fact that the similarity transformation only treats the excitation part of \eqrref{eq:casida} and does not fully account for both forward- and backward-time-ordered bubble diagrams present in $GW$, as first discussed in Ref.~\onlinecite{Lange_2018}.

More precisely, the drCCD Hamiltonian corresponds to the following treatment of the RPA matrix
\begin{equation}\label{eq:unH}
	\bar{\bH}_\RPA  = \bM^{-1} \cdot \bH_\RPA \cdot \bM,
\end{equation}
with 
\begin{align}
    \bM & = 
    \begin{pmatrix}
        \bI & \bt 
        \\
        \bO & \bI
    \end{pmatrix}, 
    &
    \bM^{-1} & = 
    \begin{pmatrix}
        \bI & -\bt
        \\
        \bO & \bI
    \end{pmatrix}.
\end{align}
This transformation leads to a new eigenvalue equation of the form $\bar{\bH}_\RPA \cdot \bar{\bR} = \bar{\bR} \cdot \bE$ with
\begin{equation}
\begin{split}
	\bar{\bH}_\RPA 
	& = 
	\begin{pmatrix}
        \bA +  \bt \cdot \bB^*  & \bB +\bA \cdot \bt + \bt \cdot \bA^* + \bt \cdot \bB^* \cdot \bt 
        \\
        -\bB^*   & - \bB^* \cdot \bt - \bA^*
    \end{pmatrix} 
    \\
    & = 
    \begin{pmatrix}
        \bA +  \bt \cdot \bB^*  & \bO
        \\
        -\bB^*   & - \bB^* \cdot \bt - \bA^*
    \end{pmatrix},
\end{split} \label{eq:simM}
\end{equation}
and
\begin{equation}
    \bar{\bR} 
    = \bM^{-1} \cdot \bR
    = 
    \begin{pmatrix}
        \bX - \bt \cdot \bY & \bY^* - \bt \cdot \bX^* 
        \\
        \bY &  \bX^*
    \end{pmatrix}
    = 
    \begin{pmatrix}
        \bar{\bX} & \bO 
        \\
        \bY & \bX^*
    \end{pmatrix}, \label{eq:simMR}
\end{equation}
where $\bar{\bX} = \bX - \bt \cdot \bY$.
The upper right block of $\bar{\bH}_\RPA$ vanishes as it represents the drCCD amplitude equations given by \eqrref{eq:rCCD_ampl}. 
Additionally, it can be shown that 
\begin{align}\label{eq:RPA_TYX-1}
	\bt = \bY^* \cdot {\bX^{-1}}^*.
\end{align}
We note that the formulation presented here to connect RPA to drCCD differs from those found in the literature by a complex conjugation.\cite{Scuseria_2008,Scuseria_2013,Berkelbach_2018} 
Specifically, other works introduce the following parametrization 
\begin{align}\label{eq:RPA_TYX-1_com}
	\bT = \bY \cdot \bX^{-1}
\end{align}
as a first step to block-diagonalize the RPA matrix. The resulting matrix $\bT$ corresponds to the deexcitation operator $\hT_2^\dagger$ rather than the excitation operator $\hT_2$ [see \eqrref{eq:T2}].
Numerical differences between the common derivation [see \eqrref{eq:RPA_TYX-1_com}] and \eqrref{eq:RPA_TYX-1} arise only when complex orbitals are employed, as the two formulations differ only in a complex conjugation. This is the case, for example, in the presence of a magnetic field as used in Ref.~\onlinecite{Holzer_2019}. Formally, however, the current derivation is consistent in defining the $\hT_2$ as an excitation operator and not as a deexcitation operator.

In order to account for the contributions necessary to replicate the block diagonal structure of Eq.~\eqref{eq:unH}, a second similarity transformation is introduced
\begin{align}
    \bar{\bM} & = 
    \begin{pmatrix}
        \bI & \bO
        \\
        \blam & \bI
    \end{pmatrix}, 
    &
    \bar{\bM}^{-1} & = 
    \begin{pmatrix}
        \bI &  \bO
        \\
        -\blam & \bI
    \end{pmatrix}.
\end{align}
The transformed problem reads $\widetilde{\bH}_\RPA \cdot \widetilde{\bR} = \widetilde{\bR} \cdot \bE$ with
\begin{equation} \label{eq:simRPA}
\begin{split} 
    \widetilde{\bH}_\RPA 
    & = \bar{\bM}^{-1} \cdot \bar{\bH}_\RPA \cdot \bar{\bM}
    \\
    & = 
    \begin{pmatrix}
        \bA +  \bt  \cdot \bB^*  & \bO 
        \\
        -\left [ \bB^*  + \blam \cdot \qty(\bA + \bA^* + \bB^* \cdot \bt + \bt \cdot \bB^*  ) \right] & - \bB^* \cdot \bt - \bA^*
    \end{pmatrix},
\end{split}
\end{equation}
and 
\begin{equation}
    \widetilde{\bR} 
    = \bar{\bM}^{-1} \cdot \bar{\bR}  
    = 
    \begin{pmatrix}
        \bar{\bX} & \bO 
        \\
        -\blam\cdot\bar{\bX}+\bY & \bX^*
    \end{pmatrix}.
\end{equation}
The lower left block of \eqrref{eq:simRPA} vanishes as it is simply the negative of the $\lambda$-amplitude equations given by \eqrref{eq:drCCD_lambda}. 
Accordingly, it can be shown that the $\lambda$ amplitudes are connected to the RPA parameters using $\blam = \bY \cdot \bar{\bX}^{-1}$. 
The final form for the doubly similarity-transformed RPA matrix in a block-diagonal form is given by
\begin{subequations}
\begin{align}
    \widetilde{\bH}_\RPA & = 
    \begin{pmatrix}
        \bA +  \bt \cdot\bB^*  & \bO 
        \\
        \bO & -  \qty( \bA^* + \bB^* \cdot \bt )
    \end{pmatrix},
    \\
    \widetilde{\bR} & = 
    \begin{pmatrix}
        \bar{\bX} & \bO 
        \\
        \bO & \bX^*
    \end{pmatrix}.
\end{align}
\end{subequations}
We note that the matrix $\bA$ is Hermitian ($\bA^* = \T{\bA}$, where the superscript $\T{}$ denotes the transpose of the matrix) and the matrices $\bB$ and $\bt$ are symmetric ($\bB = \T{\bB}$ and $\bt = \T{\bt}$). The deexcitation block can thus be rewritten as
\begin{equation}
	\bA^* + \bB^* \cdot \bt = \T{(\bA + \bt \cdot\bB^* )}.
\end{equation}
As a result, the similarity-transformed RPA problem can be recast in the form of an expectation value
\begin{equation} \label{eq:dsRPA}
	\T{\begin{pmatrix}
			\bX^* & \bO 
			\\
			\bO & \bar{\bX}
	\end{pmatrix}}
	\cdot
	\begin{pmatrix}
		\bA +  \bt \cdot\bB^* & \bO 
		\\
		\bO & - \T{(\bA +  \bt \cdot\bB^* )}
	\end{pmatrix}
	\cdot
	\begin{pmatrix}
		\bar{\bX} & \bO 
		\\
		\bO & \bX^*
	\end{pmatrix}  
	=
	\begin{pmatrix}
		\bOm & \bO 
		\\
		\bO & -\bOm
	\end{pmatrix},
\end{equation} 
which can be further simplified to
\begin{equation} \label{eq:casida_simple_blockdiag}
	{\bX}^\dagger \cdot \qty( \bA +  \bt \cdot\bB^* ) \cdot \bar{\bX} = \bOm.
\end{equation}
This equation demonstrates that $\bar{\bX}$ is the right eigenvector of the transformed RPA problem corresponding to the left eigenvector $\bX^\dagger$, where the superscript $^\dagger$ denotes the Hermitian conjugate.

This block-diagonalization reduces the dimensionality of the original RPA problem by eliminating the coupling between excitations and deexcitations. 
Additionally, it establishes an exact relationship between the original RPA eigenvectors, $\bX$ and $\bY$, the left and right eigenvectors, $\bX$ and $\bar{\bX}$ of the resulting non-Hermitian matrix, and the drCCD parameters, $\bt$ and $\blam$. 

It should be noted that the second similarity transformation is only needed to fully block diagonalize the RPA matrix, but the RPA problem can be represented in a reduced dimensionality using only the drCCD amplitudes [see Eqs.~\eqref{eq:simM} and \eqref{eq:simMR}].
Specifically, \eqrref{eq:casida_simple_blockdiag} can be alternatively derived using an excitation-energy EOM-CC treatment.\cite{Berkelbach_2018,Rishi_2020} An EOM-singles (EOM-S) ansatz is used on the drCCD Hamiltonian, which yields the following right and left eigenvalue problems
\begin{subequations} \label{eq:eigenval_equations}
	\begin{align}
		\bra{\Phi^a_i} \comm{\wtH_\dr }{ \hR_{1,\mu}} \ket{\Phi_0} & = \Omega_\mu \bra{\Phi^a_i}\hR_{1,\mu}\ket{\Phi_0}, \label{eq:eom_s_right} \\ 
		\bra{\Phi_0}\hL_{1,\mu}(\wtH_\dr-\Ec^\drCCD)  \ket{\Phi_i^a} &=  \bra{\Phi_0}\hL_{1,\mu}\ket{\Phi_i^a}\Omega_\mu.\label{eq:eom_s_left}
	\end{align}
\end{subequations}
These can be rewritten in matrix form as
\begin{subequations} \label{eq:eigenval_equations}
	\begin{align}
		\qty( \bA + \bt \cdot \bB^* ) \cdot \bR_1 = \bR_1 \cdot \bOm, \\
		\bm{L}_1 \cdot \qty( \bA + \bt \cdot \bB^* ) = \bOm \cdot \bm{L}_1. 
	\end{align}
\end{subequations}
In this way, the right-EOM operator, $\hR_{1,\mu}  = \sum_{ia} r_{i,\mu}^a \cre{a}\ani{i}$, can be identified to the matrix $\bar{\bX}$, and the left-EOM operator, $\hL_{1,\mu} = \sum_{ia} l_{a,\mu}^i \cre{i}\ani{a}$, to the matrix ${\bX}^\dagger$. 
Note that the parametrization in \eqrref{eq:RPA_TYX-1} directly influences the identification of the EOM eigenvectors to the RPA parameters. 
Previous formulations in the literature\cite{Berkelbach_2018,Rishi_2020} are unclear whether $\hT^\dagger_2$ is used instead of $\hT_2$. 
This hinders the correct identification of left and right eigenvectors, and numerical differences are not limited to a complex conjugation. 
Specifically, the right-EOM vector is associated with $\bX$ instead of $\bar{\bX}$ and the other way around for the left-EOM vector, when $\hT^\dagger_2$ is employed.

\subsection{$G_0W_0$ through the EOM-CC lens}
\label{subsec:GWEOM}
As we shall discuss in detail in a forthcoming publication, the double similarity transformation described in Sec.~\ref{subsec:blockdiag} corresponds to the direct-ring approximation of an extended coupled-cluster doubles (ECCD) treatment \cite{Arponen_1983,Arponen_1982,Arponen_1983,Arponen_1987a,Arponen_1987b} 
\begin{align}
	\wtH_\text{ECCD} & = e^{\hat{\Lambda}_2} e^{-\hT_2} \hat{H}_\text{N} e^{\hT_2} e^{-\hat{\Lambda}_2} .
\end{align}
The resulting approximated Hamiltonian, referred to here as the $\ldr$CCD transformation, is given by
\begin{equation}
	\wtH_\text{ECCD}\approx \wtH_\ldr = \wtH_\dr + \comm{ \hLam_2}{\wtH_\dr}, \label{eq:H_lamdar}
\end{equation}
where the second similarity-transformation truncates exactly after the first-order commutator within the direct-ring approximation. 
The $\lambda$-direct-ring transformation includes all deexcitations (i.e., backward time-ordered bubble diagrams) that are missing in the standard drCCD similarity transformation compared to a unitary transformation, as discussed in Sec.~\ref{subsec:blockdiag}.

\subsubsection{The IP/EA-EOM-$\lambda$-drCCD expectation value}
It has been shown that the quasiparticle equation [see Eq.~\eqref{eq:GW_std}] can be recast as an eigenvalue problem resulting from an IP/EA equation-of-motion (IP/EA-EOM) approach acting on a block-diagonalized Hamiltonian, as mentioned in Sec.~\ref{subsec:blockdiag}. \cite{Bintrim_2021,Quintero_2022}
Here, we make use of $\wtH_\ldr$, as defined in Eq.~\eqref{eq:H_lamdar}.
Specifically, the excitation space is chosen as the singles and doubles IP excitations (i.e., 1h and 2h1p, respectively) and EA deexcitations (i.e., 1p and 2p1h, respectively)
\begin{equation}
\begin{split}
    \hR^\IPEA 
		& = \sum_I R_I^{\IPEA} \hat{c}^{\IPEA}_I = \sum_I R_I^{\IP} \hat{c}^{\IP}_I + \sum_I R_I^{\EA} \left( \hat{c}^{\EA}_I\right)^\dagger \\
        & = R_1^\IP \hat{c}^{\IP}_1 + R_2^\IP \hat{c}^{\IP}_2 + R^\EA_1 \qty( \hat{c}^{\EA}_1)^\dagger + R^\EA_2 \qty( \hat{c}^{\EA}_2)^\dagger
        \\
        & = \sum_i r_i \ani{i} + \sum_{ijb} r_{ji}^b \cre{b} \ani{j} \ani{i}
        \\
        & + \sum_a r_a \ani{a}  + \sum_{ajb} r_{ba}^j \cre{j} \ani{b} \ani{a}.
\end{split}
\end{equation}
Here, the collective operator $\hat{c}^\IPEA_I$ correspond to either the IP excitation operator, $\hat{c}^{\IP}_I$, or EA deexcitation operator, $\left (\hat{c}^{\EA}_I\right )^\dagger$. They create the determinants involved in the IP/EA-EOM formalism when acting on the reference determinant, $\hat{c}_I \ket{\Phi_0} = \ket{\Phi_I}$, with
\begin{equation}
	\{ \ket{\Phi_I} \} = \{ \ket{\Phi_i}, \ket{\Phi^a}, \ket{\Phi^{ba}_j}, \ket{\Phi^b_{ji}} \}.
\end{equation}
In addition, the index pair $bj$ in the doubles amplitudes $r_{ji}^b$ and $r_{ba}^j$ follows the ring approximation, whereas the remaining index does not. 
As a result, these indices are formally nonequivalent, in contrast to the standard EOM-CC formulation.
The resulting right and left non-Hermitian eigenvalue problems read
\begin{gather}
	  \bra{\Phi_0}  \comm{  \hat{c}^{\IPEA,\dagger}_I } { \comm{ \wtH_{\ldr} }{ \hR^\IPEA } }_+ \ket{\Phi_0} \nonumber  = \\
	   \epsilon^\IPEA \bra{\Phi_0} \comm{\hat{c}^{\IPEA,\dagger}_I}{\hR^\IPEA}_{+} \ket{\Phi_0} \label{eq:EOM_right}
\end{gather}
and 
\begin{gather}
	\bra{\Phi_0}  \comm{ \hL^\IPEA  } { \comm{ \wtH_{\ldr} }{\hat{c}^{\IPEA}_I } }_+ \ket{\Phi_0} \nonumber  = \\
	\epsilon^\IPEA \bra{\Phi_0} \comm{\hL^\IPEA }{\hat{c}^{\IPEA}_I}_{+} \ket{\Phi_0}. \label{eq:EOM_left}
\end{gather}
The anticommutator is represented by $\comm{\cdot}{\cdot}_+$ and the eigenvalues by $\{\epsilon^\IPEA\}$.

The total energy of the electron-attached or -detached state is expressed as the sum of the HF reference energy and the correlation energy at the drCCD level, augmented by the corresponding quasiparticle energy:
\begin{equation} \label{eq:IPEA_totene}
	E^\IPEA_p = E^\HF_0 + \Ec^\drCCD \pm \epsilon^\IPEA. 
\end{equation}
This result follows naturally from the EOM-CC formalism and coincides exactly with the definition given in Eq.~\eqref{eq:G0W0_totene}.

The eigenvalues can also be expressed in an expectation-value form, as follows:
\begin{align}
    \epsilon^\IPEA = \bra{\Phi_0} \comm{  \hL^\IPEA }{ \comm{ \wtH_{\ldr} }{ \hR^\IPEA } }_+ \ket{\Phi_0}.
\end{align}
By enforcing a decoupling in the 1h and 1p spaces, the EOM vector for the collective IP and EA singles block reduces to having only one non-zero component
\begin{align}
    \hR^{\text{IP/EA}}_{1,p}\approx \sum_q r_q \ani{q} \delta_{pq} = r_p \ani{p}. \label{eq:diagonal_r}
\end{align} 
This treatment corresponds to the diagonal approximation of $G_0W_0$. The index $p$ explicitly enumerates the solution for clarity.
Within this approximation, the $\GOWO$ quasiparticle energies exactly correspond to the IP/EA-EOM eigenvalues as $\epsilon_p^\IPEA=-\epsilon_p^{G_0W_0}$. 

It is again noted that the use of $\wtH_\dr$ instead of $\wtH_\ldr$ would not reproduce the $G_0W_0$ quasiparticle energies, as it omits certain contributions from the deexcitation space.\cite{Tolle_2022} 
These missing terms are recovered through the second similarity transformation involving the deexcitation operator $\hLam_2$, which effectively eliminates the remaining off-diagonal block of the RPA matrix. 

Details regarding the working equations can be found in Appendix \ref{app:ipea_eom_lrCCD}. 
Additionally, exact relations between the $\GOWO$ and IP/EOM-$\ldr$CCD parameters can be found in Appendix \ref{app:connection}. 

\subsubsection{The Lagrangian treatment of $G_0W_0$}

To derive an analytic expression for the derivatives of the IP/EA-EOM eigenvalues, we introduce the following Lagrangian:
\begin{equation} \label{eq:IPEA_Lagrange}
\begin{split}
    \cL^\IPEA =
    & \pm \bra{\Phi_0} \comm{  \hL^\IPEA }{ \comm{ \wtH_{\ldr} }{ \hR^\IPEA } }_+ \ket{\Phi_0} 
    \\
    & \mp \epsilon^\IPEA \qty( \bra{\Phi_0} \comm{\hL^\IPEA }{ \hR^\IPEA }_+ \ket{\Phi_0} - 1 ) 
    \\
    & + \bra{\Phi_0} \qty( 1 + \hZ_2 ) \wtH_\dr \ket{\Phi_0} 
    \\
    & + \bra{\Phi_0} \qty( 1 + \hLam_2 ) \comm{\wtH_\dr }{ \hXi_2 } \ket{\Phi_0},
\end{split}
\end{equation}
which, by construction, reproduces the total correlation energy of the electron-detached or -attached state
\begin{equation}
    \Ec^\IPEA = \Ec^\drCCD \pm \epsilon^\IPEA
\end{equation}
from \eqrref{eq:IPEA_totene}. 
The reference correlation energy contribution is contained in the third term of the Lagrangian.
Regarding the sign of the first term in the Lagrangian, the EOM expectation value is derived as an excitation, and, as such, is added for the IP case. 
For the EA case, on the other hand, the expectation value corresponds to a deexcitation and must be subtracted.

The Lagrange multipliers introduced in Eq.~\eqref{eq:IPEA_Lagrange} are as follows: (i) the EOM eigenvalue $\epsilon^\IPEA$ enforces the biorthonormality condition between the left- and right-EOM vectors in the second term; (ii) the deexcitation operator
\begin{equation}
	\hZ_2 
	= \frac{1}{4} \sum_{ijab} \zeta^{ij}_{ab} \cre{i} \cre{j} \ani{b} \ani{a} 
\end{equation}
includes the drCCD-amplitude equations for $\hT_2$ [see \eqrref{eq:rCCD_ampl}] in the third term; and (iii) the excitation operator 
\begin{equation}
	\hXi_2 
	= \frac{1}{4} \sum_{ijab} \xi_{ij}^{ab} \cre{a} \cre{b} \ani{j} \ani{i} 
\end{equation}
the $\lambda$-amplitude equations for $\hLam_2$ [see \eqrref{eq:RPA_lambdaamps}] in the fourth term.

Enforcing the stationarity conditions on the Lagrangian [see Eq.~\eqref{eq:IPEA_Lagrange}] results in the following set of equations:
\begin{enumerate}
    \item Stationarity with respect to the Lagrange multipliers $\zeta^{ij}_{ab}$, 
    \begin{equation}
        \pdv{\cL^\IPEA}{\zeta^{ij}_{ab}} \weq 0,
    \end{equation}
    reproduces the pseudo-linear drCCD amplitude equations [see Eq.~\eqref{eq:rCCD_ampl}].
    \item Stationarity with respect to the Lagrange multipliers $\xi_{ij}^{ab}$, 
    \begin{equation}
        \pdv{\cL^\IPEA}{\xi^{ij}_{ab}} \weq 0,
    \end{equation}
    reproduces the linear drCCD $\lambda$-amplitude equations [see Eq.~\eqref{eq:RPA_lambdaamps}].
    \item Stationarity with respect to the left-EOM amplitudes $l_I$,
    \begin{equation}
    	\pdv{\cL^\IPEA}{l_{I}} \weq 0,
    \end{equation}
    reproduces the right eigenvalue problem [see \eqrref{eq:EOM_right}].
    \item Stationarity with respect to the right-EOM amplitudes $r_I$,
    \begin{equation}
        \pdv{\cL^\IPEA}{r_{I}} \weq 0,
    \end{equation}
    reproduces the left eigenvalue problem [\eqrref{eq:EOM_left}].
    \item Stationarity with respect to the Lagrange multiplier $\epsilon^\IPEA$,
    \begin{equation}
    	\pdv{\cL^\IPEA}{\epsilon^\IPEA} \weq 0,
    \end{equation}
    reproduces the biorthonormality condition
    \begin{equation}
    	\bra{\Phi_0} \comm{\hL^\IPEA }{ \hR^\IPEA }_+\ket{\Phi_0} = 1.
    \end{equation}
    \item Stationarity with respect to the $\lambda$-amplitudes, 
    \begin{equation}
        \pdv{\cL^\IPEA}{\lambda^{ij}_{ab}} \weq 0,
    \end{equation}
    yields the inhomogeneous linear $\xi$ amplitude equations
    \begin{multline}\label{eq:ksi_amps}
		\mp\bra{\Phi_0} \comm{ \hL^\IPEA }{ \comm{ \pdv{\wtH_{\ldr}}{\lambda^{ij}_{ab}} }{ \hR^\IPEA } }_+\ket{\Phi_0} =
		\\ 
		\bra{\Phi_{ij}^{ab}} \comm{ \wtH_\dr }{ \hXi_2 }\ket{\Phi_0}.
    \end{multline}
    \item Stationarity with respect to the drCCD amplitudes 
    \begin{equation}
        \pdv{\cL^\IPEA}{t_{ij}^{ab}} \weq 0
    \end{equation}
    yields the inhomogeneous linear $\zeta$ amplitude equations
    \begin{multline} \label{eq:zeta_amps}
		\mp\bra{\Phi_0} \comm{ \hL^\IPEA }{ \comm{ \pdv{\wtH_{\ldr}}{t_{ij}^{ab}} }{ \hR^\IPEA } }_+ \ket{\Phi_0} 
		\\
		- \bra{\Phi_0} \qty(1 + \hLam_2 ) \comm{ \pdv{\wtH_\dr}{t_{ij}^{ab}} }{ \hXi_2 } \ket{\Phi_0} =
		\bra{\Phi_0} \qty(1 + \hZ_2 ) \pdv{\wtH_\dr}{t_{ij}^{ab}} \ket{\Phi_0}.
    \end{multline}
\end{enumerate}
Since \eqrref{eq:zeta_amps} includes contributions from the $\xi$ amplitudes, the set of linear equations in \eqrref{eq:ksi_amps} has to be solved first, and is thus decoupled.
The diagrammatic representation and the derivation of working equations are presented in Appendix \ref{app:ipea_eom_lrCCD}.

Compared to the drUCCD approach employed in Ref.~\onlinecite{Tolle_2025}, the $\ldr$CCD formulation avoids the need for a numerical truncation of the otherwise non-terminating Baker-Campbell-Hausdorff (BCH) expansion inherent to the unitary CC framework. 
This advantage, however, comes at the cost of introducing additional sets of wavefunction parameters, namely, the left-EOM vector, the $\lambda$ amplitudes, and the $\xi$ amplitudes.

After recasting \eqrref{eq:IPEA_Lagrange} as a function of the contributions of the one- and two-electron reduced density matrices,
\begin{equation}
    \cL^\IPEA = \sum_{pq} \gamma_{pq} f_{pq} + \frac{1}{2} \sum_{pqrs} \Gamma_{pqrs} \ERI{pq}{rs},
\end{equation}
and a subsequent Hermitization, i.e., 
\begin{equation}
    \bar{\cL}^\IPEA = \frac{\cL^\IPEA + (\cL^\IPEA)^*}{2},
\end{equation}
to derive Hermitian reduced densities 
\begin{align}
    \bar{\gamma}_{pq} & = \frac{ {\gamma}_{pq} + {\gamma}_{qp}^*}{2},
    &
    \bar{\Gamma}_{pqrs} & = \frac{ {\Gamma}_{pqrs} + {\Gamma}_{rspq}^*}{2},
\end{align}
the $z$-vector method can be applied to account for the orbital relaxation. \cite{Handy_1984}
The resulting Lagrangian is 
\begin{equation}
	{\cL}^\IPEA_\text{rel} 
	= \bar{\cL}^\IPEA 
	+ \sum_{pq}w_{pq} \qty( f_{pq} - \delta_{pq} \epsilon_p ) 
	+ \sum_{pq}I_{pq} \qty( S_{pq} - \delta_{pq} ).
\end{equation}
The Lagrange multipliers $w_{pq}$ and $I_{pq}$ are introduced for the diagonal form of the Fock matrix and the orbital orthonormality $S_{pq} = \braket{\phi_p|\phi_q} = \delta_{pq}$, respectively. 
They can be assumed Hermitian with no loss of generality. 
The diagonal approximation that results in the decoupling of the singles-singles block in the EOM problem implies that the method is not invariant under rotations within the occupied-occupied and virtual-virtual orbital subspaces. 
As a consequence, the $w_{pq}$ amplitudes acquire non-zero components in these blocks, even in the absence of a frozen-core approximation. 
Working equations for the evaluation of $w_{pq}$ and $I_{pq}$ are provided in Ref.~\onlinecite{Kitsaras_thes}.

Incorporating the $w_{pq}$ contributions in the one-body reduced density matrix, $\gamma_{pq}^\text{rel} = \bar{\gamma}_{pq} + w_{pq}$, yields a final expression for the derivative of the energy with respect to a parameter $\kappa$ that depends only on the partial derivatives of the mean-field Fock matrix elements, the two-electron integrals, and the overlap integrals:
\begin{equation}  
\begin{split}
    \dv{E^\IPEA}{\kappa} 
    & = \pdv{\cL^\IPEA}{\kappa} 
    \\
    & = \sum_{pq} \gamma_{pq}^\text{rel} \pdv{f_{pq}}{\kappa}  + \sum_{pq} I_{pq} \pdv{S_{pq}}{\kappa} 
    + \frac{1}{2} \sum_{pqrs}\bar{\Gamma}_{pqrs} \pdv{\ERI{pq}{rs}}{\kappa}.
\end{split}
\end{equation}
Different choices for the parameter $\kappa$ lead to expressing first-order properties as analytic first-order derivatives of the energy. In the case of $\kappa$ being nuclear displacements, geometrical gradients are derived to use in a geometry optimization calculation.

\begin{table*}[!ht]
	\begin{ruledtabular}
		
		\caption{Vertical and adiabatic ionization potentials (in \si{\eV}) using the aug-cc-pVTZ basis set. 
			The IRREP of the HOMO is reported. 
			For the vertical IPs, the optimal geometry of the neutral molecule at the respective level of theory has been used. 
			For the adiabatic IPs, the energy of the ionized state at the optimal geometry calculated at the respective level of theory has been subtracted from the optimal energy of the neutral molecule at the respective level of theory. 
			The MAE and MSE are reported with respect to the IP-EOM-CC3 results.}
		\label{tab:IP_va}
		\begin{tabular}{lcrrrrrrrrrr}
			& & \multicolumn{5}{c}{Vertical IPs} & \multicolumn{5}{c}{Adiabatic IPs}  \\
			\cline{3-7} \cline{8-12}
			\multicolumn{1}{c}{System} & \multicolumn{1}{c}{HOMO}  & \multicolumn{1}{c}{ADC(2)} & \multicolumn{1}{c}{CC2} & \multicolumn{1}{c}{$\GOWO$} & \multicolumn{1}{c}{CCSD} & \multicolumn{1}{c}{CC3}  & \multicolumn{1}{c}{ADC(2)} & \multicolumn{1}{c}{CC2} & \multicolumn{1}{c}{$\GOWO$} & \multicolumn{1}{c}{CCSD} & \multicolumn{1}{c}{CC3}  \\ \hline
			\ce{H2} & $\sigma_g^+$ & 16.350 & 16.330 & 18.036 & 16.397 & 16.397                      & 15.415 & 15.387 & 15.621 & 15.520 & 15.520  \\
			\ce{LiH} & $\sigma^+$ & 8.024 & 8.002 & 8.233 & 8.005 & 8.005                          & 7.783 & 7.757 & 8.024 & 7.728 & 7.729 \\ 
			\ce{BH3} & $e'$& 13.290 & 13.251 & 13.716 & 13.353 & 13.314                     & 12.223 & 12.193 & 12.620 & 12.315 & 12.285 \\ 
			\ce{Li2} & $\sigma_g^+$& 5.178 & 5.155 & 5.348 & 5.235 & 5.235                          & 5.073 & 5.057 & 5.240 & 5.121 & 5.121 \\ 
			\ce{CH4} & $t_2$ & 14.115 & 14.056 & 14.797 & 14.404 & 14.369                     & 12.490 & 12.428 & 13.110 & 12.825 & 12.758 \\ 
			\ce{NH3} & $a'$ & 10.185 & 10.163 & 11.162 & 10.857 & 10.891                     & 9.441 & 9.416 & 10.414 & 10.121 & 10.135 \\ 
			\ce{H2O} & $b_1$ & 11.519 & 11.542 & 12.916 & 12.615 & 12.666                     & 11.328 & 11.347 & 12.841 & 12.516 & 12.565 \\ 
			\ce{HF} & $\pi$ & 14.643 & 14.719 & 16.273 & 16.045 & 16.132      & 14.320 & 14.389 & 16.154 & 15.896 & 15.976 \\
			\ce{BN} & $\pi$ & 10.812 & 10.856 & 11.769 & 12.005 & 12.071                      & 10.700 & 10.730 & 11.722 & 11.892 & 11.914 \\ 
			\ce{BeO} & $\pi$ & 8.237 & 7.784 & 9.976 & 10.041 & 9.744                         & 7.472 & 6.982 & 9.768 & 9.897 & 8.755 \\ 
			\ce{LiF} & $\pi$ & 9.665 & 9.773 & 11.432 & 11.427 & \fnm[2]11.391                       & 8.785 & \fnm[1]  & 10.965 & 10.980 & \fnm[2]10.892 \\ 
			\ce{CO} & $\sigma^+$ & 14.000 & 13.766 & 14.721 & 14.179 & 13.866                      & 13.990 & 13.762 & 14.685 & 14.159 & 13.860 \\ 
			\ce{N2} & $\pi_u$ & 17.108 & 16.999 & 17.267 & 17.314 & 16.851                      & 16.899 & 16.771 & 16.963 & 17.051 & 16.461 \\ 
			\ce{BF} & $\sigma_g^+$& 11.024 & 10.955 & 11.266 & 11.235 & 11.102                      & 10.931 & 10.864 & 11.165 & 11.130 & 11.012 \\ 
			\ce{H2S} & $b_1$ & 10.179 & 10.148 & 10.508 & 10.414 & 10.381                     & 10.164 & 10.132 & 10.503 & 10.404 & 10.370 \\ 
			\ce{HCl}   & $\pi$ & 12.408 & 12.389 & 12.789 & 12.704 & 12.664                   & 12.373 & 12.355 & 12.772 & 12.680 & 12.641 \\ 
			\ce{F2} & $\pi_g$ & 14.124 & 14.156 & 16.122 & 15.562 & 15.756                      & 14.101 & 14.149 & 15.854 & 15.397 & 15.591 \\ 
			\hline
			MAE &  &0.635 & 0.652 & 0.359 & 0.107 &  & 0.667 & 0.603 & 0.307 & 0.163  &  \\ 
			MSE &  & -0.587 & -0.635 & 0.323 & 0.056  &  & -0.594 & -0.561 & 0.284 & 0.120 & \\
		\end{tabular}
		\fnt[1]{A geometry optimization for \ce{LiF+} at the IP-EOM-CC2 level was not possible due to the deteriorating quality of the reference CC2 wavefunction for interatomic distances near the expected optimal geometry for the \ce{LiF+} cation. In addition, the LiH results have been excluded when calculating the MAE and MSE in all cases due to apparent qualitative differences between the different levels of theory. }
		\fnt[2]{Prediction at the IP-EOM-CC3 level for the \ce{LiF+} are non consistent. For this reason, predictions at the CCSDT level were used. Further details can be found in the \SupInf.  }
	\end{ruledtabular}
\end{table*}

\begin{table*}[!ht]
	\begin{ruledtabular}
		
		\caption{Interatomic distance (in \si{\angstrom}) in the optimal geometry for the neutral $R_\text{neu}$ and cationic $R_\text{cat}$ linear molecules using the aug-cc-pVTZ basis set. The MAE and MSE are reported with respect to the CC3 and IP-EOM-CC3 results.}
		\label{tab:R_diat}
		\begin{tabular}{lrrrrrrrrrr}
			&  \multicolumn{5}{c}{$R_\text{neu}$} &  \multicolumn{5}{c}{$R_\text{cat}$} \\
			\cline{2-6} \cline{7-11}
			\multicolumn{1}{c}{System}  & \multicolumn{1}{c}{ADC(2)} & \multicolumn{1}{c}{CC2} & \multicolumn{1}{c}{$\GOWO$} & \multicolumn{1}{c}{CCSD} & \multicolumn{1}{c}{CC3} & \multicolumn{1}{c}{ADC(2)} & \multicolumn{1}{c}{CC2} & \multicolumn{1}{c}{$\GOWO$} & \multicolumn{1}{c}{CCSD} & \multicolumn{1}{c}{CC3} \\ \hline
			\ce{H2} & 0.7374 & 0.7377 & 0.7354 & 0.7430 & 0.7430 & 1.0627 & 1.0657 & 1.0578 & 1.0581 &  1.0581  \\ 
			\ce{LiH} & 1.5879 & 1.5883 & 1.5719 & 1.5924 & 1.5922 & 2.0544 & 2.0599 & 1.9661 & 2.1635 & 2.1633 \\ 
			\ce{Li2} & 2.7125 & 2.7117 & 2.6725 & 2.6623 & 2.6600 & 3.1636 & 3.1431 & 3.1278 & 3.1076 & 3.1030 \\  
			\ce{HF} & 0.9201 & 0.9225 & 0.9097 & 0.9164 & 0.9196 & 1.0671 & 1.0753 & 0.9799 & 0.9989 & 1.0075 \\ 
			\ce{BN} & 1.3212 & 1.2972 & 1.2700 & 1.2713 & 1.2765 & 1.3691 & 1.3657 & 1.3058 & 1.3362 & 1.3489 \\
			\ce{BeO} & 1.3452 & 1.4101 & 1.3077 & 1.3200 & 1.3621 & 1.6012 & 1.6097 & 1.4026 & 1.3975 & 1.6061 \\ 
			\ce{LiF} & 1.5817 & 1.5880 & 1.5641 & 1.5722 &  \fnm[2]1.5762 & 2.7302 & \fnm[1] & 1.9711 & 1.9600 &  \fnm[2]2.0344\\
			\ce{CO} & 1.1342 & 1.1442 & 1.1150 & 1.1241 & 1.1335 & 1.1203 & 1.1351 & 1.0923 & 1.1067 & 1.1234 \\ 
			\ce{N2} & 1.1097 & 1.1167 & 1.0830 & 1.0929 & 1.1007 & 1.1712 & 1.1853 & 1.1489 & 1.1560 & 1.1869 \\ 
			\ce{BF} & 1.2640 & 1.2699 & 1.2521 & 1.2629 & 1.2686 & 1.2109 & 1.2163 & 1.1994 & 1.2075 & 1.2160 \\ 
			\ce{HCl}   & 1.2710 & 1.2717 & 1.2649 & 1.2728 & 1.2751 & 1.3212 & 1.3222 & 1.2979 & 1.3138 & 1.3156 \\ 
			\ce{F2} & 1.3977 & 1.4152 & 1.3641 & 1.3918 & 1.4137 & 1.3545 & 1.3872 & 1.2608 & 1.2997 & 1.3145 \\ 
			\hline
			MAE & 0.0137 & 0.0148 & 0.0196 & 0.0087  & & 0.0853 & 0.0302 & 0.0597 & 0.0318  & \\
			
			MSE & 0.0051 & 0.0127 & -0.0176 & -0.0083  &  & 0.0624 & 0.0111 & -0.0556 & -0.0310 &  \\
			
		\end{tabular}
		\fnt[1]{A geometry optimization for \ce{LiF+} at the IP-EOM-CC2 level was not possible due to the deteriorating quality of the reference CC2 wavefunction for interatomic distances near the expected optimal geometry for the \ce{LiF+} cation. In addition, the LiH results have been excluded when calculating the MAE and MSE in all cases due to apparent qualitative differences between the different levels of theory.} 
		\fnt[2]{Prediction at the IP-EOM-CC3 level for the \ce{LiF+} are non consistent. For this reason, predictions at the CCSDT level were used. Further details can be found in the \SupInf.  }
	\end{ruledtabular}
\end{table*}

\begin{table}[!ht]
	\centering
	\begin{ruledtabular}
		
		\caption{Geometric parameters in the optimal geometry for the neutral and cationic polyatomic molecules using the aug-cc-pVTZ basis set.
			Distances are given in \si{\angstrom} and angles in \si{\degree}.
			The point group (PG) of the molecular symmetry is reported in the last column.}
		\label{tab:R_polyat}
		\begin{tabular}{lrrrrrcc}
			
			System & \multicolumn{1}{c}{ADC(2)} & \multicolumn{1}{c}{CC2} & \multicolumn{1}{c}{$\GOWO$} & \multicolumn{1}{c}{CCSD} & \multicolumn{1}{c}{CC3} & \multicolumn{1}{c}{Param.} & \multicolumn{1}{c}{PG} \\ \hline 
			\ce{BH3} & 1.1811 & 1.1814 & 1.1756 & 1.1840 & 1.1848 & $R$ & $\mathcal{D}_{3{h}}$ \fnm[1] \\ \hline
			\multirow{ 3}{*}{\ce{BH3+}} & 1.4140 & 1.4096 & 1.4420 & 1.4395 & 1.4365 & $R_1$ & \multirow{ 3}{*}{$\mathcal{C}_{2{v}}$\fnm[2]} \\ 
			~ & 1.1749 & 1.1765 & 1.1636 & 1.1776 & 1.1801 & $R_2$ & ~ \\ 
			~ & 164.62 & 164.71 & 163.07 & 163.07 & 163.38 & $\omega$ & ~ \\ \hline 
			\ce{CH4} & 1.0841 & 1.0846 & 1.0808 & 1.0863 & 1.0879 & $R$ & $\mathcal{T}_{{d}}$\fnm[3] \\ \hline
			\multirow{ 4}{*}{\ce{CH4+}} & 1.1760 & 1.1781 & 1.1746 & 1.1782 & 1.1838 & $R_1$ & \multirow{ 4}{*}{$\mathcal{C}_{2{v}}$\fnm[4]} \\ 
			~ & 55.15 & 55.04 & 56.08 & 56.14 & 55.34 & $\omega_1$ & ~ \\ 
			~ & 1.0796 & 1.0812 & 1.0670 & 1.0783 & 1.0805 & $R_2$ & ~ \\ 
			~ & 125.43 & 125.42 & 125.33 & 125.01 & 125.54 & $\omega_2$ & ~ \\ \hline
			\multirow{ 2}{*}{\ce{NH3}} & 1.0095 & 1.0107 & 1.0024 & 1.0095 & 1.0122 & $R$ & \multirow{ 2}{*}{$\mathcal{C}_{3{v}}$\fnm[5]} \\ 
			~ & 111.80 & 111.94 & 111.49 & 111.90 & 112.15 & $\omega$ & ~ \\ \hline
			\ce{NH3+} & 1.0275 & 1.0293 & 1.0068 & 1.0187 & 1.0209 & $R$ & $\mathcal{D}_{3{h}}$\fnm[6] \\ \hline
			\multirow{ 2}{*}{\ce{H2O}} & 0.9588 & 0.9607 & 0.9484 & 0.9561 & 0.9591 & $R$ & \multirow{ 2}{*}{$\mathcal{C}_{2{v}}$\fnm[7]} \\ 
			~ & 104.27 & 104.06 & 105.01 & 104.59 & 104.33 & $\omega$ & ~ \\ \hline
			\multirow{ 2}{*}{\ce{H2O+}} & 1.0245 & 1.0285 & 0.9798 & 0.9972 & 1.0009 & $R$ & \multirow{ 2}{*}{$\mathcal{C}_{2{v}}$\fnm[7]} \\ 
			~ & 108.73 & 108.20 & 110.59 & 109.38 & 109.38 & $\omega$ & ~ \\ \hline
			\multirow{ 2}{*}{\ce{H2S}} & 1.3319 & 1.3326 & 1.3264 & 1.3351 & 1.3375 & $R$ & \multirow{ 2}{*}{$\mathcal{C}_{2{v}}$\fnm[8]} \\ 
			~ & 91.93 & 91.87 & 92.51 & 92.27 & 92.00 & $\omega$ & ~ \\ \hline
			\multirow{ 2}{*}{\ce{H2S+}} & 1.3558 & 1.3575 & 1.3394 & 1.3550 & 1.3579 & $R$ & \multirow{ 2}{*}{$\mathcal{C}_{2{v}}$\fnm[8]} \\ 
			~ & 92.69 & 92.63 & 93.29 & 92.99 & 92.71 & $\omega$ & ~ \\ 
		\end{tabular}
		\fnt[1]{Planar symmetric. $R$ is the \ce{B-H} distance.}
		\fnt[2]{Planar with one nonequivalent \ce{B-H} bond. $R_1$ is the nonequivalent \ce{B-H} distance, $R_2$ is the \ce{B-H} distance of the equivalent \ce{H} atomic centers, $\omega$ is the angle \ce{H-B-H} angle between the equivalent \ce{H} atomic centers.} 
		\fnt[3]{Tetrahedral symmetric. $R$ is the \ce{C-H} distance.}
		\fnt[4]{Two pairs of equivalent \ce{H} atomic centers. $R_1$ is the \ce{C-H} distance of the first pair, $\omega_1$ is the \ce{H-C-H} angle of the first pair, $R_2$ is the \ce{C-H} distance of the second pair, $\omega_2$ is the \ce{H-C-H} angle of the second pair.}
		\fnt[5]{$R$ is the \ce{N-H} distance and $\omega$ is the angle of the \ce{N-H} with respect to the $C_3$ rotational axis.}
		\fnt[6]{Planar symmetric. $R$ is the \ce{N-H} distance.}
		\fnt[7]{$R$ is the \ce{O-H} distance, $\omega$ is the \ce{H-O-H} angle. The symmetry does not change between the neutral and cationic systems.}
		\fnt[8]{$R$ is the \ce{S-H} distance, $\omega$ is the \ce{H-S-H} angle. The symmetry does not change between the neutral and cationic systems.}
	\end{ruledtabular}
\end{table}
\section{Computational details}
For the purpose of the current work, the drCCD approximation (energy and fully analytic properties), and the $G_0W_0$ approximation through the IP/EA-EOM-$\lambda$-drCCD formulation (energy and fully analytic properties) were implemented in the \textsc{qcumbre} program package.\cite{qcumbre} 
The integrals and the SCF solution are provided by the \textsc{mint} module \cite{mint} of the \textsc{cfour} program package.\cite{Matthews_2020,cfour} 
The geometry optimizer of \textsc{cfour} was also used to get the equilibrium geometries. 

The verification of the RPA and $G_0W_0$ energies was achieved through comparison to the \textsc{quack} program package.\cite{QuAcK} 
Analytic one-electron properties were tested in the case of dipole moments against numerical derivatives with finite electric-field calculations. 
Accordingly, analytic geometric gradients were tested against numerical differentiation by finite displacements of the nuclear coordinates, as well as the implementation reported in Ref.~\onlinecite{Tolle_2025}.

The implemented method was applied on the $GW$20 set \cite{Lewis_2019,Loos_2020a,Monino_2023} (composed by the 20 smallest systems of the $GW$100 set \cite{vanSetten_2015}) where the He, Ne, and Ar atomic systems were excluded. 
Results were generated in different levels of theory for comparison and benchmarking purposes. Specifically, the optimal geometry of the neutral system was calculated at the ground-state ADC(2), CC2, $G_0W_0$, CCSD, and CC3 levels of theory. 
For the purposes of the current study, only closed-shell configurations were targeted, even if for some systems a triplet state may be lower in energy. 
To target the lowest electronic state of the cationic systems, calculations were performed at the excited-state ADC(2), EOM-CC2, $\GOWO$, EOM-CCSD, and EOM-CC3 levels of theory. 
For EOM-CCSD, the corresponding IP-EOM-CC implementation was used, while excitations in continuum orbitals were probed to target the respective cationic states in the case of ADC(2), EOM-CC2, and EOM-CC3. \cite{Stanton_1999} 
All electrons were included for the calculation of correlation effects using the aug-cc-pVTZ basis set.\cite{dunning1989a,kendall1992a,prascher2011a,woon1993a} 
The scheme to generate $\GOWO$ quasiparticle energies, as described in the previous paragraphs, corresponds to an $\eta$ parameter of $0$ on top of canonical restricted HF orbitals. 
In the following discussion, the RPA ground-state energy for the neutral system calculated via the drCCD scheme can be assumed when referring to $\GOWO$ results. 

All calculations were performed using the \textsc{qcumbre}\cite{qcumbre} and \textsc{cfour} program packages.\cite{Matthews_2020,cfour,mint} 
A double convergence criterion of $10^{-7} \Eh$ for the energy difference and $10^{-7}$ for the DIIS error was used for the HF calculation. Accordingly, a $10^{-7}$ criterion was used for the norm of the residual vector in the linear solver for the drCCD and $\lambda$ amplitudes, and in the Davidson algorithm for the excited-state calculations. 
For the geometry optimizations, a convergence criterion of $10^{-6} \Eh a_0^{-1} $ was used for the gradient norm in the quasi-Newton algorithm.

\section{Results}
In the following paragraphs, results regarding the closed-shell states of the neutral systems and the corresponding lowest in energy states of the cationic systems in the molecular subset of the $GW$20 set are presented.  
Vertical IPs were calculated as the energy difference between the ionized-state and ground-state energy at the geometry optimized for the neutral system at the corresponding level of theory. 
Instead, for the calculation of the adiabatic IPs, the same ground-state energy was used in combination with the ionized-state energy calculated at the geometry optimized for the cationic system at the corresponding level of theory. 
The IPs are collected in Table \ref{tab:IP_va}, where the irreducible representation (IRREP) of the highest occupied molecular orbital (HOMO) is also reported. 
In the present calculation, the lowest-energy cationic state consistently corresponds to removing an electron from the HOMO. 
In Tables \ref{tab:R_diat} and \ref{tab:R_polyat}, the geometric parameters of the optimal geometries of the neutral and the cationic systems are presented. 
Intermolecular distances for diatomic molecules are reported in Table \ref{tab:R_diat}. 
The point group (PG) of the neutral and cationic systems in the case of polyatomic molecules, accompanied by a description of the reported geometrical parameters, can be found in Table \ref{tab:R_polyat}. 
In Tables \ref{tab:IP_va} and \ref{tab:R_diat}, the mean absolute error (MAE) and mean signed error (MSE) have been calculated with respect to the CC3 and IP-EOM-CC3 results. 
The only exception is \ce{LiF} for which we employed CCSDT reference values.
This point is discussed in detail in the \SupInf.

Regarding the IPs (Table~\ref{tab:IP_va}), the relative performance of the methods remains essentially consistent between vertical and adiabatic values.
ADC(2) and CC2 yield comparable results, with MAEs of approximately \SIrange{0.6}{0.7}{\eV} and negative MSEs of similar magnitude.
The $G_0W_0$ approximation performs noticeably better, producing MAEs in the range of \SIrange{0.3}{0.4}{\eV} and positive MSEs of comparable size. Similar superior performance of $G_0W_0$ compared to ADC(2) has also been observed in Ref.~\onlinecite{Tolle_2025} without employing the diagonal approximation as used in the current work.
CCSD delivers the most accurate results, with small positive MSEs and MAEs between \SIrange{0.1}{0.2}{\eV}.
The VIP of \ce{H2}, however, is known to be poorly described, likely due to the self-screening problem inherent to the $GW$ approximation.\cite{Nelson_2007,Romaniello_2009a,Aryasetiawan_2012,Wetherell_2018}
In contrast, the AIP appears largely unaffected.

For the equilibrium geometries of the diatomic molecules (Table \ref{tab:R_diat}), ADC(2), CC2, and CCSD all outperform $G_0W_0$.
While the errors in the ground-state bond lengths are generally small (\SIrange{0.01}{0.02}{\angstrom}), the deviations in the equilibrium geometries of the corresponding cations are larger by a factor of two to three for all methods.
Consistent with observations for neutral excited states, \cite{Budzak_2017,Loos_2025} CCSD tends to underestimate bond lengths in the cationic states, whereas ADC(2) and CC2 systematically overestimate them.
Similarly, $G_0W_0$ exhibits a systematic underestimation of cationic bond lengths, in contrast to BSE@$G_0W_0$, which has been shown to overestimate bond-length changes upon excitation in the case of neutral excitations. \cite{Tolle_2025b}
Analogous trends are observed for the larger polyatomic species in the test set (Table~\ref{tab:R_polyat}).

\section{Conclusion}
\label{sec:conclusion}
In this work, we have presented the implementation of the ionization-potential and electron-affinity equation-of-motion $\lambda$-direct-ring coupled-cluster doubles (IP/EA-EOM-$\lambda$-drCCD) formalism, enabling the computation of fully analytic $G_0W_0$ energies and nuclear gradients within a standard CC Lagrangian framework.
The approach has been implemented in the \textsc{qcumbre} program package and rigorously verified against numerical derivatives, as well as existing analytic implementations of $GW$ gradients. \cite{Tolle_2025}

Benchmark calculations were performed on the molecular subset of the $GW$20 test set.
Both vertical and adiabatic IPs were obtained at various levels of theory and compared to CC3 and IP-EOM-CC3 reference data.
The results confirm that the newly implemented $G_0W_0$ formalism yields IPs with MAEs in the range of \SIrange{0.3}{0.4}{\eV}, thereby outperforming second-order methods such as ADC(2) and CC2.
The equilibrium geometries of both neutral and cationic species are accurately reproduced, with typical bond-length deviations on the order of \SI{0.01}{\angstrom} for neutrals and slightly larger discrepancies for cations.
Systematic trends were observed across methods: CCSD and $G_0W_0$ tend to underestimate bond lengths in cationic states, whereas ADC(2) and CC2 systematically overestimate them.

These results demonstrate that the IP/EA-EOM-$\lambda$-drCCD formulation provides new route for obtaining fully analytic $G_0W_0$ properties, by leveraging the connection between Green's function and coupled-cluster theories.
The achieved agreement with high-level wavefunction benchmarks highlights the accuracy and robustness of the approach, while its analytic nature opens the door to geometry optimization and property evaluation for larger and more complex systems.
In future work, we plan to conduct a more extensive benchmark study on a broader set of IPs and EAs, in order to further assess the transferability and accuracy of the present formulation across diverse chemical environments.
Extensions toward self-consistent $GW$ variants and BSE within the same analytic formulation are also under active development.

\acknowledgements{
This project has received funding from the European Research Council (ERC) under the European Union's Horizon 2020 research and innovation programme (Grant agreement No.~863481).
J.~T.~acknowledges funding from the Fonds der Chemischen Industrie (FCI) via a Liebig fellowship and support by the Cluster of Excellence ``CUI: Advanced Imaging of Matter'' of the Deutsche Forschungsgemeinschaft (DFG) (EXC 2056, funding ID 390715994).
This work used the HPC resources from CALMIP (Toulouse) under allocation 2025-18005. 
M.-P.~K.~thanks Dr.~Laura Grazioli for her support with generating the ADC(2) results.

\section*{Data availability statement}
The data that support the findings of this study are available within the article and its supplementary material.

\section*{Appendices}
\appendix
\input{appendix.tex}

\section*{References}

\bibliography{biblio}

\end{document}

%% file: appendix.tex
\section{Diagrammatic rules for the direct-ring approximation and additional aspects} 
\label{app:rules}

In the diagrammatic representation, we employ the following conventions and graphical elements.
\begin{itemize}
\item Operator $\hT_2$:
\begin{center}
	\scalebox{0.7}{
		\begin{fmffile}{t2_dep}
			\begin{fmfgraph*}(100,60)
				\fmfstraight
				\fmftop{tl1,t1,tr1,tl2,t2,tr2}     
				\fmfbottom{bl1,b1,br1,bl2,b2,br2}     
				\fmffreeze
				\fmf{fermion}{tl1,b1,tr1}
				\fmf{fermion}{tl2,b2,tr2}
				\fmf{plain}{bl1,br2}
			\end{fmfgraph*}
		\end{fmffile}
	}
\end{center}
\item Operator $\hLam_2$:
\begin{center}
	\scalebox{0.7}{
		\begin{fmffile}{lam2_dep}
			\begin{fmfgraph*}(100,60)
				\fmfstraight
				\fmfbottom{tl1,t1,tr1,tl2,t2,tr2}     
				\fmftop{bl1,b1,br1,bl2,b2,br2}     
				\fmffreeze
				\fmf{fermion}{tl1,b1,tr1}
				\fmf{fermion}{tl2,b2,tr2}
				\fmf{plain}{bl1,br2}
			\end{fmfgraph*}
		\end{fmffile}
	}
\end{center}
\item Operator $\hXi_2$:
\begin{center}
	\scalebox{0.7}{
		\begin{fmffile}{ksi2_dep}
			\begin{fmfgraph*}(100,60)
				\fmfstraight
				\fmftop{tl1,t1,tr1,tl2,t2,tr2}     
				\fmfbottom{bl1,b1,br1,bl2,b2,br2}     
				\fmffreeze
				\fmf{fermion}{tl1,b1,tr1}
				\fmf{fermion}{tl2,b2,tr2}
				\fmfset{zigzag_width}{0.5mm}
				\fmf{zigzag}{bl1,br2}
			\end{fmfgraph*}
		\end{fmffile}
	}
\end{center}
\item Operator $\hZ_2$:
\begin{center}
	\scalebox{0.7}{
		\begin{fmffile}{zeta2_dep}
			\begin{fmfgraph*}(100,60)
				\fmfstraight
				\fmfbottom{tl1,t1,tr1,tl2,t2,tr2}     
				\fmftop{bl1,b1,br1,bl2,b2,br2}     
				\fmffreeze
				\fmf{fermion}{tl1,b1,tr1}
				\fmf{fermion}{tl2,b2,tr2}
				\fmfset{zigzag_width}{0.5mm}
				\fmf{zigzag}{bl1,br2}
			\end{fmfgraph*}
		\end{fmffile}
	}
\end{center}
\item The left and right EOM operators, $\hL$ and $\hR$, are depicted by bold lines, such as
\begin{center}
	\scalebox{0.7}{
		\begin{fmffile}{eom_dep}
			\begin{fmfgraph*}(100,60)
				\fmfstraight
				\fmftop{tl1,t1,tr1,tl2,t2,tr2}     
				\fmfbottom{bl1,b1,br1,bl2,b2,br2}     
				\fmffreeze
				\fmf{plain}{t1,b1}
				\fmf{fermion}{tl2,b2,tr2}
				\fmf{plain,width=3}{bl1,br2}
			\end{fmfgraph*}
		\end{fmffile}
	}
\end{center}
\item The Fock operator and the two-electron interaction are depicted by dashed lines, such as
\begin{center}
\scalebox{0.7}{
	\begin{fmffile}{fock_dep}
		\begin{fmfgraph*}(100,60)
			\fmfstraight
			\fmftop{tle,tl,t1,tr,tre}     
			\fmfbottom{vle,vl,v1,vr,vre}  
			\fmfright{r1}
			\fmf{phantom}{t1,m1,v1}  
			\fmf{plain}{t1,m1}  
			\fmf{plain}{m1,v1}
			\fmffreeze
			\fmf{dashes}{m1,r1}
		\end{fmfgraph*}
\end{fmffile}} 
\end{center}
\item The vertices with open circles are used to denote the indices of the one- and two-body reduced density matrices. Lines connected to open circles are to be treated as external lines starting from the vertex. 
For example, $\gamma_{ab}$ is depicted as
\begin{center}
	\scalebox{0.7}{
		\begin{fmffile}{dense_dep}
			\begin{fmfgraph*}(100,60)
				\fmfstraight
				\fmftop{tle,tl,t1,tr,tre}     
				\fmfbottom{vle,vl,v1,vr,vre}  
				\fmfright{r1}
				\fmf{phantom}{t1,m1,v1}  
				\fmf{fermion,label=$a$}{m1,t1}  
				\fmf{fermion,label=$b$}{v1,m1}
				\fmffreeze
				\fmf{dashes}{m1,r1}
				\fmfv{decor.shape=circle,decor.filled=empty,decor.size=2thick}{m1}
			\end{fmfgraph*}
	\end{fmffile}} 
\end{center} 
\item Otherwise, the standard convention can be assumed.\cite{Crawford_2000,Shavitt_2009}
\end{itemize}
In the direct ring approximation, the following rules are imposed in addition to the standard CC diagrammatic rules.\cite{Crawford_2000,Shavitt_2009}

\begin{enumerate}
	\item \textbf{Loop connectivity:} Indices subject to the ring approximation must connect the same initial and final vertices, i.e., they must form a closed loop: \\
	\begin{center}
	\scalebox{0.7}{
	\begin{fmffile}{loop}
		\begin{fmfgraph*}(100,60)
			\fmfstraight
			\fmftop{tl1,t1,tr1,tl2,t2,tr2,tl3,t3,tr3}     
			\fmfbottom{bl1,b1,br1,bl2,b2,br2,bl3,b3,br3}     

			\fmffreeze
			\fmf{fermion, left=0.1}{t2,b2} 
			\fmf{fermion, left=0.1}{b2,t2}
			\fmf{plain}{tl1,tr2}
			\fmf{plain}{bl2,br3}
		\end{fmfgraph*}
	\end{fmffile}
	}
	\end{center}
	The Fock operator is exempt from this rule.
	\item \textbf{Symmetrization of non-equivalent external indices:} For external, non-equivalent indices obeying the ring approximation, symmetrization (but not antisymmetrization) is applied separately to each particle-particle or hole-hole index.
For example, for two external index pairs $ia$ and $jb$, the symmetrization is performed as 
	\begin{equation}
		P(ia|jb) M_{iajb} = M_{iajb} + M_{jbia}.
	\end{equation}
	Individual antisymmetrizations over occupied [$P(ij)$] or virtual [$P(ab)$] indices are omitted.
Consequently, in projections onto IP/EA doubly excited determinants ${ \ket{\Phi^{b}_{ji}}, \ket{\Phi^{ba}_j} }$, indices belonging to the same particle or hole spaces are treated as non-equivalent. 
	\item \textbf{Equivalent indices in loop pairs:} Indices equivalent under the ring approximation are only considered as loop pairs, not individually.
When one vertex corresponds to a two-electron integral, each such pair contributes a factor of $\tfrac{1}{2}$ (rather than $\tfrac{1}{4}$) owing to the direct approximation.
For example, the following diagram
	\begin{center}
		\scalebox{0.7}{
		\begin{fmffile}{Ener_ijab1}
			\begin{fmfgraph*}(100,60)
				\fmfstraight
				\fmfbottom{t1,ta,t2,ti,tb,t3,tj,t4}     
				\fmftop{v1,v2,v3,v4,v5,v6,v7,v8}  
				\fmf{fermion,right=0.15, label= $a$}{t2,v3}  
				\fmf{fermion,right=0.15, label= $i$}{v3,t2}  
				\fmf{fermion,right=0.15, label= $b$}{t3,v6}
				\fmf{fermion,right=0.15, label= $j$}{v6,t3}
				\fmf{dashes}{v3,v6}
				\fmf{plain}{ta,tj}
			\end{fmfgraph*}
		\end{fmffile}} 
	\end{center}
	represents the drCCD correlation energy and is interpreted as $\frac{1}{2}\sum_{ijab}\ERI{ij}{ab}t^{ab}_{ij}$. 
	\item \textbf{Non-antisymmetrized two-electron integrals:} In the direct ring approximation, two-electron diagrams depict the non-antisymmetrized integrals $\ERI{pq}{rs}$ rather than $\aERI{pq}{rs}$.
Consequently, the following two diagrams are not related via anti-symmetry
	\begin{center}
		\scalebox{0.7}{
			\begin{fmffile}{anti1}
				\begin{fmfgraph*}(80,60)
					\fmfstraight
					\fmfbottom{bl1,b1,br1,bl2,b2,br2}     
					\fmftop{tl1,t1,tr1,tl2,t2,tr2}   
					\fmf{phantom}{t1,m1,b1}  
					\fmf{phantom}{t2,m2,b2}
					\fmffreeze
					\fmf{fermion}{tl1,m1}
					\fmf{fermion}{m1,tr1}  
					\fmf{fermion}{bl2,m2}
					\fmf{fermion}{m2,br2}  
					\fmf{dashes}{m1,m2}
				\end{fmfgraph*}
		\end{fmffile}}  \scalebox{0.7}{
		\begin{fmffile}{anti2}
		\begin{fmfgraph*}(80,60)
			\fmfstraight
			\fmfbottom{bl1,b1,br1,bl2,b2,br2}     
			\fmftop{tl1,t1,tr1,tl2,t2,tr2}   
			\fmf{phantom}{t1,m1,b1}  
			\fmf{phantom}{t2,m2,b2}
			\fmffreeze
			\fmf{fermion}{t1,m1}
			\fmf{fermion}{m1,b1}  
			\fmf{fermion}{b2,m2}
			\fmf{fermion}{m2,t2}  
			\fmf{dashes}{m1,m2}
		\end{fmfgraph*}
	\end{fmffile}}
	\end{center}
	Similar to the second rule, two-body vertices may be mirrored (corresponding to the exchange of electrons 1 and 2), but the individual vertex lines cannot be exchanged independently.
	\item \textbf{Additional factors in density matrix diagrams:} For diagrams contributing to the density matrices, special care must be taken for the \textit{ovvo} ($iabj$) block of the two-body reduced density matrix $\Gamma_{pqrs}$.
	\begin{center}
		\scalebox{0.7}{
			\begin{fmffile}{ovvo_dens}
				\begin{fmfgraph*}(80,60)
					\fmfstraight
					\fmfbottom{bl1,b1,br1,bl2,b2,br2}     
					\fmftop{tl1,t1,tr1,tl2,t2,tr2}   
					\fmf{phantom}{t1,m1,b1}  
					\fmf{phantom}{t2,m2,b2}
					\fmffreeze
					\fmf{fermion}{tl1,m1}
					\fmf{fermion}{m1,tr1}  
					\fmf{fermion}{bl2,m2}
					\fmf{fermion}{m2,br2}  
					\fmf{dashes}{m1,m2}
					\fmfv{decor.shape=circle,decor.filled=empty,decor.size=2thick}{m1,m2}
				\end{fmfgraph*}
		\end{fmffile}} 
	\end{center}
	Formally, the energy contribution is multiplied by a factor of $4$ for each of the combinations $iabj$, $aijb$, $aibj$, and $iajb$, resulting in $4 \cdot \frac{1}{2} \sum_{iabj} \Gamma_{iabj}\ERI{ia}{bj}$. However, within the ring approximation, the $aibj$ and $iajb$ blocks are absent, which introduces an additional factor of $\frac{1}{2}$ in the contributions from $\Gamma_{iabj}\ERI{ia}{bj}$.
	\item \textbf{Sign conventions for left and right EOM operators:} 
	When the right EOM operator acts on the left ($\hR^{\EA\dagger}_1$, $\hR^{\EA\dagger}_2$) or the left EOM operator acts on the right ($\hL^{\EA\dagger}_1$, $\hL^{\EA\dagger}_2$), an additional factor of $-1$ must be included.
This arises from the second term of the commutator.
It is applied in addition to the standard sign factor $(-1)^K$, where $K = l + h$ is the sum of the number of loops ($l$) and internal holes ($h$). 
For example, in the following diagrams:
\begin{center}\scalebox{0.7}{
	\begin{fmffile}{r1ipfij}
		\begin{fmfgraph*}(100,60)
			\fmfstraight
			\fmftop{tle,tl,t1,tr,tre}     
			\fmfbottom{vle,vl,v1,vr,vre}  
			\fmfright{r1}
			\fmf{phantom}{t1,m1,v1}  
			\fmf{fermion}{t1,m1}  
			\fmf{fermion, label= $l$}{m1,v1}
			\fmffreeze
			\fmf{dashes}{m1,r1}
			\fmf{plain,width=3}{vr,vl}
			\fmfv{label=$i$,label.angle=90}{t1}
		\end{fmfgraph*}
\end{fmffile}} 
\scalebox{0.7}{
	\begin{fmffile}{r1ipfai}
		\begin{fmfgraph*}(100,60)
			\fmfstraight
			\fmftop{tle,tl,t1,tr,tre}     
			\fmfbottom{vle,vl,v1,vr,vre}  
			\fmfright{r1}
			\fmf{phantom}{t1,m1,v1}  
			\fmffreeze
			\fmf{fermion}{tr,m1}  
			\fmf{fermion, label= $d$}{m1,tl}
			
			\fmf{dashes}{m1,r1}
			\fmf{plain,width=3}{t1,tle}
			\fmfv{label=$i$,label.angle=90}{tr}
		\end{fmfgraph*}
\end{fmffile}} 
\end{center}
the overall sign in both cases is $-1$.
In the first diagram, there is one internal hole and no loop ($K = 0 + 1$), while in the second, there are neither loops nor internal holes ($K = 0$), but the $\hR_1^{\EA\dagger}$ operator acts on the left, introducing the additional negative sign.
\end{enumerate}

\section{drCCD working equations \label{app:rCCD}}
The drCCD amplitude equations and the corresponding $\lambda$-amplitude equations are presented here in a diagrammatic and algebraic form. 

\subsection{drCCD amplitude equations}


\begin{longtable}{c  c}
	
	\multicolumn{2}{c}{$\bra{\Phi^{ab}_{ij}}  \wtH_\dr \ket{\Phi_0}$} \\ \hline
	\\
	\scalebox{0.5}{
		\begin{fmffile}{abij}
			\begin{fmfgraph*}(100,60)
				\fmfstraight
				\fmftop{t1,ta,t2,ti,tb,t3,tj,t4}     
				\fmfbottom{v1,v2,v3,v4,v5,v6,v7,v8}      
				\fmf{fermion}{ti,v3}  
				\fmf{fermion}{v3,ta} 
				\fmf{fermion}{tj,v6}  
				\fmf{fermion}{v6,tb} 
				\fmf{dashes}{v3,v6}
				\fmfv{label=$i$,label.angle=90,label.dist=3}{ti}
				\fmfv{label=$j$,label.angle=90,label.dist=2}{tj}
				\fmfv{label=$a$,label.angle=90,label.dist=3}{ta}
				\fmfv{label=$b$,label.angle=90,label.dist=3}{tb}
			\end{fmfgraph*}
	\end{fmffile}} 
	& \parbox{170pt}{\begin{equation*}\braket{ab|ij}\end{equation*}}  \\ \\

	\scalebox{0.5}{
		\begin{fmffile}{FT2v}
			\begin{fmfgraph*}(100,60)
				\fmfstraight
				\fmftop{t1,ta,t2,ti,tb,t3,tj,t4}     
				\fmfbottom{v1,v2,bcl,v4,v5,bcr,v7,v8}
				\fmfright{r}
				\fmf{fermion}{ti,bcl}  
				\fmf{fermion}{bcl,ta} 
				\fmf{fermion}{tb,bcr} 
				\fmf{fermion}{bcr,mc,tj} 
				\fmf{plain}{v2,v7}
				\fmfv{label=$i$,label.angle=90,label.dist=3}{ti}
				\fmfv{label=$b$,label.angle=90,label.dist=2}{tj}
				\fmfv{label=$a$,label.angle=90,label.dist=3}{ta}
				\fmfv{label=$j$,label.angle=90,label.dist=3}{tb}
				\fmffreeze
				\fmf{phantom,label.side=left,label=$c$}{mc,bcr}
				\fmf{dashes}{mc,r}
			\end{fmfgraph*}
	\end{fmffile}} & 
	\parbox{170pt}{\begin{equation*}\sum_c f_{bc} t^{ac}_{ij}+\sum_c f_{ac} t^{cb}_{ij}\end{equation*}}   \\ \\
	
	\scalebox{0.5}{
		\begin{fmffile}{FT2o}
			\begin{fmfgraph*}(100,60)
				\fmfstraight
				\fmftop{t1,ta,t2,ti,tb,t3,tj,t4}     
				\fmfbottom{v1,v2,bcl,v4,v5,bcr,v7,v8}
				\fmfright{r}
				\fmf{fermion}{ti,bcl}  
				\fmf{fermion}{bcl,ta} 
				\fmf{fermion}{bcr,tb} 
				\fmf{fermion}{tj,mc,bcr} 
				\fmf{plain}{v2,v7}
				\fmfv{label=$i$,label.angle=90,label.dist=3}{ti}
				\fmfv{label=$j$,label.angle=90,label.dist=2}{tj}
				\fmfv{label=$a$,label.angle=90,label.dist=3}{ta}
				\fmfv{label=$b$,label.angle=90,label.dist=3}{tb}
				\fmffreeze
				\fmf{phantom,label.side=left,label=$k$}{mc,bcr}
				\fmf{dashes}{mc,r}
			\end{fmfgraph*}
	\end{fmffile}} & 
	\parbox{170pt}{\begin{equation*}-\sum_k f_{kj} t^{ab}_{ik}-\sum_k f_{ki} t^{ab}_{kj}\end{equation*}}  \\ \\
	
	\scalebox{0.5}{
		\begin{fmffile}{T2Viabj}
			\begin{fmfgraph*}(100,60)
				\fmfstraight
				\fmftop{t1,ta,t2,ti,tm1,tm2,tm3,tb,t3,tj,t4}     
				\fmfbottom{v1,v2,bcl,v4,v5,bcr,v7,v8,bm1,bm2,bm3}
				\fmf{plain}{v2,v7}
				\fmf{fermion}{ti,bcl}  
				\fmf{fermion}{bcl,ta} 
				\fmf{phantom}{bcr,mc1,tm2}
				\fmf{phantom}{bm1,mc2,t3}
				\fmf{phantom}{tj,bm1}  
				\fmf{phantom}{bm1,tb}

				\fmffreeze
				\fmf{fermion, left=0.5,label=$c$,label.side=left}{bcr,mc1} 
				\fmf{fermion, left=0.5,label=$k$,label.side=left}{mc1,bcr} 
				\fmf{fermion}{mc2,tb}
				\fmf{fermion}{tj,mc2}
				\fmf{dashes}{mc1,mc2}
				\fmfv{label=$i$,label.angle=90,label.dist=3}{ti}
				\fmfv{label=$j$,label.angle=90,label.dist=2}{tj}
				\fmfv{label=$a$,label.angle=90,label.dist=3}{ta}
				\fmfv{label=$b$,label.angle=90,label.dist=3}{tb}
			\end{fmfgraph*}
	\end{fmffile}} & 
	\parbox{170pt}{\begin{equation*}P(ia|jb)\sum_{kc}t^{ac}_{ik}\braket{kb|cj} \end{equation*}}  \\ \\
	
	\scalebox{0.5}{
		\begin{fmffile}{T2ViabjT2}
			\begin{fmfgraph*}(100,60)
				\fmfstraight
				\fmftop{t1,ta,t2,ti,tm1,tm2,tm3,tmgap,tm4,tm5,tm6,tj,t3,tb,t4}     
				\fmfbottom{v1,v2,bcl1,v4,v5,bcr1,v7,v8,bm1,bcl2,bm3,bm4,bcr2,bm5,bm6}
				\fmf{plain}{v2,v7}
				\fmf{plain}{bm1,bm5}
				\fmf{fermion}{ti,bcl1}  
				\fmf{fermion}{bcl1,ta}
				\fmf{fermion}{tj,bcr2}  
				\fmf{fermion}{bcr2,tb}
				\fmf{phantom}{bcr1,mc1,tm2}
				\fmf{phantom}{bcl2,mc2,tm5}
				\fmf{phantom}{tj,bm1}  
				\fmf{phantom}{bm1,tb}

				\fmffreeze
				\fmf{fermion, left=0.5,label=$c$,label.side=left,label.dist=3}{bcr1,mc1} 
				\fmf{fermion, left=0.5,label=$k$,label.side=right,label.dist=3}{mc1,bcr1} 
				\fmf{fermion, left=0.5,label=$d$,label.side=right,label.dist=3}{bcl2,mc2} 
				\fmf{fermion, left=0.5,label=$l$,label.side=left,label.dist=3}{mc2,bcl2} 
				\fmf{dashes}{mc1,mc2}
				\fmfv{label=$i$,label.angle=90,label.dist=3}{ti}
				\fmfv{label=$j$,label.angle=90,label.dist=2}{tj}
				\fmfv{label=$a$,label.angle=90,label.dist=3}{ta}
				\fmfv{label=$b$,label.angle=90,label.dist=3}{tb}
			\end{fmfgraph*}
	\end{fmffile}} & 
	\parbox{170pt}{\begin{gather*}\frac{1}{2} P(ia|jb) \sum_{klcd}t^{ac}_{ik}\braket{kl|cd}t^{db}_{lj} = \\ \sum_{klcd}t^{ac}_{ik}\braket{kl|cd}t^{db}_{lj}\end{gather*}}  

\end{longtable}

\subsection{$\lambda$-drCCD amplitude equations}
\begin{longtable}{c  c}
	
	\multicolumn{2}{c}{$\bra{\Phi_0} \qty(1 + \hLam_2 ) \pdv{\wtH_\dr}{t^{ab}_{ij}} \ket{\Phi_0}$} \\ \hline
	\\
	\scalebox{0.5}{
		\begin{fmffile}{ijab}
			\begin{fmfgraph*}(100,60)
				\fmfstraight
				\fmfbottom{t1,ta,t2,ti,tb,t3,tj,t4}     
				\fmftop{v1,v2,v3,v4,v5,v6,v7,v8}      
				\fmf{fermion}{ti,v3}  
				\fmf{fermion}{v3,ta} 
				\fmf{fermion}{tj,v6}  
				\fmf{fermion}{v6,tb} 
				\fmf{dashes}{v3,v6}
				\fmfv{label=$a$,label.angle=-90,label.dist=3}{ti}
				\fmfv{label=$b$,label.angle=-90,label.dist=2}{tj}
				\fmfv{label=$i$,label.angle=-90,label.dist=3}{ta}
				\fmfv{label=$j$,label.angle=-90,label.dist=3}{tb}
			\end{fmfgraph*}
	\end{fmffile}} 
	& \parbox{170pt}{\begin{equation*}  \color{black}\braket{ij|ab}\end{equation*}}  \\ \\

	\scalebox{0.5}{
		\begin{fmffile}{FL2v}
			\begin{fmfgraph*}(100,60)
				\fmfstraight
				\fmfbottom{t1,ta,t2,ti,tb,t3,tj,t4}     
				\fmftop{v1,v2,bcl,v4,v5,bcr,v7,v8}
				\fmfright{r}
				\fmf{fermion}{ti,bcl}  
				\fmf{fermion}{bcl,ta} 
				\fmf{fermion}{tb,bcr} 
				\fmf{fermion}{tj,mc,bcr} 
				\fmf{plain}{v2,v7}
				\fmfv{label=$a$,label.angle=-90,label.dist=3}{ti}
				\fmfv{label=$b$,label.angle=-90,label.dist=2}{tj}
				\fmfv{label=$i$,label.angle=-90,label.dist=3}{ta}
				\fmfv{label=$j$,label.angle=-90,label.dist=3}{tb}
				\fmffreeze
				\fmf{phantom,label.side=right,label=$c$}{mc,bcr}
				\fmf{dashes}{mc,r}
			\end{fmfgraph*}
	\end{fmffile}}
	& 
	\parbox{170pt}{\begin{equation*}\sum_c f_{cb} \lambda_{ac}^{ij}+\sum_c f_{ca} \lambda_{cb}^{ij}\end{equation*}}   \\ \\
	
	\scalebox{0.5}{
		\begin{fmffile}{FL2o}
			\begin{fmfgraph*}(100,60)
				\fmfstraight
				\fmfbottom{t1,ta,t2,ti,tj,t3,tb,t4}     
				\fmftop{v1,v2,bcl,v4,v5,bcr,v7,v8}
				\fmfright{r}
				\fmf{fermion}{ti,bcl}  
				\fmf{fermion}{bcl,ta} 
				\fmf{fermion}{bcr,mc,tb} 
				\fmf{fermion}{tj,bcr} 
				\fmf{plain}{v2,v7}
				\fmfv{label=$a$,label.angle=-90,label.dist=3}{ti}
				\fmfv{label=$b$,label.angle=-90,label.dist=2}{tj}
				\fmfv{label=$i$,label.angle=-90,label.dist=3}{ta}
				\fmfv{label=$j$,label.angle=-90,label.dist=3}{tb}
				\fmffreeze
				\fmf{phantom,label.side=right,label=$k$}{mc,bcr}
				\fmf{dashes}{mc,r}
			\end{fmfgraph*}
	\end{fmffile}}
	& 
	\parbox{170pt}{\begin{equation*}-\sum_k f_{jk} \lambda_{ab}^{ik}-\sum_k f_{ik} \lambda_{ab}^{kj}\end{equation*}}  \\ \\
	
	\scalebox{0.5}{
		\begin{fmffile}{L2Viabj}
			\begin{fmfgraph*}(100,60)
				\fmfstraight
				\fmfbottom{t1,ta,t2,ti,tm1,tm2,tm3,tb,t3,tj,t4}     
				\fmftop{v1,v2,bcl,v4,v5,bcr,v7,v8,bm1,bm2,bm3}
				\fmf{plain}{v2,v7}
				\fmf{fermion}{ti,bcl}  
				\fmf{fermion}{bcl,ta} 
				\fmf{phantom}{bcr,mc1,tm2}
				\fmf{phantom}{bm1,mc2,t3}
				\fmf{phantom}{tj,bm1}  
				\fmf{phantom}{bm1,tb}

				\fmffreeze
				\fmf{fermion, left=0.5,label=$k$,label.side=left}{bcr,mc1} 
				\fmf{fermion, left=0.5,label=$c$,label.side=left}{mc1,bcr} 
				\fmf{fermion}{mc2,tb}
				\fmf{fermion}{tj,mc2}
				\fmf{dashes}{mc1,mc2}
				\fmfv{label=$a$,label.angle=-90,label.dist=3}{ti}
				\fmfv{label=$b$,label.angle=-90,label.dist=2}{tj}
				\fmfv{label=$i$,label.angle=-90,label.dist=3}{ta}
				\fmfv{label=$j$,label.angle=-90,label.dist=3}{tb}
			\end{fmfgraph*}
	\end{fmffile}} & 
	\parbox{170pt}{\begin{equation*}P(ia|jb) \sum_{kc}\lambda_{ac}^{ik}\braket{cj|kb} \end{equation*}}  \\ \\
	
	\scalebox{0.5}{
		\begin{fmffile}{L2T2Viabj}
			\begin{fmfgraph*}(100,60)
				\fmfstraight
				\fmfbottom{t1,ta,t2,ti,tm1,tm2,tm3,tmgap,tm4,tm5,tm6,tj,t3,tb,t4}     
				\fmftop{v1,v2,bcl1,v4,v5,bcr1,v7,v8,bm1,bcl2,bm3,bm4,bcr2,bm5,bm6}
				\fmf{plain}{v2,v7}
				\fmf{dashes}{bcl2,bcr2}
				\fmf{fermion}{ti,bcl1}  
				\fmf{fermion}{bcl1,ta}
				\fmf{fermion}{tj,bcr2}  
				\fmf{fermion}{bcr2,tb}
				\fmf{phantom}{bcr1,mc1,tm2}
				\fmf{phantom}{bcl2,mc2,tm5}
				\fmf{phantom}{tj,bm1}  
				\fmf{phantom}{bm1,tb}

				\fmffreeze
				\fmf{fermion, left=0.15,label=$k$,label.side=left,label.dist=4}{bcr1,tm2} 
				\fmf{fermion, left=0.15,label=$c$,label.side=left,label.dist=4}{tm2,bcr1} 
				\fmf{fermion, left=0.15,label=$l$,label.side=left,label.dist=4}{bcl2,tm5} 
				\fmf{fermion, left=0.15,label=$d$,label.side=left,label.dist=4}{tm5,bcl2} 
				\fmf{plain}{tm1,tm6}
				\fmfv{label=$a$,label.angle=-90,label.dist=3}{ti}
				\fmfv{label=$b$,label.angle=-90,label.dist=2}{tj}
				\fmfv{label=$i$,label.angle=-90,label.dist=3}{ta}
				\fmfv{label=$j$,label.angle=-90,label.dist=3}{tb}
			\end{fmfgraph*}
	\end{fmffile}} & 
	\parbox{170pt}{\begin{equation*}P(ia|jb) \sum_{klcd}\lambda^{ik}_{ac} t^{cd}_{kl}\braket{lj|db} \end{equation*} }  

\end{longtable}

\subsection{One-electron density matrices}
\begin{longtable}{c  c}
	\multicolumn{2}{c}{$\gamma_{ij}$} \\ \hline
	\\
	\scalebox{0.5}{
		\begin{fmffile}{d_ij}
			\begin{fmfgraph*}(100,60)
				\fmfstraight
				\fmftop{t1,ta,t2,ti,tb,t3,tj,t4}     
				\fmfbottom{v1,v2,v3,v4,v5,v6,v7,v8}  
				\fmfright{r1}
				\fmf{fermion,left=0.15, label= $k$}{t2,v3}  
				\fmf{fermion,left=0.15, label= $c$}{v3,t2}  
				\fmf{fermion,left=0.15, label= $d$}{v6,t3}
				\fmf{phantom}{t3,m1,v6}
				\fmffreeze
				\fmfshift{(6.0,0)}{m1}
				\fmf{fermion,left=0.09, label= $j$}{t3,m1}
				\fmf{fermion,left=0.09, label= $i$}{m1,v6}
				\fmf{dashes}{m1,r1}
				\fmf{plain}{v2,v7}
				\fmf{plain}{ta,tj}
				\fmfv{decor.shape=circle,decor.filled=empty,decor.size=2thick}{m1}
			\end{fmfgraph*}
	\end{fmffile}} 
	& \parbox{170pt}{\begin{equation*}- \sum_{ckd}t_{ki}^{cd} \lambda^{kj}_{cd}\end{equation*}}  \\ \\

	\multicolumn{2}{c}{\parbox{170pt}{\begin{equation*}\gamma_{ab}\end{equation*}}} \\ \hline
	\\
	\scalebox{0.5}{
		\begin{fmffile}{d_ab}
			\begin{fmfgraph*}(100,60)
				\fmfstraight
				\fmftop{t1,ta,t2,ti,tb,t3,tj,t4}     
				\fmfbottom{v1,v2,v3,v4,v5,v6,v7,v8}  
				\fmfright{r1}
				\fmf{fermion,right=0.15, label= $k$}{t2,v3}  
				\fmf{fermion,right=0.15, label= $c$}{v3,t2}  
				\fmf{fermion,right=0.15, label= $l$}{t3,v6}
				\fmf{phantom}{t3,m1,v6}
				\fmffreeze
				\fmfshift{(6.0,0)}{m1}
				\fmf{fermion,right=0.09, label= $b$}{v6,m1}
				\fmf{fermion,right=0.09, label= $a$}{m1,t3}
				\fmf{dashes}{m1,r1}
				\fmf{plain}{v2,v7}
				\fmf{plain}{ta,tj}
				\fmfv{decor.shape=circle,decor.filled=empty,decor.size=2thick}{m1}
			\end{fmfgraph*}
	\end{fmffile}} 
	& \parbox{170pt}{\begin{equation*}  \sum_{ckl}t_{kl}^{cb} \lambda^{kl}_{ca}\end{equation*}}  
\end{longtable}

\subsection{Two-electron matrices}
\begin{longtable}{c  c}
	
	\multicolumn{2}{c}{$\Gamma_{ijab}$} \\ \hline
	\\
	\scalebox{0.5}{
		\begin{fmffile}{D_ijab1}
			\begin{fmfgraph*}(100,60)
				\fmfstraight
				\fmfbottom{t1,ta,t2,ti,tb,t3,tj,t4}     
				\fmftop{v1,v2,v3,v4,v5,v6,v7,v8}  
				\fmf{fermion,right=0.15, label= $a$}{t2,v3}  
				\fmf{fermion,right=0.15, label= $i$}{v3,t2}  
				\fmf{fermion,right=0.15, label= $b$}{t3,v6}
				\fmf{fermion,right=0.15, label= $j$}{v6,t3}
				\fmf{dashes}{v3,v6}
				\fmf{plain}{ta,tj}
				\fmfv{decor.shape=circle,decor.filled=empty,decor.size=2thick}{v3,v6}
			\end{fmfgraph*}
	\end{fmffile}} 
	& \parbox{170pt}{\begin{equation*}   t^{ab}_{ij}\end{equation*}}  \\ \\
	
	\scalebox{0.5}{
		\begin{fmffile}{D_ijab2}
			\begin{fmfgraph*}(100,60)
				\fmfstraight
				\fmfbottom{t1,t2,tcl1,t3,t4,t5,tcr1,t6,t7,t8,tcl2,t9,t10,t11,tcr2,t12,t13}     
				\fmftop{l1,l2,lcl1,l3,l4,l5,lcr1,l6,l7,l8,lcl2,l9,l10,l11,lcr2,l12,l13}
				\fmf{phantom}{tcr1,mc1,lcr1}
				\fmf{phantom}{tcl2,mc2,lcl2}
				\fmffreeze
				\fmf{fermion,right=0.15, label= $c$}{tcl1,lcl1}  
				\fmf{fermion,right=0.15, label= $k$}{lcl1,tcl1}  
				\fmf{fermion,right=0.15, label= $d$}{tcr2,lcr2}  
				\fmf{fermion,right=0.15, label= $l$}{lcr2,tcr2}  
				\fmf{fermion,right=0.5, label= $a$,label.side=left,label.dist=3}{tcr1,mc1}  
				\fmf{fermion,right=0.5, label= $i$,label.side=right,label.dist=3}{mc1,tcr1}  
				\fmf{fermion,right=0.5, label= $b$,label.side=right,label.dist=3}{tcl2,mc2}  
				\fmf{fermion,right=0.5, label= $j$,label.side=left,label.dist=3}{mc2,tcl2}  
				\fmf{dashes}{mc1,mc2}
				\fmf{plain}{t2,t6}
				\fmf{plain}{t8,t12}
				\fmf{plain}{l2,l12}
				\fmfv{decor.shape=circle,decor.filled=empty,decor.size=2thick}{mc1,mc2}
			\end{fmfgraph*}
	\end{fmffile}} 
	& \parbox{170pt}{\begin{equation*} \sum_{klcd}  t^{ca}_{ki} t^{bd}_{jl} \lambda^{kl}_{cd}\end{equation*}}  \\ \\
	
	\multicolumn{2}{c}{\parbox{170pt}{\begin{equation*}\Gamma_{abij}\end{equation*}}} \\ \hline
	\\
	\scalebox{0.5}{
		\begin{fmffile}{D_abij}
			\begin{fmfgraph*}(100,60)
				\fmfstraight
				\fmftop{t1,ta,t2,ti,tb,t3,tj,t4}     
				\fmfbottom{v1,v2,v3,v4,v5,v6,v7,v8}  
				\fmf{fermion,left=0.15, label= $i$}{t2,v3}  
				\fmf{fermion,left=0.15, label= $a$}{v3,t2}  
				\fmf{fermion,left=0.15, label= $j$}{t3,v6}
				\fmf{fermion,left=0.15, label= $b$}{v6,t3}
				\fmf{dashes}{v3,v6}
				\fmf{plain}{ta,tj}
				\fmfv{decor.shape=circle,decor.filled=empty,decor.size=2thick}{v3,v6}
			\end{fmfgraph*}
	\end{fmffile}} 
	& \parbox{170pt}{\begin{equation*} \lambda_{ab}^{ij}\end{equation*}}  \\ \\
	
	\multicolumn{2}{c}{\parbox{170pt}{\begin{equation*}\Gamma_{iabj}\end{equation*}}} \\ \hline
	\\
	\scalebox{0.5}{
		\begin{fmffile}{D_iabj}
			\begin{fmfgraph*}(100,60)
				\fmfstraight
				\fmfbottom{t1,ta,t2,ti,tm1,tm2,tm3,tb,t3,tj,t4}     
				\fmftop{v1,v2,bcl,v4,v5,bcr,v7,v8,bm1,bm2,bm3}
				\fmf{phantom}{bcr,mc1,tm2}
				\fmf{phantom}{bm1,mc2,t3}
				\fmf{phantom}{tj,bm1}  
				\fmf{phantom}{bm1,tb}

				\fmffreeze
				\fmf{fermion, left=0.15,label=$c$,label.side=left}{t2,bcl}  
				\fmf{fermion, left=0.15,label=$k$,label.side=left}{bcl,t2} 
				\fmf{fermion, left=0.5,label=$j$,label.side=left}{bcr,mc1} 
				\fmf{fermion, left=0.5,label=$a$,label.side=left}{mc1,bcr} 
				\fmf{fermion, left=0.5,label=$i$,label.side=left}{mc2,t3}
				\fmf{fermion, left=0.5,label=$b$,label.side=left}{t3,mc2}
				\fmf{plain}{ta,tj}
				\fmf{plain}{v2,v7}
				\fmf{dashes}{mc1,mc2}
				\fmfv{decor.shape=circle,decor.filled=empty,decor.size=2thick}{mc1,mc2}
			\end{fmfgraph*}
	\end{fmffile}} 
	& \parbox{170pt}{\begin{equation*} \frac{1}{2} \sum_{kc}  \lambda^{kj}_{ca} t_{ki}^{cb}\end{equation*}}

\end{longtable}

\section{EA/IP-EOM-$\lambda$-drCCD working equations}
\label{app:ipea_eom_lrCCD}

\subsection{Right-EOM eigenvalue working equations}
The right eigenvalue problem is divided into the singles IP and EA blocks, and the doubles IP and EA blocks, as follows:
\begin{equation}
	\begin{aligned}
		r_i \epsilon^\IPEA
		& =  \bra{\Phi_i} \comm{ \wtH_{\ldr} }{ \hR^\IPEA } \ket{\Phi_0} \\
		r_a \epsilon^\IPEA
		& =  \bra{\Phi_0} \comm{ \wtH_{\ldr} }{ \hR^\IPEA } \ket{\Phi^a} \\
		r_{ji}^b \epsilon^\IPEA
		& =  \bra{\Phi_{ji}^b} \comm{ \wtH_{\ldr} }{ \hR^\IPEA }\ket{\Phi_0} \\
		r_{ba}^j \epsilon^\IPEA
		& =  \bra{\Phi_0} \comm{ \wtH_{\ldr} }{ \hR^\IPEA } \ket{\Phi^{ba}_j} \\
	\end{aligned}
\end{equation}
The evaluation of the required matrix elements is presented in the following tables.
\begin{longtable}{c  c}
	
	\multicolumn{2}{c}{$\bra{\Phi_i} \comm{ \wtH_{\ldr} }{ \hR^\IPEA } \ket{\Phi_0}$} \\ \hline
	\\
	\scalebox{0.5}{
		\begin{fmffile}{r1ipfij}
			\begin{fmfgraph*}(100,60)
				\fmfstraight
				\fmftop{tle,tl,t1,tr,tre}     
				\fmfbottom{vle,vl,v1,vr,vre}  
				\fmfright{r1}
				\fmf{phantom}{t1,m1,v1}  
				\fmf{fermion}{t1,m1}  
				\fmf{fermion, label= $l$}{m1,v1}
				\fmffreeze
				\fmf{dashes}{m1,r1}
				\fmf{plain,width=3}{vr,vl}
				\fmfv{label=$i$,label.angle=90}{t1}
			\end{fmfgraph*}
	\end{fmffile}} 
	& \parbox{170pt}{\begin{equation*}-\sum_l f_{li} r_k\end{equation*}}  \\ \\
	
	\scalebox{0.5}{
		\begin{fmffile}{r1ipfai}
			\begin{fmfgraph*}(100,60)
				\fmfstraight
				\fmftop{tle,tl,t1,tr,tre}     
				\fmfbottom{vle,vl,v1,vr,vre}  
				\fmfright{r1}
				\fmf{phantom}{t1,m1,v1}  
				\fmffreeze
				\fmf{fermion}{tr,m1}  
				\fmf{fermion, label= $d$}{m1,tl}
				
				\fmf{dashes}{m1,r1}
				\fmf{plain,width=3}{t1,tle}
				\fmfv{label=$i$,label.angle=90}{tr}
			\end{fmfgraph*}
	\end{fmffile}} 
	& \parbox{170pt}{\begin{equation*}-\sum_d f_{di} r_d\end{equation*}}  \\ \\
	
	\scalebox{0.5}{
		\begin{fmffile}{r1ipvijka}
			\begin{fmfgraph*}(100,60)
				\fmfstraight
				\fmftop{tle,tl,t1,tr,tl2,t2,tr2,tre}     
				\fmfbottom{vle,vl,v1,vr,vl2,v2,vr2,vre}  
				\fmf{phantom}{t1,m1,v1}
				\fmf{phantom}{t2,m2,v2}
				\fmffreeze
				\fmf{fermion}{t2,m2}  
				\fmf{fermion, label= $m$}{m2,v2}
				\fmf{fermion,left=0.5, label= $l$}{m1,v1}
				\fmf{fermion,left=0.5, label= $d$}{v1,m1}
				\fmf{dashes}{m1,m2}
				\fmf{plain,width=3}{vl,vr2}
				\fmfv{label=$i$,label.angle=90}{t2}
			\end{fmfgraph*}
	\end{fmffile}} 
	& \parbox{170pt}{\begin{equation*}-\sum_{dlm} \braket{lm|di} r_{lm}^d\end{equation*}}  \\ \\
	
	\scalebox{0.5}{
		\begin{fmffile}{r1iplvijka}
			\begin{fmfgraph*}(100,60)
				\fmfstraight
				\fmftop{tle,tl,t1,tr,tl2,t2,tr2,tl3,t3,tr3,tre}     
				\fmfbottom{vle,vl,v1,vr,vl2,v2,vr2,vl3,v3,vr3,vre}  
				\fmf{phantom}{t3,m3,v3}
				\fmf{phantom}{t2,m2,v2}
				\fmffreeze
				\fmf{fermion}{t3,m3}  
				\fmf{fermion, label= $m$}{m3,v3}
				\fmf{fermion,left=0.1, label= $l$}{t1,v1}
				\fmf{fermion,left=0.1, label= $d$}{v1,t1}
				\fmf{fermion,left=0.5, label= $e$}{m2,t2}
				\fmf{fermion,left=0.5, label= $n$}{t2,m2}
				\fmf{dashes}{m3,m2}
				\fmf{plain,width=3}{vl,vr3}
				\fmf{plain}{tl,tr2}
				\fmfv{label=$i$,label.angle=90}{t3}
			\end{fmfgraph*}
	\end{fmffile}} 
	& \parbox{170pt}{\begin{equation*}-\sum_{dlm}\sum_{en}\lambda^{ln}_{ed}\braket{em|ni}r_{lm}^d\end{equation*}}  \\ \\
	
	\scalebox{0.5}{
		\begin{fmffile}{r1ipltvijka}
			\begin{fmfgraph*}(100,60)
				\fmfstraight
				\fmftop{tle,tl,t1,tr,tl2,t2,tr2,tl3,t3,tr3,tl4,t4,tr4,tre}     
				\fmfbottom{vle,vl,v1,vr,vl2,v2,vr2,vl3,v3,vr3,vl4,v4,vr4,vre}  
				\fmf{phantom}{t3,m3,v3}
				\fmf{phantom}{t4,m4,v4}
				\fmffreeze
				\fmfshift{(0.0,+10.0)}{vl2,v2,vr2,vl3,v3,vr3, m4, m3}
				\fmf{fermion}{t4,m4}  
				\fmf{fermion, label= $m$, label.side=left}{m4,v4}
				\fmf{fermion,left=0.25, label= $o$, label.dist=4,label.side=left}{t2,v2}
				\fmf{fermion,left=0.25, label= $f$, label.dist=4,label.side=right}{v2,t2}
				\fmf{fermion,left=0.1, label= $l$, label.dist=4,label.side=left}{t1,v1}
				\fmf{fermion,left=0.1, label= $d$, label.dist=4,label.side=left}{v1,t1}
				\fmf{fermion,left=0.6, label= $n$, label.dist=4,label.side=left}{m3,v3}
				\fmf{fermion,left=0.6, label= $e$, label.dist=4,label.side=left}{v3,m3}
				\fmf{dashes}{m3,m4}
				\fmf{plain}{vl2,vr3}
				\fmf{plain}{tl,tr2}
				\fmf{plain,width=3}{vl,vr4}
				\fmfv{label=$i$,label.angle=90}{t4}
			\end{fmfgraph*}
	\end{fmffile}} 
	& \parbox{170pt}{\begin{equation*}-\sum_{dlm}\sum_{en} \sum_{fo} \lambda^{lo}_{df}t_{on}^{fe}\braket{nm|ei}r_{lm}^d\end{equation*}}  \\ \\
	
	\scalebox{0.5}{
		\begin{fmffile}{r1ipvabij}
			\begin{fmfgraph*}(100,60)
				\fmfstraight
				\fmftop{tle,tl,t1,tr,tl2,t2,tr2,tre}     
				\fmfbottom{vle,vl,v1,vr,vl2,v2,vr2,vre}  
				\fmf{phantom}{t1,m1,v1}
				\fmf{phantom}{t2,m2,v2}
				\fmffreeze
				\fmf{fermion}{tr2,m2}  
				\fmf{fermion, label= $e$,label.side=left}{m2,tl2}
				\fmf{fermion,left=0.5, label= $d$}{m1,t1}
				\fmf{fermion,left=0.5, label= $l$}{t1,m1}
				\fmf{dashes}{m1,m2}
				\fmf{plain,width=3}{tl,t2}
				\fmfv{label=$i$,label.angle=90}{tr2}
			\end{fmfgraph*}
	\end{fmffile}} 
	& \parbox{170pt}{\begin{equation*}-\sum_{edl}  \braket{de|li}r_{de}^l\end{equation*}}  \\ \\
	
	\scalebox{0.5}{
		\begin{fmffile}{r1iptvabij}
			\begin{fmfgraph*}(100,60)
				\fmfstraight
				\fmftop{tle,tl,t1,tr,tl2,t2,tr2,tl3,t3,tr3,tre}     
				\fmfbottom{vle,vl,v1,vr,vl2,v2,vr2,vl3,v3,vr3,vre}  
				\fmf{phantom}{t2,m2,v2}
				\fmf{phantom}{t3,m3,v3}
				\fmffreeze
				\fmf{fermion}{tr3,m3}  
				\fmf{fermion, label= $e$,label.side=left}{m3,tl3}
				\fmf{fermion,left=0.5, label= $m$}{m2,v2}
				\fmf{fermion,left=0.5, label= $f$}{v2,m2}
				\fmf{fermion,left=0.1, label= $d$}{v1,t1}
				\fmf{fermion,left=0.1, label= $l$}{t1,v1}
				\fmf{dashes}{m3,m2}
				\fmf{plain,width=3}{tl,t3}
				\fmf{plain}{vl,vr2}
				\fmfv{label=$i$,label.angle=90}{tr3}
			\end{fmfgraph*}
	\end{fmffile}} 
	& \parbox{170pt}{\begin{equation*}-\sum_{edl} \sum_{fm} t_{lm}^{df}  \braket{de|li}r_{de}^l\end{equation*}} 
	
\end{longtable}

\begin{longtable}{c  c}
	\multicolumn{2}{c}{$\bra{\Phi_0} \comm{ \wtH_{\ldr} }{ \hR^\IPEA } \ket{\Phi^a}$} \\ \hline
	\\
	\scalebox{0.5}{
		\begin{fmffile}{r1eafab}
			\begin{fmfgraph*}(100,60)
				\fmfstraight
				\fmfbottom{tle,tl,t1,tr,tre}     
				\fmftop{vle,vl,v1,vr,vre}  
				\fmfright{r1}
				\fmf{phantom}{t1,m1,v1}  
				\fmf{fermion}{t1,m1}  
				\fmf{fermion, label= $d$}{m1,v1}
				\fmffreeze
				\fmf{dashes}{m1,r1}
				\fmf{plain,width=3}{vr,vl}
				\fmfv{label=$a$,label.angle=-90}{t1}
			\end{fmfgraph*}
	\end{fmffile}} 
	& \parbox{170pt}{\begin{equation*}-\sum_d f_{da} r_d\end{equation*}}  \\ \\
	
	\scalebox{0.5}{
		\begin{fmffile}{r1eafai}
			\begin{fmfgraph*}(100,60)
				\fmfstraight
				\fmfbottom{tle,tl,t1,tr,tre}     
				\fmftop{vle,vl,v1,vr,vre}  
				\fmfright{r1}
				\fmf{phantom}{t1,m1,v1}  
				\fmffreeze
				\fmf{fermion}{tr,m1}  
				\fmf{fermion, label= $l$}{m1,tl}
				
				\fmf{dashes}{m1,r1}
				\fmf{plain,width=3}{t1,tle}
				\fmfv{label=$a$,label.angle=-90}{tr}
			\end{fmfgraph*}
	\end{fmffile}} 
	& \parbox{170pt}{\begin{equation*}-\sum_k f_{la} r_l\end{equation*}}  \\ \\
	
	\scalebox{0.5}{
		\begin{fmffile}{r1eavabci}
			\begin{fmfgraph*}(100,60)
				\fmfstraight
				\fmfbottom{tle,tl,t1,tr,tl2,t2,tr2,tre}     
				\fmftop{vle,vl,v1,vr,vl2,v2,vr2,vre}  
				\fmf{phantom}{t1,m1,v1}
				\fmf{phantom}{t2,m2,v2}
				\fmffreeze
				\fmf{fermion}{t2,m2}  
				\fmf{fermion, label= $e$}{m2,v2}
				\fmf{fermion,left=0.5, label= $d$}{m1,v1}
				\fmf{fermion,left=0.5, label= $l$}{v1,m1}
				\fmf{dashes}{m1,m2}
				\fmf{plain,width=3}{vl,vr2}
				\fmfv{label=$a$,label.angle=-90}{t2}
			\end{fmfgraph*}
	\end{fmffile}} 
	& \parbox{170pt}{\begin{equation*}-\sum_{edl} \braket{de|la} r_{de}^l\end{equation*}}  \\ \\
	
	\scalebox{0.5}{
		\begin{fmffile}{r1eatvabci}
			\begin{fmfgraph*}(100,60)
				\fmfstraight
				\fmfbottom{tle,tl,t1,tr,tl2,t2,tr2,tl3,t3,tr3,tre}     
				\fmftop{vle,vl,v1,vr,vl2,v2,vr2,vl3,v3,vr3,vre}  
				\fmf{phantom}{t3,m3,v3}
				\fmf{phantom}{t2,m2,v2}
				\fmffreeze
				\fmf{fermion}{t3,m3}  
				\fmf{fermion, label= $e$}{m3,v3}
				\fmf{fermion,left=0.1, label= $d$}{t1,v1}
				\fmf{fermion,left=0.1, label= $l$}{v1,t1}
				\fmf{fermion,left=0.5, label= $m$}{m2,t2}
				\fmf{fermion,left=0.5, label= $f$}{t2,m2}
				\fmf{dashes}{m3,m2}
				\fmf{plain,width=3}{vl,vr3}
				\fmf{plain}{tl,tr2}
				\fmfv{label=$a$,label.angle=-90}{t3}
			\end{fmfgraph*}
	\end{fmffile}} 
	& \parbox{170pt}{\begin{equation*}-\sum_{edl}\sum_{fm} t^{ef}_{lm}\braket{me|fa} r_{de}^l\end{equation*}}  \\ \\

	\scalebox{0.5}{
		\begin{fmffile}{r1eavijab}
			\begin{fmfgraph*}(100,60)
				\fmfstraight
				\fmfbottom{tle,tl,t1,tr,tl2,t2,tr2,tre}     
				\fmftop{vle,vl,v1,vr,vl2,v2,vr2,vre}  
				\fmf{phantom}{t1,m1,v1}
				\fmf{phantom}{t2,m2,v2}
				\fmffreeze
				\fmf{fermion}{tr2,m2}  
				\fmf{fermion, label= $m$,label.side=right}{m2,tl2}
				\fmf{fermion,left=0.5, label= $l$}{m1,t1}
				\fmf{fermion,left=0.5, label= $d$}{t1,m1}
				\fmf{dashes}{m1,m2}
				\fmf{plain,width=3}{tl,t2}
				\fmfv{label=$a$,label.angle=-90}{tr2}
			\end{fmfgraph*}
	\end{fmffile}} 
	& \parbox{170pt}{\begin{equation*}-\sum_{dlm}  \braket{lm|da}r_{lm}^d\end{equation*}}  \\ \\
	
	\scalebox{0.5}{
		\begin{fmffile}{r1ealvijab}
			\begin{fmfgraph*}(100,60)
				\fmfstraight
				\fmfbottom{tle,tl,t1,tr,tl2,t2,tr2,tl3,t3,tr3,tre}     
				\fmftop{vle,vl,v1,vr,vl2,v2,vr2,vl3,v3,vr3,vre}  
				\fmf{phantom}{t2,m2,v2}
				\fmf{phantom}{t3,m3,v3}
				\fmffreeze
				\fmf{fermion}{tr3,m3}  
				\fmf{fermion, label= $m$,label.side=right}{m3,tl3}
				\fmf{fermion,left=0.5, label= $e$}{m2,v2}
				\fmf{fermion,left=0.5, label= $n$}{v2,m2}
				\fmf{fermion,left=0.1, label= $l$}{v1,t1}
				\fmf{fermion,left=0.1, label= $d$}{t1,v1}
				\fmf{dashes}{m3,m2}
				\fmf{plain,width=3}{tl,t3}
				\fmf{plain}{vl,vr2}
				\fmfv{label=$a$,label.angle=-90}{tr3}
			\end{fmfgraph*}
	\end{fmffile}} 
	& \parbox{170pt}{\begin{equation*}-\sum_{dlm} \sum_{en} \lambda_{de}^{ln}  \braket{em|na}r_{lm}^d\end{equation*}}  \\ \\
	
	\scalebox{0.5}{
		\begin{fmffile}{r1ealtvijab}
			\begin{fmfgraph*}(100,60)
				\fmfstraight
				\fmftop{tle,tl,t1,tr,tl2,t2,tr2,t,tl3,t3,tr3,tl4,t4,tr4,tre}     
				\fmfbottom{vle,vl,v1,vr,vl2,v2,vr2,v,vl3,v3,vr3,vl4,v4,vr4,vre}  
				\fmf{phantom}{t2,m2,v2}
				\fmf{phantom}{t3,m3,v3}
				\fmffreeze
				\fmfshift{(5,0)}{v3}
				\fmf{fermion}{v,m3}  
				\fmf{fermion, label= $m$, label.side=left}{m3,v3}
				\fmf{fermion,left=0.1, label= $d$, label.dist=4,label.side=left}{v4,t4}
				\fmf{fermion,left=0.1, label= $l$, label.dist=4,label.side=left}{t4,v4}
				\fmf{fermion,left=0.1, label= $n$, label.dist=4,label.side=left}{t1,v1}
				\fmf{fermion,left=0.1, label= $e$, label.dist=4,label.side=left}{v1,t1}
				\fmf{fermion,left=0.5, label= $o$, label.dist=4,label.side=right}{m2,v2}
				\fmf{fermion,left=0.5, label= $f$, label.dist=4,label.side=left}{v2,m2}
				\fmf{dashes}{m2,m3}
				\fmf{plain}{vl,vr2}
				\fmf{plain}{tl,tr4}
				\fmf{plain,width=3}{vl3,vr4}
				\fmfv{label=$a$,label.angle=-90}{v}
			\end{fmfgraph*}
	\end{fmffile}} 
	& \parbox{170pt}{\begin{equation*}-\sum_{dlm}\sum_{en} \sum_{fo} \lambda^{nl}_{ed}t_{no}^{ef}\braket{om|fa}r_{lm}^d\end{equation*}}

\end{longtable}

\begin{longtable}{c  c}
	\multicolumn{2}{c}{$\bra{\Phi_{ji}^b} \comm{ \wtH_{\ldr} }{ \hR^\IPEA }\ket{\Phi_0}$} \\ \hline
	\\

	\scalebox{0.5}{
		\begin{fmffile}{r2ipfo}
			\begin{fmfgraph*}(100,60)
				\fmfstraight
				\fmftop{tle,tl,t1,tr,tl2,t2,tr2,tre}     
				\fmfbottom{vle,vl,v1,vr,vl2,v2,vr2,vre}  
				\fmfright{r1}
				\fmf{phantom}{t2,m2,v2}
				\fmffreeze
				\fmf{fermion}{v1,tl}
				\fmf{fermion}{tr,v1}
				\fmf{fermion}{t2,m2}  
				\fmf{fermion, label= $l$, label.side=left}{m2,v2}
				\fmf{dashes}{m2,r1}
				\fmf{plain,width=3}{vl,vr2}
				\fmfv{label=$b$,label.angle=90}{tl}
				\fmfv{label=$j$,label.angle=90}{tr}
				\fmfv{label=$i$,label.angle=90}{t2}
			\end{fmfgraph*}
	\end{fmffile}} 
	& \parbox{170pt}{\begin{equation*}-\sum_{l} f_{li}r_{jl}^{b}-\sum_{l} f_{lj}r_{li}^{b}\end{equation*}}  \\ \\
	
	\scalebox{0.5}{
		\begin{fmffile}{r2ipfv}
			\begin{fmfgraph*}(100,60)
				\fmfstraight
				\fmftop{tle,tl,t1,tr,tl2,t2,tr2,tre}     
				\fmfbottom{vle,vl,v1,vr,vl2,v2,vr2,vre}  
				\fmfleft{l1}
				\fmf{phantom}{tl,m1,v1}
				\fmffreeze
				\fmf{fermion}{m1,tl}
				\fmf{fermion}{tr,v1}
				\fmf{fermion}{t2,v2}  
				\fmf{fermion, label= $d$, label.side=left}{v1,m1}
				\fmf{dashes}{m1,l1}
				\fmf{plain,width=3}{vl,vr2}
				\fmfv{label=$b$,label.angle=90}{tl}
				\fmfv{label=$j$,label.angle=90}{tr}
				\fmfv{label=$i$,label.angle=90}{t2}
			\end{fmfgraph*}
	\end{fmffile}} 
	& \parbox{170pt}{\begin{equation*}\sum_{d} f_{bd}r_{ji}^{d}\end{equation*}}  \\ \\
	
	\scalebox{0.5}{
		\begin{fmffile}{r2ipvabij}
			\begin{fmfgraph*}(100,60)
				\fmfstraight
				\fmftop{tle,tl,t1,tr,t,tl2,t2,tr2,tre}     
				\fmfbottom{vle,vl,v1,vr,v,vl2,v2,vr2,vre}  
				\fmf{phantom}{t1,m1,v1}
				\fmf{phantom}{t2,m2,v2}
				
				\fmffreeze
				\fmf{fermion}{m1,tl}
				\fmf{fermion}{tr,m1}
				\fmf{fermion}{tr2,m2}
				\fmf{fermion, label= $d$, label.side=left}{m2,tl2}
				\fmf{dashes}{m1,m2}
				\fmf{plain,width=3}{t,t2}
				\fmfv{label=$b$,label.angle=90}{tl}
				\fmfv{label=$j$,label.angle=90}{tr}
				\fmfv{label=$i$,label.angle=90}{tr2}
			\end{fmfgraph*}
	\end{fmffile}} 
	& \parbox{170pt}{\begin{equation*}-\sum_{d} \braket{bd|ji}r_{d}\end{equation*}}  \\ \\
	
	\scalebox{0.5}{
		\begin{fmffile}{r2iptvabij}
			\begin{fmfgraph*}(100,60)
				\fmfstraight
				\fmftop{tle,tl,t1,tr,tl2,t2,tr2,tl3,t3,tr3,tre}     
				\fmfbottom{vle,vl,v1,vr,vl2,v2,vr2,vl3,v3,vr3,vre}  
				\fmf{phantom}{t2,m2,v2}
				\fmf{phantom}{t3,m3,v3}
				
				\fmffreeze
				\fmf{fermion}{v1,tl}
				\fmf{fermion}{tr,v1}
				\fmf{fermion}{tr3,m3}
				\fmf{fermion, label= $d$, label.side=left}{m3,tl3}
				\fmf{fermion,left=0.5, label= $l$}{m2,v2}
				\fmf{fermion,left=0.5, label= $e$}{v2,m2}
				\fmf{dashes}{m3,m2}
				\fmf{plain,width=3}{tr2,t3}
				\fmf{plain}{vl,vr2}
				\fmfv{label=$b$,label.angle=90}{tl}
				\fmfv{label=$j$,label.angle=90}{tr}
				\fmfv{label=$i$,label.angle=90}{tr3}
			\end{fmfgraph*}
	\end{fmffile}} 
	& \parbox{170pt}{\begin{equation*}-\sum_{d} \sum_{el}t_{jl}^{be} \braket{ld|ei}r_{d}\end{equation*}}  \\ \\
	
	\scalebox{0.5}{
		\begin{fmffile}{r2ipvijka}
			\begin{fmfgraph*}(100,60)
				\fmfstraight
				\fmftop{tle,tl,t1,tr,tl2,t2,tr2,tre}     
				\fmfbottom{vle,vl,v1,vr,vl2,v2,vr2,vre}  
				\fmf{phantom}{t1,m1,v1}
				\fmf{phantom}{t2,m2,v2}
				
				\fmffreeze
				\fmf{fermion}{m1,tl}
				\fmf{fermion}{tr,m1}
				\fmf{fermion}{t2,m2}
				\fmf{fermion, label= $l$, label.side=left}{m2,v2}
				\fmf{dashes}{m1,m2}
				\fmf{plain,width=3}{vr2,vl2}
				\fmfv{label=$b$,label.angle=90}{tl}
				\fmfv{label=$j$,label.angle=90}{tr}
				\fmfv{label=$i$,label.angle=90}{t2}
			\end{fmfgraph*}
	\end{fmffile}} 
	& \parbox{170pt}{\begin{equation*}-\sum_{l} \braket{bl|ji}r_{l}\end{equation*}}  \\ \\
	
	\scalebox{0.5}{
		\begin{fmffile}{r2iptvijka}
			\begin{fmfgraph*}(100,60)
				\fmfstraight
				\fmftop{tle,tl,t1,tr,tl2,t2,tr2,tl3,t3,tr3,tre}     
				\fmfbottom{vle,vl,v1,vr,vl2,v2,vr2,vl3,v3,vr3,vre}  
				\fmf{phantom}{t2,m2,v2}
				\fmf{phantom}{t3,m3,v3}
				
				\fmffreeze
				\fmf{fermion}{v1,tl}
				\fmf{fermion}{tr,v1}
				\fmf{fermion}{t3,m3}
				\fmf{fermion, label= $l$, label.side=left}{m3,v3}
				\fmf{fermion,left=0.5, label= $m$}{m2,v2}
				\fmf{fermion,left=0.5, label= $d$}{v2,m2}
				\fmf{dashes}{m3,m2}
				\fmf{plain,width=3}{vl3,vr3}
				\fmf{plain}{vl,vr2}
				\fmfv{label=$b$,label.angle=90}{tl}
				\fmfv{label=$j$,label.angle=90}{tr}
				\fmfv{label=$i$,label.angle=90}{t3}
			\end{fmfgraph*}
	\end{fmffile}} 
	& \parbox{170pt}{\begin{equation*}-\sum_{l} \sum_{md}t_{jm}^{bd} \braket{ml|di}r_{l}\end{equation*}}  \\ \\
	
	\scalebox{0.5}{
		\begin{fmffile}{r2ipr2}
			\begin{fmfgraph*}(100,60)
				\fmfstraight
				\fmftop{tle,tl,t1,tr,tl2,t2,tr2,tl3,t3,tr3,tre}     
				\fmfbottom{vle,vl,v1,vr,vl2,v2,vr2,vl3,v3,vr3,vre}  
				\fmf{phantom}{t1,m1,v1}
				\fmf{phantom}{t2,m2,v2}
				
				\fmffreeze
				\fmf{fermion}{m1,tl}
				\fmf{fermion}{tr,m1}
				\fmf{fermion}{t3,v3}
				\fmf{fermion,left=0.5, label= $l$}{m2,v2}
				\fmf{fermion,left=0.5, label= $d$}{v2,m2}
				\fmf{dashes}{m1,m2}
				\fmf{plain,width=3}{vl2,vr3}
				\fmfv{label=$b$,label.angle=90}{tl}
				\fmfv{label=$j$,label.angle=90}{tr}
				\fmfv{label=$i$,label.angle=90}{t3}
			\end{fmfgraph*}
	\end{fmffile}} 
	& \parbox{170pt}{\begin{equation*} \sum_{dl} \braket{bl|jd}r_{li}^d\end{equation*}}  \\ \\
	
	\scalebox{0.5}{
		\begin{fmffile}{r2ipr2t}
			\begin{fmfgraph*}(100,60)
				\fmfstraight
				\fmftop{tle,tl,t1,tr,tl2,t2,tr2,tl3,t3,tr3,tl4,t4,tr4,tre}     
				\fmfbottom{vle,vl,v1,vr,vl2,v2,vr2,vl3,v3,vr3,vl4,v4,vr4,vre}  
				\fmf{phantom}{t3,m3,v3}
				\fmf{phantom}{t2,m2,v2}
				
				\fmffreeze
				\fmf{fermion}{v1,tl}
				\fmf{fermion}{tr,v1}
				\fmf{fermion}{t4,v4}
				\fmf{fermion,left=0.5, label= $l$,label.dist=3,label.side=left}{m3,v3}
				\fmf{fermion,left=0.5, label= $d$,label.dist=3,label.side=right}{v3,m3}
				\fmf{fermion,left=0.5, label= $m$,label.dist=3,label.side=right}{m2,v2}
				\fmf{fermion,left=0.5, label= $e$,label.dist=3,label.side=left}{v2,m2}
				\fmf{dashes}{m3,m2}
				\fmf{plain,width=3}{vl3,vr4}
				\fmf{plain}{vl,vr2}
				\fmfv{label=$b$,label.angle=90}{tl}
				\fmfv{label=$j$,label.angle=90}{tr}
				\fmfv{label=$i$,label.angle=90}{t4}
			\end{fmfgraph*}
	\end{fmffile}} 
	& \parbox{170pt}{\begin{equation*} \sum_{dl} \sum_{em} t_{jm}^{be} \braket{ml|ed}r_{li}^d\end{equation*}}  \\ \\
	
	\scalebox{0.5}{
		\begin{fmffile}{r2ipcc1}
			\begin{fmfgraph*}(130,60)
				\fmfstraight
				\fmftop{tle,tl,t1,tr,tl2,t2,tr2,tl3,t3,tr3,tl4,t4,tr4,tre}     
				\fmfbottom{vle,vl,v1,vr,vl2,v2,vr2,vl3,v3,vr3,vl4,v4,vr4,vre}  
				\fmf{phantom}{t1,m1,v1}
				\fmf{phantom}{t2,m2,v2}
				
				\fmffreeze
				\fmf{fermion}{m1,tl}
				\fmf{fermion}{tr,m1}
				\fmf{fermion}{t4,v4}
				\fmf{fermion,left=0.1, label= $l$}{t3,v3}
				\fmf{fermion,left=0.1, label= $d$}{v3,t3}
				\fmf{fermion,left=0.5, label= $m$,label.dist=3,label.side=right}{t2,m2}
				\fmf{fermion,left=0.5, label= $e$,label.dist=3,label.side=left}{m2,t2}
				\fmf{dashes}{m1,m2}
				\fmf{plain,width=3}{vl3,vr4}
				\fmf{plain}{tl2,tr3}
				\fmfv{label=$b$,label.angle=90}{tl}
				\fmfv{label=$j$,label.angle=90}{tr}
				\fmfv{label=$i$,label.angle=90}{t4}
			\end{fmfgraph*}
	\end{fmffile}} 
	& \parbox{170pt}{\begin{equation*} \sum_{dl} \sum_{em}\color{red}  \braket{be|jm} \color{black} \lambda_{ed}^{ml} r_{li}^d\end{equation*}}  \\ \\
	
	\scalebox{0.5}{
		\begin{fmffile}{r2ipcc2}
			\begin{fmfgraph*}(130,60)
				\fmfstraight
				\fmftop{tle,tl,t1,tr,tl2,t2,tr2,tl3,t3,tr3,tl4,t4,tr4,tre}     
				\fmfbottom{vle,vl,v1,vr,vl2,v2,vr2,vl3,v3,vr3,vl4,v4,vr4,vre}  
				\fmfleft{l1}
				\fmf{phantom}{tl,m1,v1}
				\fmf{phantom}{t2,m2,v2}
				
				\fmffreeze
				\fmf{fermion}{m1,tl}
				\fmf{fermion}{tr,v1}
				\fmf{fermion}{t4,v4}
				\fmf{fermion,left=0.1, label= $l$,label.dist=3}{t3,v3}
				\fmf{fermion,left=0.1, label= $d$,label.dist=3}{v3,t3}
				\fmf{fermion,left=0.1, label= $m$,label.dist=3}{t2,v2}
				\fmf{fermion,left=0.1, label= $e$,label.dist=3}{v2,t2}
				\fmf{fermion, label= $f$,label.dist=3}{v1,m1}
				\fmf{dashes}{m1,l1}
				\fmf{plain,width=3}{vl3,vr4}
				\fmf{plain}{tl2,tr3}
				\fmf{plain}{vl,vr2}
				\fmfv{label=$b$,label.angle=90}{tl}
				\fmfv{label=$j$,label.angle=90}{tr}
				\fmfv{label=$i$,label.angle=90}{t4}
			\end{fmfgraph*}
	\end{fmffile}} 
	& \parbox{170pt}{\begin{equation*} \sum_{dl} \sum_{em} \color{red} \sum_f t_{jm}^{fe} f_{bf} \color{black}\lambda_{ed}^{ml} r_{li}^d\end{equation*}}  \\ \\
	
	\scalebox{0.5}{
		\begin{fmffile}{r2ipcc3}
			\begin{fmfgraph*}(130,60)
				\fmfstraight
				\fmftop{tle,tl,t1,tr,tl2,t2,tr2,tl3,t3,tr3,tl4,t4,tr4,tre}     
				\fmfbottom{vle,vl,v1,vr,vl2,v2,vr2,vl3,v3,vr3,vl4,v4,vr4,vre}  
				\fmfleft{l1}
				\fmf{phantom}{tl,m1,v1}
				\fmf{phantom}{t2,m2,v2}
				
				\fmffreeze
				\fmf{fermion}{tl,m1}
				\fmf{fermion}{v1,tr}
				\fmf{fermion}{t4,v4}
				\fmf{fermion,left=0.1, label= $l$,label.dist=3}{t3,v3}
				\fmf{fermion,left=0.1, label= $d$,label.dist=3}{v3,t3}
				\fmf{fermion,left=0.1, label= $m$,label.dist=3}{t2,v2}
				\fmf{fermion,left=0.1, label= $e$,label.dist=3}{v2,t2}
				\fmf{fermion, label= $n$,label.dist=3, label.side=right}{m1,v1}
				\fmf{dashes}{m1,l1}
				\fmf{plain,width=3}{vl3,vr4}
				\fmf{plain}{tl2,tr3}
				\fmf{plain}{vl,vr2}
				\fmfv{label=$j$,label.angle=90}{tl}
				\fmfv{label=$b$,label.angle=90}{tr}
				\fmfv{label=$i$,label.angle=90}{t4}
			\end{fmfgraph*}
	\end{fmffile}} 
	& \parbox{170pt}{\begin{equation*} -\sum_{dl} \sum_{em} \color{red}  \sum_n t_{nm}^{be} f_{nj}  \color{black}\lambda_{ed}^{ml} r_{li}^d\end{equation*}}  \\ \\
	
	\scalebox{0.5}{
		\begin{fmffile}{r2ipcc4}
			\begin{fmfgraph*}(130,60)
				\fmfstraight
				\fmftop{tle,tl,t1,tr,tl2,t2,tr2,tl3,t3,tr3,tl4,t4,tr4,tre}     
				\fmfbottom{vle,vl,v1,vr,vl2,v2,vr2,vl3,v3,vr3,vl4,v4,vr4,vre}  
				\fmf{phantom}{tr,m1,v1}
				\fmf{phantom}{t2,m2,v2}
				
				\fmffreeze
				\fmfshift{(10,0)}{m1}
				\fmfshift{(-6,0)}{m2}
				\fmf{fermion}{v1,tl}
				\fmf{fermion}{tr,v1}
				\fmf{fermion}{t4,v4}
				\fmf{fermion,left=0.1, label= $l$,label.dist=3}{t3,v3}
				\fmf{fermion,left=0.1, label= $d$,label.dist=3}{v3,t3}
				\fmf{fermion,left=0.1, label= $m$,label.dist=3}{t2,v2}
				\fmf{fermion,left=0.09, label= $f$,label.dist=3}{v2,m2}
				\fmf{fermion,left=0.09, label= $e$,label.dist=3}{m2,t2}
				\fmf{dashes}{m1,m2}
				\fmf{plain,width=3}{vl3,vr4}
				\fmf{plain}{tl2,tr3}
				\fmf{plain}{vl,vr2}
				\fmfv{label=$b$,label.angle=90}{tl}
				\fmfv{label=$j$,label.angle=90}{tr}
				\fmfv{label=$i$,label.angle=90}{t4}
			\end{fmfgraph*}
	\end{fmffile}} 
	& \parbox{170pt}{\begin{equation*} \sum_{dl} \sum_{em} \color{red} \sum_f t_{jm}^{bf} f_{ef} \color{black} \lambda_{ed}^{ml} r_{li}^d\end{equation*}}  \\ \\
	
	\scalebox{0.5}{
		\begin{fmffile}{r2ipcc5}
			\begin{fmfgraph*}(130,60)
				\fmfstraight
				\fmftop{tle,tl,t1,tr,tl2,t2,tr2,tl3,t3,tr3,tl4,t4,tr4,tre}     
				\fmfbottom{vle,vl,v1,vr,vl2,v2,vr2,vl3,v3,vr3,vl4,v4,vr4,vre}  
				\fmf{phantom}{tr,m1,v1}
				\fmf{phantom}{t2,m2,v2}
				
				\fmffreeze
				\fmfshift{(10,0)}{m1}
				\fmfshift{(-6,0)}{m2}
				\fmf{fermion}{v1,tl}
				\fmf{fermion}{tr,v1}
				\fmf{fermion}{t4,v4}
				\fmf{fermion,left=0.1, label= $l$,label.dist=3}{t3,v3}
				\fmf{fermion,left=0.1, label= $d$,label.dist=3}{v3,t3}
				\fmf{fermion,right=0.1, label= $e$,label.dist=3}{v2,t2}
				\fmf{fermion,right=0.09, label= $m$,label.dist=3}{t2,m2}
				\fmf{fermion,right=0.09, label= $n$,label.dist=3}{m2,v2}
				\fmf{dashes}{m1,m2}
				\fmf{plain,width=3}{vl3,vr4}
				\fmf{plain}{tl2,tr3}
				\fmf{plain}{vl,vr2}
				\fmfv{label=$b$,label.angle=90}{tl}
				\fmfv{label=$j$,label.angle=90}{tr}
				\fmfv{label=$i$,label.angle=90}{t4}
			\end{fmfgraph*}
	\end{fmffile}} 
	& \parbox{170pt}{\begin{equation*} -\sum_{dl} \sum_{em}\color{red}  \sum_n t_{jn}^{be} f_{nm} \color{black} \lambda_{ed}^{ml} r_{li}^d\end{equation*}}  \\ \\
	
	\scalebox{0.5}{
		\begin{fmffile}{r2ipcc6}
			\begin{fmfgraph*}(150,60)
				\fmfstraight
				\fmftop{tle,tl0,t0,tr0,tl,t1,tr,tl2,t2,tr2,tl3,t3,tr3,tl4,t4,tr4,tre}     
				\fmfbottom{vle,vl0,v0,vr0,vl,v1,vr,vl2,v2,vr2,vl3,v3,vr3,vl4,v4,vr4,vre}  
				\fmf{phantom}{t1,m1,v1}
				\fmf{phantom}{t2,m2,v2}
				
				\fmffreeze
				\fmf{fermion}{v0,tl0}
				\fmf{fermion}{tr0,v0}
				\fmf{fermion}{t4,v4}
				\fmf{fermion,left=0.1, label= $l$,label.dist=3}{t3,v3}
				\fmf{fermion,left=0.1, label= $d$,label.dist=3}{v3,t3}
				\fmf{fermion,left=0.5, label= $e$,label.dist=3}{m2,t2}
				\fmf{fermion,left=0.5, label= $m$,label.dist=3}{t2,m2}
				\fmf{fermion,left=0.5, label= $f$,label.dist=3}{v1,m1}
				\fmf{fermion,left=0.5, label= $n$,label.dist=3}{m1,v1}
				\fmf{dashes}{m1,m2}
				\fmf{plain,width=3}{vl3,vr4}
				\fmf{plain}{tl2,tr3}
				\fmf{plain}{vl0,vr}
				\fmfv{label=$b$,label.angle=90}{tl0}
				\fmfv{label=$j$,label.angle=90}{tr0}
				\fmfv{label=$i$,label.angle=90}{t4}
			\end{fmfgraph*}
	\end{fmffile}} 
	& \parbox{170pt}{\begin{equation*} \sum_{dl} \sum_{em} \color{red} \sum_{fn} t_{jn}^{bf} \braket{ne|fm} \color{black} \lambda_{ed}^{ml} r_{li}^d\end{equation*}}  \\ \\
	
	\scalebox{0.5}{
		\begin{fmffile}{r2ipcc7}
			\begin{fmfgraph*}(150,60)
				\fmfstraight
				\fmftop{tle,tl0,t0,tr0,tl,t1,tr,tl2,t2,tr2,tl3,t3,tr3,tl4,t4,tr4,tre}     
				\fmfbottom{vle,vl0,v0,vr0,vl,v1,vr,vl2,v2,vr2,vl3,v3,vr3,vl4,v4,vr4,vre}  
				\fmf{phantom}{t1,m1,v1}
				\fmf{phantom}{t0,m0,v0}
				
				\fmffreeze
				\fmf{fermion}{m0,tl0}
				\fmf{fermion}{tr0,m0}
				\fmf{fermion}{t4,v4}
				\fmf{fermion,left=0.1, label= $l$,label.dist=3}{t3,v3}
				\fmf{fermion,left=0.1, label= $d$,label.dist=3}{v3,t3}
				\fmf{fermion,left=0.1, label= $e$,label.dist=3}{v2,t2}
				\fmf{fermion,left=0.1, label= $m$,label.dist=3}{t2,v2}
				\fmf{fermion,left=0.5, label= $f$,label.dist=3}{v1,m1}
				\fmf{fermion,left=0.5, label= $n$,label.dist=3}{m1,v1}
				\fmf{dashes}{m1,m0}
				\fmf{plain,width=3}{vl3,vr4}
				\fmf{plain}{tl2,tr3}
				\fmf{plain}{vl,vr2}
				\fmfv{label=$b$,label.angle=90}{tl0}
				\fmfv{label=$j$,label.angle=90}{tr0}
				\fmfv{label=$i$,label.angle=90}{t4}
			\end{fmfgraph*}
	\end{fmffile}} 
	
	& \parbox{170pt}{\begin{equation*} \sum_{dl} \sum_{em} \color{red} \sum_{fn} t_{nm}^{fe} \braket{bn|jf} \color{black} \lambda_{ed}^{ml} r_{li}^d\end{equation*}}  \\ \\
	
	\scalebox{0.5}{
		\begin{fmffile}{r2ipcc8}
			\begin{fmfgraph*}(170,60)
				\fmfstraight
				\fmftop{tle,tl5,t5,tr5,tl0,t0,tr0,tl,t1,tr,tl2,t2,tr2,tl3,t3,tr3,tl4,t4,tr4,tre}     
				\fmfbottom{vle,vl5,v5,vr5,vl0,v0,vr0,vl,v1,vr,vl2,v2,vr2,vl3,v3,vr3,vl4,v4,vr4,vre}  
				\fmf{phantom}{t1,m1,v1}
				\fmf{phantom}{t0,m0,v0}
				
				\fmffreeze
				\fmf{fermion}{v5,tl5}
				\fmf{fermion}{tr5,v5}
				\fmf{fermion}{t4,v4}
				\fmf{fermion,left=0.1, label= $l$,label.dist=3}{t3,v3}
				\fmf{fermion,left=0.1, label= $d$,label.dist=3}{v3,t3}
				\fmf{fermion,left=0.1, label= $e$,label.dist=3}{v2,t2}
				\fmf{fermion,left=0.1, label= $m$,label.dist=3}{t2,v2}
				\fmf{fermion,left=0.5, label= $f$,label.dist=3,label.side=right}{v1,m1}
				\fmf{fermion,left=0.5, label= $n$,label.dist=3}{m1,v1}
				\fmf{fermion,left=0.5, label= $g$,label.dist=3}{v0,m0}
				\fmf{fermion,left=0.5, label= $o$,label.dist=3,label.side=right}{m0,v0}
				\fmf{dashes}{m1,m0}
				\fmf{plain,width=3}{vl3,vr4}
				\fmf{plain}{tl2,tr3}
				\fmf{plain}{vl,vr2}
				\fmf{plain}{vl5,vr0}
				\fmfv{label=$b$,label.angle=90}{tl5}
				\fmfv{label=$j$,label.angle=90}{tr5}
				\fmfv{label=$i$,label.angle=90}{t4}
			\end{fmfgraph*}
	\end{fmffile}} 
	
	& \parbox{170pt}{\begin{equation*} \sum_{dl} \sum_{em}\color{red}  \sum_{fn} \sum_{go} t_{jo}^{bg} \braket{on|gf} t^{fe}_{nm} \color{black} \lambda_{ed}^{ml} r_{li}^d\end{equation*}}

\end{longtable}

\begin{longtable}{c  c}
	\multicolumn{2}{c}{$\bra{\Phi_0} \comm{ \wtH_{\ldr} }{ \hR^\IPEA } \ket{\Phi^{ba}_j}$} \\ \hline
	\\

	\scalebox{0.5}{
		\begin{fmffile}{r2eafo}
			\begin{fmfgraph*}(100,60)
				\fmfstraight
				\fmfbottom{tle,tl,t1,tr,tl2,t2,tr2,tre}     
				\fmftop{vle,vl,v1,vr,vl2,v2,vr2,vre}  
				\fmfright{r1}
				\fmf{phantom}{t2,m2,v2}
				\fmffreeze
				\fmf{fermion}{v1,tl}
				\fmf{fermion}{tr,v1}
				\fmf{fermion}{t2,m2}  
				\fmf{fermion, label= $d$, label.side=left}{m2,v2}
				\fmf{dashes}{m2,r1}
				\fmf{plain,width=3}{vl,vr2}
				\fmfv{label=$j$,label.angle=-90}{tl}
				\fmfv{label=$b$,label.angle=-90}{tr}
				\fmfv{label=$a$,label.angle=-90}{t2}
			\end{fmfgraph*}
	\end{fmffile}} 
	& \parbox{170pt}{\begin{equation*}-\sum_{d} f_{da}r_{bd}^{j}-\sum_{d} f_{db}r_{da}^{j}\end{equation*}}  \\ \\
	
	\scalebox{0.5}{
		\begin{fmffile}{r2eafv}
			\begin{fmfgraph*}(100,60)
				\fmfstraight
				\fmfbottom{tle,tl,t1,tr,tl2,t2,tr2,tre}     
				\fmftop{vle,vl,v1,vr,vl2,v2,vr2,vre}  
				\fmfleft{l1}
				\fmf{phantom}{tl,m1,v1}
				\fmffreeze
				\fmf{fermion}{m1,tl}
				\fmf{fermion}{tr,v1}
				\fmf{fermion}{t2,v2}  
				\fmf{fermion, label= $l$, label.side=right}{v1,m1}
				\fmf{dashes}{m1,l1}
				\fmf{plain,width=3}{vl,vr2}
				\fmfv{label=$j$,label.angle=-90}{tl}
				\fmfv{label=$b$,label.angle=-90}{tr}
				\fmfv{label=$a$,label.angle=-90}{t2}
			\end{fmfgraph*}
	\end{fmffile}} 
	& \parbox{170pt}{\begin{equation*}\sum_{l} f_{lj}r_{ba}^{l}\end{equation*}}  \\ \\
	
	\scalebox{0.5}{
		\begin{fmffile}{r2eavabij}
			\begin{fmfgraph*}(100,60)
				\fmfstraight
				\fmfbottom{tle,tl,t1,tr,t,tl2,t2,tr2,tre}     
				\fmftop{vle,vl,v1,vr,v,vl2,v2,vr2,vre}  
				\fmf{phantom}{t1,m1,v1}
				\fmf{phantom}{t2,m2,v2}
				
				\fmffreeze
				\fmf{fermion}{m1,tl}
				\fmf{fermion}{tr,m1}
				\fmf{fermion}{tr2,m2}
				\fmf{fermion, label= $l$, label.side=right}{m2,tl2}
				\fmf{dashes}{m1,m2}
				\fmf{plain,width=3}{t,t2}
				\fmfv{label=$j$,label.angle=-90}{tl}
				\fmfv{label=$b$,label.angle=-90}{tr}
				\fmfv{label=$a$,label.angle=-90}{tr2}
			\end{fmfgraph*}
	\end{fmffile}} 
	& \parbox{170pt}{\begin{equation*}-\sum_{l} \braket{jl|ba}r_{l}\end{equation*}}  \\ \\
	
	\scalebox{0.5}{
		\begin{fmffile}{r2eatvabij}
			\begin{fmfgraph*}(100,60)
				\fmfstraight
				\fmfbottom{tle,tl,t1,tr,tl2,t2,tr2,tl3,t3,tr3,tre}     
				\fmftop{vle,vl,v1,vr,vl2,v2,vr2,vl3,v3,vr3,vre}  
				\fmf{phantom}{t2,m2,v2}
				\fmf{phantom}{t3,m3,v3}
				
				\fmffreeze
				\fmf{fermion}{v1,tl}
				\fmf{fermion}{tr,v1}
				\fmf{fermion}{tr3,m3}
				\fmf{fermion, label= $l$, label.side=right}{m3,tl3}
				\fmf{fermion,left=0.5, label= $d$}{m2,v2}
				\fmf{fermion,left=0.5, label= $m$}{v2,m2}
				\fmf{dashes}{m3,m2}
				\fmf{plain,width=3}{tr2,t3}
				\fmf{plain}{vl,vr2}
				\fmfv{label=$j$,label.angle=-90}{tl}
				\fmfv{label=$b$,label.angle=-90}{tr}
				\fmfv{label=$a$,label.angle=-90}{tr3}
			\end{fmfgraph*}
	\end{fmffile}} 
	& \parbox{170pt}{\begin{equation*}-\sum_{l} \sum_{dm}\lambda^{jm}_{bd} \braket{dl|ma}r_{l}\end{equation*}}  \\ \\
	
	\scalebox{0.5}{
		\begin{fmffile}{r2ealtvabij}
			\begin{fmfgraph*}(100,60)
				\fmfstraight
				\fmfbottom{tle,tl0,t0,tr0,tl,t1,tr,tl2,t2,tr2,t,tl3,t3,tr3,tre}     
				\fmftop{vle,vl0,v0,vr0,vl,v1,vr,vl2,v2,vr2,v,vl3,v3,vr3,vre}  
				\fmf{phantom}{t2,m2,v2}
				\fmf{phantom}{t3,m3,v3}
				
				\fmffreeze
				\fmf{fermion}{v0,tl0}
				\fmf{fermion}{tr0,v0}
				\fmf{fermion}{tr3,m3}
				\fmf{fermion, label= $l$, label.side=right}{m3,tl3}
				\fmf{fermion,left=0.5, label= $d$,label.side=left,label.dist=3}{t2,m2}
				\fmf{fermion,left=0.5, label= $m$,label.side=right,label.dist=3}{m2,t2}
				\fmf{fermion,left=0.1, label= $e$,label.side=left,label.dist=3}{t1,v1}
				\fmf{fermion,left=0.1, label= $n$,label.side=left,label.dist=3}{v1,t1}
				\fmf{dashes}{m3,m2}
				\fmf{plain,width=3}{t,t3}
				\fmf{plain}{vl0,vr}
				\fmf{plain}{tl,tr2}
				\fmfv{label=$j$,label.angle=-90}{tl0}
				\fmfv{label=$b$,label.angle=-90}{tr0}
				\fmfv{label=$a$,label.angle=-90}{tr3}
			\end{fmfgraph*}
	\end{fmffile}} 
	& \parbox{170pt}{\begin{equation*}-\sum_{l} \sum_{dm} \sum_{en}\lambda^{jn}_{be} t^{ed}_{nm}\braket{ml|da}r_{l}\end{equation*}}  \\ \\
	
	\scalebox{0.5}{
		\begin{fmffile}{r2eavijka}
			\begin{fmfgraph*}(100,60)
				\fmfstraight
				\fmfbottom{tle,tl,t1,tr,tl2,t2,tr2,tre}     
				\fmftop{vle,vl,v1,vr,vl2,v2,vr2,vre}  
				\fmf{phantom}{t1,m1,v1}
				\fmf{phantom}{t2,m2,v2}
				
				\fmffreeze
				\fmf{fermion}{m1,tl}
				\fmf{fermion}{tr,m1}
				\fmf{fermion}{t2,m2}
				\fmf{fermion, label= $d$, label.side=right}{m2,v2}
				\fmf{dashes}{m1,m2}
				\fmf{plain,width=3}{vr2,vl2}
				\fmfv{label=$j$,label.angle=-90}{tl}
				\fmfv{label=$b$,label.angle=-90}{tr}
				\fmfv{label=$a$,label.angle=-90}{t2}
			\end{fmfgraph*}
	\end{fmffile}} 
	& \parbox{170pt}{\begin{equation*}-\sum_{d} \braket{jd|ba}r_{d}\end{equation*}}  \\ \\
	
	\scalebox{0.5}{
		\begin{fmffile}{r2eatvijka}
			\begin{fmfgraph*}(100,60)
				\fmfstraight
				\fmfbottom{tle,tl,t1,tr,tl2,t2,tr2,tl3,t3,tr3,tre}     
				\fmftop{vle,vl,v1,vr,vl2,v2,vr2,vl3,v3,vr3,vre}  
				\fmf{phantom}{t2,m2,v2}
				\fmf{phantom}{t3,m3,v3}
				
				\fmffreeze
				\fmf{fermion}{v1,tl}
				\fmf{fermion}{tr,v1}
				\fmf{fermion}{t3,m3}
				\fmf{fermion, label= $d$, label.side=right}{m3,v3}
				\fmf{fermion,left=0.5, label= $e$}{m2,v2}
				\fmf{fermion,left=0.5, label= $l$}{v2,m2}
				\fmf{dashes}{m3,m2}
				\fmf{plain,width=3}{vl3,vr3}
				\fmf{plain}{vl,vr2}
				\fmfv{label=$j$,label.angle=-90}{tl}
				\fmfv{label=$b$,label.angle=-90}{tr}
				\fmfv{label=$a$,label.angle=-90}{t3}
			\end{fmfgraph*}
	\end{fmffile}} 
	& \parbox{170pt}{\begin{equation*}-\sum_{d} \sum_{el}\lambda_{be}^{jl} \braket{ed|la}r_{d}\end{equation*}}  \\ \\
	
	\scalebox{0.5}{
		\begin{fmffile}{r2ealtvijka}
			\begin{fmfgraph*}(100,60)
				\fmfstraight
				\fmfbottom{tle,tl0,t0,tr0,tl,t1,tr,tl2,t2,tr2,tl3,t3,tr3,tre}     
				\fmftop{vle,vl0,v0,vr0,vl,v1,vr,vl2,v2,vr2,vl3,v3,vr3,vre}  
				\fmf{phantom}{t2,m2,v2}
				\fmf{phantom}{t3,m3,v3}
				
				\fmffreeze
				\fmf{fermion}{v0,tl0}
				\fmf{fermion}{tr0,v0}
				\fmf{fermion}{t3,m3}
				\fmf{fermion, label= $d$, label.side=right}{m3,v3}
				\fmf{fermion,left=0.5, label= $e$,label.side=left,label.dist=3}{t2,m2}
				\fmf{fermion,left=0.5, label= $l$,label.side=right,label.dist=3}{m2,t2}
				\fmf{fermion,left=0.1, label= $f$,label.side=left,label.dist=3}{t1,v1}
				\fmf{fermion,left=0.1, label= $m$,label.side=left,label.dist=3}{v1,t1}
				\fmf{dashes}{m3,m2}
				\fmf{plain,width=3}{vr3,vl3}
				\fmf{plain}{vl0,vr}
				\fmf{plain}{tl,tr2}
				\fmfv{label=$j$,label.angle=-90}{tl0}
				\fmfv{label=$b$,label.angle=-90}{tr0}
				\fmfv{label=$a$,label.angle=-90}{t3}
			\end{fmfgraph*}
	\end{fmffile}} 
	& \parbox{170pt}{\begin{equation*}-\sum_{d} \sum_{el} \sum_{fm}\lambda^{jm}_{bd} t^{de}_{mn}\braket{ld|ea}r_{d}\end{equation*}} \\ \\
	
	\scalebox{0.5}{
		\begin{fmffile}{r2ear2}
			\begin{fmfgraph*}(100,60)
				\fmfstraight
				\fmfbottom{tle,tl,t1,tr,tl2,t2,tr2,tl3,t3,tr3,tre}     
				\fmftop{vle,vl,v1,vr,vl2,v2,vr2,vl3,v3,vr3,vre}  
				\fmf{phantom}{t1,m1,v1}
				\fmf{phantom}{t2,m2,v2}
				
				\fmffreeze
				\fmf{fermion}{m1,tl}
				\fmf{fermion}{tr,m1}
				\fmf{fermion}{t3,v3}
				\fmf{fermion,left=0.5, label= $d$}{m2,v2}
				\fmf{fermion,left=0.5, label= $l$}{v2,m2}
				\fmf{dashes}{m1,m2}
				\fmf{plain,width=3}{vl2,vr3}
				\fmfv{label=$j$,label.angle=-90}{tl}
				\fmfv{label=$b$,label.angle=-90}{tr}
				\fmfv{label=$a$,label.angle=-90}{t3}
			\end{fmfgraph*}
	\end{fmffile}} 
	& \parbox{170pt}{\begin{equation*} -\sum_{dl} \braket{jd|bl}r^{l}_{da}\end{equation*}}  \\ \\
	
	\scalebox{0.5}{
		\begin{fmffile}{r2eatr2}
			\begin{fmfgraph*}(130,60)
				\fmfstraight
				\fmfbottom{tle,tl,t1,tr,tl2,t2,tr2,tl3,t3,tr3,tl4,t4,tr4,tre}     
				\fmftop{vle,vl,v1,vr,vl2,v2,vr2,vl3,v3,vr3,vl4,v4,vr4,vre}  
				\fmf{phantom}{t1,m1,v1}
				\fmf{phantom}{t2,m2,v2}
				
				\fmffreeze
				\fmf{fermion}{m1,tl}
				\fmf{fermion}{tr,m1}
				\fmf{fermion}{t4,v4}
				\fmf{fermion,left=0.1, label= $d$}{t3,v3}
				\fmf{fermion,left=0.1, label= $l$}{v3,t3}
				\fmf{fermion,left=0.5, label= $e$,label.dist=3,label.side=right}{t2,m2}
				\fmf{fermion,left=0.5, label= $m$,label.dist=3,label.side=left}{m2,t2}
				\fmf{dashes}{m1,m2}
				\fmf{plain,width=3}{vl3,vr4}
				\fmf{plain}{tl2,tr3}
				\fmfv{label=$j$,label.angle=-90}{tl}
				\fmfv{label=$b$,label.angle=-90}{tr}
				\fmfv{label=$a$,label.angle=-90}{t4}
			\end{fmfgraph*}
	\end{fmffile}} 
	& \parbox{170pt}{\begin{equation*} -\sum_{dl} \sum_{em}\braket{jm|be}  t_{ml}^{ed} r^{l}_{da}\end{equation*}}  \\ \\
	
	\scalebox{0.5}{
		\begin{fmffile}{r2eacc1}
			\begin{fmfgraph*}(100,60)
				\fmfstraight
				\fmfbottom{tle,tl,t1,tr,tl2,t2,tr2,tl3,t3,tr3,tl4,t4,tr4,tre}     
				\fmftop{vle,vl,v1,vr,vl2,v2,vr2,vl3,v3,vr3,vl4,v4,vr4,vre}  
				\fmf{phantom}{t3,m3,v3}
				\fmf{phantom}{t2,m2,v2}
				
				\fmffreeze
				\fmf{fermion}{v1,tl}
				\fmf{fermion}{tr,v1}
				\fmf{fermion}{t4,v4}
				\fmf{fermion,left=0.5, label= $d$,label.dist=3,label.side=right}{m3,v3}
				\fmf{fermion,left=0.5, label= $l$,label.dist=3,label.side=left}{v3,m3}
				\fmf{fermion,left=0.5, label= $e$,label.dist=3,label.side=left}{m2,v2}
				\fmf{fermion,left=0.5, label= $m$,label.dist=3,label.side=right}{v2,m2}
				\fmf{dashes}{m3,m2}
				\fmf{plain,width=3}{vl3,vr4}
				\fmf{plain}{vl,vr2}
				\fmfv{label=$j$,label.angle=-90}{tl}
				\fmfv{label=$b$,label.angle=-90}{tr}
				\fmfv{label=$a$,label.angle=-90}{t4}
			\end{fmfgraph*}
	\end{fmffile}} 
	& \parbox{170pt}{\begin{equation*} -\sum_{dl} \sum_{em} \lambda_{be}^{jm} \color{red} \braket{ed|ml} \color{black}  r^{l}_{da}\end{equation*}}  \\ \\

	\scalebox{0.5}{
		\begin{fmffile}{r2eacc2}
			\begin{fmfgraph*}(130,60)
				\fmfstraight
				\fmfbottom{tle,tl,t1,tr,tl2,t2,tr2,tl3,t3,tr3,tl4,t4,tr4,tre}     
				\fmftop{vle,vl,v1,vr,vl2,v2,vr2,vl3,v3,vr3,vl4,v4,vr4,vre}  
				\fmf{phantom}{tr,m1,v1}
				\fmf{phantom}{t2,m2,v2}
				
				\fmffreeze
				\fmfshift{(10,0)}{m1}
				\fmfshift{(-6,0)}{m2}
				\fmf{fermion}{v1,tl}
				\fmf{fermion}{tr,v1}
				\fmf{fermion}{t4,v4}
				\fmf{fermion,right=0.1, label= $d$,label.dist=3}{t3,v3}
				\fmf{fermion,right=0.1, label= $l$,label.dist=3}{v3,t3}
				\fmf{fermion,right=0.1, label= $e$,label.dist=3}{t2,v2}
				\fmf{fermion,right=0.09, label= $m$,label.dist=3}{v2,m2}
				\fmf{fermion,right=0.09, label= $n$,label.dist=3}{m2,t2}
				\fmf{dashes}{m1,m2}
				\fmf{plain,width=3}{vl3,vr4}
				\fmf{plain}{tl2,tr3}
				\fmf{plain}{vl,vr2}
				\fmfv{label=$j$,label.angle=-90}{tl}
				\fmfv{label=$b$,label.angle=-90}{tr}
				\fmfv{label=$a$,label.angle=-90}{t4}
			\end{fmfgraph*}
	\end{fmffile}} 
	& \parbox{170pt}{\begin{equation*} \sum_{dl} \sum_{em} \lambda_{be}^{jm} \color{red} \sum_n t_{nl}^{ed} f_{nm} \color{black}   r^{l}_{da}\end{equation*}}  \\ \\
	
	\scalebox{0.5}{
		\begin{fmffile}{r2eacc3}
			\begin{fmfgraph*}(130,60)
				\fmfstraight
				\fmfbottom{tle,tl,t1,tr,tl2,t2,tr2,tl3,t3,tr3,tl4,t4,tr4,tre}     
				\fmftop{vle,vl,v1,vr,vl2,v2,vr2,vl3,v3,vr3,vl4,v4,vr4,vre}  
				\fmf{phantom}{tr,m1,v1}
				\fmf{phantom}{t2,m2,v2}
				
				\fmffreeze
				\fmfshift{(10,0)}{m1}
				\fmfshift{(-6,0)}{m2}
				\fmf{fermion}{v1,tl}
				\fmf{fermion}{tr,v1}
				\fmf{fermion}{t4,v4}
				\fmf{fermion,left=0.1, label= $d$,label.dist=3}{t3,v3}
				\fmf{fermion,left=0.1, label= $l$,label.dist=3}{v3,t3}
				\fmf{fermion,left=0.1, label= $m$,label.dist=3}{v2,t2}
				\fmf{fermion,left=0.09, label= $f$,label.dist=3}{t2,m2}
				\fmf{fermion,left=0.09, label= $e$,label.dist=3}{m2,v2}
				\fmf{dashes}{m1,m2}
				\fmf{plain,width=3}{vl3,vr4}
				\fmf{plain}{tl2,tr3}
				\fmf{plain}{vl,vr2}
				\fmfv{label=$j$,label.angle=-90}{tl}
				\fmfv{label=$n$,label.angle=-90}{tr}
				\fmfv{label=$a$,label.angle=-90}{t4}
			\end{fmfgraph*}
	\end{fmffile}} 
	& \parbox{170pt}{\begin{equation*} -\sum_{dl} \sum_{em} \lambda_{be}^{jm} \color{red} \sum_f t_{ml}^{fd} f_{ef} \color{black}   r^{l}_{da}\end{equation*}}  \\ \\
	
	\scalebox{0.5}{
		\begin{fmffile}{r2eacc4}
			\begin{fmfgraph*}(130,60)
				\fmfstraight
				\fmfbottom{tle,tl,t1,tr,tl2,t2,tr2,tl3,t3,tr3,tl4,t4,tr4,tre}     
				\fmftop{vle,vl,v1,vr,vl2,v2,vr2,vl3,v3,vr3,vl4,v4,vr4,vre}  
				\fmf{phantom}{tl4,m4,v4}
				\fmf{phantom}{t3,m3,v3}
				
				\fmffreeze
				\fmfshift{(-10,0)}{m4}
				\fmfshift{(6,0)}{m3}
				\fmf{fermion}{v1,tl}
				\fmf{fermion}{tr,v1}
				\fmf{fermion}{t4,v4}
				\fmf{fermion,left=0.1, label= $d$,label.dist=3}{t3,v3}
				\fmf{fermion,left=0.1, label= $l$,label.dist=3}{v3,m3}
				\fmf{fermion,left=0.1, label= $n$,label.dist=3}{m3,t3}
				\fmf{fermion,right=0.1, label= $e$,label.dist=3}{t2,v2}
				\fmf{fermion,right=0.1, label= $m$,label.dist=3}{v2,t2}
				\fmf{dashes}{m3,m4}
				\fmf{plain,width=3}{vl3,vr4}
				\fmf{plain}{tl2,tr3}
				\fmf{plain}{vl,vr2}
				\fmfv{label=$j$,label.angle=-90}{tl}
				\fmfv{label=$b$,label.angle=-90}{tr}
				\fmfv{label=$a$,label.angle=-90}{t4}
			\end{fmfgraph*}
	\end{fmffile}} 
	& \parbox{170pt}{\begin{equation*} \sum_{dl} \sum_{em} \lambda_{be}^{jm} \color{red} \sum_n t_{mn}^{ed} f_{nl} \color{black}   r^{l}_{da}\end{equation*}}  \\ \\
	
	\scalebox{0.5}{
		\begin{fmffile}{r2eacc5}
			\begin{fmfgraph*}(130,60)
				\fmfstraight
				\fmfbottom{tle,tl,t1,tr,tl2,t2,tr2,tl3,t3,tr3,tl4,t4,tr4,tre}     
				\fmftop{vle,vl,v1,vr,vl2,v2,vr2,vl3,v3,vr3,vl4,v4,vr4,vre}  
				\fmf{phantom}{tl4,m4,v4}
				\fmf{phantom}{t3,m3,v3}
				
				\fmffreeze
				\fmfshift{(-10,0)}{m4}
				\fmfshift{(6,0)}{m3}
				\fmf{fermion}{v1,tl}
				\fmf{fermion}{tr,v1}
				\fmf{fermion}{t4,v4}
				\fmf{fermion,right=0.1, label= $d$,label.dist=3}{t3,m3}
				\fmf{fermion,right=0.1, label= $f$,label.dist=3}{m3,v3}
				\fmf{fermion,right=0.1, label= $l$,label.dist=3}{v3,t3}
				\fmf{fermion,right=0.1, label= $e$,label.dist=3}{t2,v2}
				\fmf{fermion,right=0.1, label= $m$,label.dist=3}{v2,t2}
				\fmf{dashes}{m3,m4}
				\fmf{plain,width=3}{vl3,vr4}
				\fmf{plain}{tl2,tr3}
				\fmf{plain}{vl,vr2}
				\fmfv{label=$j$,label.angle=-90}{tl}
				\fmfv{label=$b$,label.angle=-90}{tr}
				\fmfv{label=$a$,label.angle=-90}{t4}
			\end{fmfgraph*}
	\end{fmffile}} 
	& \parbox{170pt}{\begin{equation*} -\sum_{dl} \sum_{em} \lambda_{be}^{jm} \color{red} \sum_f t_{ml}^{ed} f_{fd} \color{black}   r^{l}_{da}\end{equation*}}  \\ \\
	
	\scalebox{0.5}{
		\begin{fmffile}{r2eacc6}
			\begin{fmfgraph*}(150,60)
				\fmfstraight
				\fmfbottom{tle,tl0,t0,tr0,tl,t1,tr,tl2,t2,tr2,tl3,t3,tr3,tl4,t4,tr4,tre}     
				\fmftop{vle,vl0,v0,vr0,vl,v1,vr,vl2,v2,vr2,vl3,v3,vr3,vl4,v4,vr4,vre}  
				\fmf{phantom}{t1,m1,v1}
				\fmf{phantom}{t2,m2,v2}
				
				\fmffreeze
				\fmf{fermion}{v0,tl0}
				\fmf{fermion}{tr0,v0}
				\fmf{fermion}{t4,v4}
				\fmf{fermion,left=0.1, label= $d$,label.dist=3}{t3,v3}
				\fmf{fermion,left=0.1, label= $l$,label.dist=3}{v3,t3}
				\fmf{fermion,left=0.5, label= $n$,label.dist=3}{m2,t2}
				\fmf{fermion,left=0.5, label= $f$,label.dist=3}{t2,m2}
				\fmf{fermion,left=0.5, label= $e$,label.dist=3}{v1,m1}
				\fmf{fermion,left=0.5, label= $m$,label.dist=3}{m1,v1}
				\fmf{dashes}{m1,m2}
				\fmf{plain,width=3}{vl3,vr4}
				\fmf{plain}{tl2,tr3}
				\fmf{plain}{vl0,vr}
				\fmfv{label=$j$,label.angle=-90}{tl0}
				\fmfv{label=$b$,label.angle=-90}{tr0}
				\fmfv{label=$a$,label.angle=-90}{t4}
			\end{fmfgraph*}
	\end{fmffile}} 
	& \parbox{170pt}{\begin{equation*} -\sum_{dl} \sum_{em} \lambda_{be}^{jm} \color{red} \sum_{fn} t_{ml}^{ed} \braket{fm|ne} \color{black}   r^{l}_{da}\end{equation*}}  \\ \\
	
	\scalebox{0.5}{
		\begin{fmffile}{r2eacc7}
			\begin{fmfgraph*}(150,60)
				\fmfstraight
				\fmfbottom{tle,tl0,t0,tr0,tl,t1,tr,tl2,t2,tr2,tl3,t3,tr3,tl4,t4,tr4,tre}     
				\fmftop{vle,vl0,v0,vr0,vl,v1,vr,vl2,v2,vr2,vl3,v3,vr3,vl4,v4,vr4,vre}  
				\fmf{phantom}{t3,m3,v3}
				\fmf{phantom}{t2,m2,v2}
				
				\fmffreeze
				\fmf{fermion}{v0,tl0}
				\fmf{fermion}{tr0,v0}
				\fmf{fermion}{t4,v4}
				\fmf{fermion,left=0.5, label= $d$,label.dist=3}{m3,v3}
				\fmf{fermion,left=0.5, label= $l$,label.dist=3}{v3,m3}
				\fmf{fermion,left=0.1, label= $m$,label.dist=3}{v1,t1}
				\fmf{fermion,left=0.1, label= $e$,label.dist=3}{t1,v1}
				\fmf{fermion,left=0.5, label= $f$,label.dist=3}{t2,m2}
				\fmf{fermion,left=0.5, label= $n$,label.dist=3}{m2,t2}
				\fmf{dashes}{m3,m2}
				\fmf{plain,width=3}{vl3,vr4}
				\fmf{plain}{tl,tr2}
				\fmf{plain}{vl0,vr}
				\fmfv{label=$j$,label.angle=-90}{tl0}
				\fmfv{label=$b$,label.angle=-90}{tr0}
				\fmfv{label=$a$,label.angle=-90}{t4}
			\end{fmfgraph*}
	\end{fmffile}} 
	& \parbox{170pt}{\begin{equation*} -\sum_{dl} \sum_{em} \lambda_{be}^{jm} \color{red} \sum_{fn} t_{mn}^{ef} \braket{nd|fl} \color{black}   r^{l}_{da}\end{equation*}}  \\ \\
	
	\scalebox{0.5}{
		\begin{fmffile}{r2eacc8}
			\begin{fmfgraph*}(170,60)
				\fmfstraight
				\fmfbottom{tle,tl5,t5,tr5,tl0,t0,tr0,tl,t1,tr,tl2,t2,tr2,tl3,t3,tr3,tl4,t4,tr4,tre}     
				\fmftop{vle,vl5,v5,vr5,vl0,v0,vr0,vl,v1,vr,vl2,v2,vr2,vl3,v3,vr3,vl4,v4,vr4,vre}  
				\fmf{phantom}{t1,m1,v1}
				\fmf{phantom}{t2,m2,v2}
				
				\fmffreeze
				\fmf{fermion}{v5,tl5}
				\fmf{fermion}{tr5,v5}
				\fmf{fermion}{t4,v4}
				\fmf{fermion,left=0.1, label= $d$,label.dist=3}{t3,v3}
				\fmf{fermion,left=0.1, label= $l$,label.dist=3}{v3,t3}
				\fmf{fermion,left=0.5, label= $o$,label.dist=3}{m2,t2}
				\fmf{fermion,left=0.5, label= $g$,label.dist=3,label.side=right}{t2,m2}
				\fmf{fermion,left=0.5, label= $f$,label.dist=3,label.side=left}{t1,m1}
				\fmf{fermion,left=0.5, label= $n$,label.dist=3,label.side=right}{m1,t1}
				\fmf{fermion,left=0.1, label= $m$,label.dist=3}{v0,t0}
				\fmf{fermion,left=0.1, label= $e$,label.dist=3}{t0,v0}
				\fmf{dashes}{m1,m2}
				\fmf{plain,width=3}{vl3,vr4}
				\fmf{plain}{tl2,tr3}
				\fmf{plain}{tl0,tr}
				\fmf{plain}{vl5,vr0}
				\fmfv{label=$j$,label.angle=-90}{tl5}
				\fmfv{label=$b$,label.angle=-90}{tr5}
				\fmfv{label=$a$,label.angle=-90}{t4}
			\end{fmfgraph*}
	\end{fmffile}} 
	
	& \parbox{170pt}{\begin{equation*} -\sum_{dl} \sum_{em} \lambda_{be}^{jm} \color{red} \sum_{fn} \sum_{go} t_{mn}^{ef} \braket{no|fg} t_{ol}^{hd} \color{black}   r^{l}_{da}\end{equation*}} 
	
\end{longtable}

\subsection{Left-EOM eigenvalue working equations}
The left eigenvalue problem is divided into the singles IP and EA blocks, and the doubles IP and EA blocks, as follows:
\begin{equation}
	\begin{aligned}
		\epsilon^\IPEA l^i
		& =  \bra{\Phi_0} \comm{ \hL^\IPEA }{ \wtH_{\ldr} } \ket{\Phi_i} \\
		\epsilon^\IPEA l^a
		& =  \bra{\Phi^a} \comm{\hL^\IPEA }{ \wtH_{\ldr} } \ket{\Phi_0} \\
		\epsilon^\IPEA l^{ji}_b
		& =  \bra{\Phi_0} \comm{ \hL^\IPEA }{ \wtH_{\ldr} }\ket{\Phi_{ji}^b} \\
		\epsilon^\IPEA l^{ba}_j
		& =  \bra{\Phi^{ba}_j} \comm{ \hL^\IPEA } { \wtH_{\ldr} }\ket{\Phi_0} \\
	\end{aligned}
\end{equation}
The evaluation of the required matrix elements is presented in the following tables.

\begin{longtable}{c  c}
	
	\multicolumn{2}{c}{$\bra{\Phi_0} \comm{ \hL^\IPEA }{ \wtH_{\ldr}} ]\ket{\Phi_i}$} \\ \hline
	\\
	\scalebox{0.5}{
		\begin{fmffile}{l1ipfab}
			\begin{fmfgraph*}(100,60)
				\fmfstraight
				\fmfbottom{tle,tl,t1,tr,tre}     
				\fmftop{vle,vl,v1,vr,vre}  
				\fmfright{r1}
				\fmf{phantom}{t1,m1,v1}  
				\fmf{fermion}{m1,t1}  
				\fmf{fermion, label= $l$}{v1,m1}
				\fmffreeze
				\fmf{dashes}{m1,r1}
				\fmf{plain,width=3}{vr,vl}
				\fmfv{label=$i$,label.angle=-90}{t1}
			\end{fmfgraph*}
	\end{fmffile}} 
	& \parbox{170pt}{\begin{equation*}-\sum_l f_{il} l^l\end{equation*}}  \\ \\
	
	\scalebox{0.5}{
		\begin{fmffile}{l1ipfai}
			\begin{fmfgraph*}(100,60)
				\fmfstraight
				\fmfbottom{tle,tl,t1,tr,tre}     
				\fmftop{vle,vl,v1,vr,vre}  
				\fmfright{r1}
				\fmf{phantom}{t1,m1,v1}  
				\fmffreeze
				\fmf{fermion}{m1,tr}  
				\fmf{fermion, label= $d$}{tl,m1}
				
				\fmf{dashes}{m1,r1}
				\fmf{plain,width=3}{t1,tle}
				\fmfv{label=$i$,label.angle=-90}{tr}
			\end{fmfgraph*}
	\end{fmffile}} 
	& \parbox{170pt}{\begin{equation*}-\sum_d f_{id} l^d\end{equation*}}  \\ \\
	
	\scalebox{0.5}{
		\begin{fmffile}{l1ipvabci}
			\begin{fmfgraph*}(100,60)
				\fmfstraight
				\fmfbottom{tle,tl,t1,tr,tl2,t2,tr2,tre}     
				\fmftop{vle,vl,v1,vr,vl2,v2,vr2,vre}  
				\fmf{phantom}{t1,m1,v1}
				\fmf{phantom}{t2,m2,v2}
				\fmffreeze
				\fmf{fermion}{m2,t2}  
				\fmf{fermion, label= $m$}{v2,m2}
				\fmf{fermion,left=0.5, label= $d$}{m1,v1}
				\fmf{fermion,left=0.5, label= $l$}{v1,m1}
				\fmf{dashes}{m1,m2}
				\fmf{plain,width=3}{vl,vr2}
				\fmfv{label=$i$,label.angle=-90}{t2}
			\end{fmfgraph*}
	\end{fmffile}} 
	& \parbox{170pt}{\begin{equation*}-\sum_{dlm} \braket{di|lm} l^{lm}_d\end{equation*}}  \\ \\
\end{longtable}

\begin{longtable}{c  c}
	\multicolumn{2}{c}{$\bra{\Phi^a} \comm{ \hL^\IPEA }{ \wtH_{\ldr} } \ket{\Phi_0}$} \\ \hline
	\\
	\scalebox{0.5}{
		\begin{fmffile}{l1eafij}
			\begin{fmfgraph*}(100,60)
				\fmfstraight
				\fmftop{tle,tl,t1,tr,tre}     
				\fmfbottom{vle,vl,v1,vr,vre}  
				\fmfright{r1}
				\fmf{phantom}{t1,m1,v1}  
				\fmf{fermion}{m1,t1}  
				\fmf{fermion, label= $d$}{v1,m1}
				\fmffreeze
				\fmf{dashes}{m1,r1}
				\fmf{plain,width=3}{vr,vl}
				\fmfv{label=$a$,label.angle=90}{t1}
			\end{fmfgraph*}
	\end{fmffile}} 
	& \parbox{170pt}{\begin{equation*}-\sum_d f_{ad} l^d\end{equation*}}  \\ \\
	
	\scalebox{0.5}{
		\begin{fmffile}{l1eafai}
			\begin{fmfgraph*}(100,60)
				\fmfstraight
				\fmftop{tle,tl,t1,tr,tre}     
				\fmfbottom{vle,vl,v1,vr,vre}  
				\fmfright{r1}
				\fmf{phantom}{t1,m1,v1}  
				\fmffreeze
				\fmf{fermion}{m1,tr}  
				\fmf{fermion, label= $l$}{tl,m1}
				
				\fmf{dashes}{m1,r1}
				\fmf{plain,width=3}{t1,tle}
				\fmfv{label=$a$,label.angle=90}{tr}
			\end{fmfgraph*}
	\end{fmffile}} 
	& \parbox{170pt}{\begin{equation*}-\sum_l f_{ad} l^l\end{equation*}}  \\ \\
	
	\scalebox{0.5}{
		\begin{fmffile}{l1eavijka}
			\begin{fmfgraph*}(100,60)
				\fmfstraight
				\fmftop{tle,tl,t1,tr,tl2,t2,tr2,tre}     
				\fmfbottom{vle,vl,v1,vr,vl2,v2,vr2,vre}  
				\fmf{phantom}{t1,m1,v1}
				\fmf{phantom}{t2,m2,v2}
				\fmffreeze
				\fmf{fermion}{m2,t2}  
				\fmf{fermion, label= $e$}{v2,m2}
				\fmf{fermion,left=0.5, label= $l$}{m1,v1}
				\fmf{fermion,left=0.5, label= $d$}{v1,m1}
				\fmf{dashes}{m1,m2}
				\fmf{plain,width=3}{vl,vr2}
				\fmfv{label=$a$,label.angle=90}{t2}
			\end{fmfgraph*}
	\end{fmffile}} 
	& \parbox{170pt}{\begin{equation*}-\sum_{edl} \braket{la|de} l^{de}_l\end{equation*}}  \\ \\
\end{longtable}

\begin{longtable}{c  c}
	\multicolumn{2}{c}{$\bra{\Phi_0} \comm{ \hL^\IPEA }{ \wtH_{\ldr} } \ket{\Phi^{b}_{ji}}$} \\ \hline
	\\

	\scalebox{0.5}{
		\begin{fmffile}{l2ipfo}
			\begin{fmfgraph*}(100,60)
				\fmfstraight
				\fmfbottom{tle,tl,t1,tr,tl2,t2,tr2,tre}     
				\fmftop{vle,vl,v1,vr,vl2,v2,vr2,vre}  
				\fmfright{r1}
				\fmf{phantom}{t2,m2,v2}
				\fmffreeze
				\fmf{fermion}{v1,tl}
				\fmf{fermion}{tr,v1}
				\fmf{fermion}{m2,t2}  
				\fmf{fermion, label= $l$, label.side=right}{v2,m2}
				\fmf{dashes}{m2,r1}
				\fmf{plain,width=3}{vl,vr2}
				\fmfv{label=$j$,label.angle=-90}{tl}
				\fmfv{label=$b$,label.angle=-90}{tr}
				\fmfv{label=$i$,label.angle=-90}{t2}
			\end{fmfgraph*}
	\end{fmffile}} 
	& \parbox{170pt}{\begin{equation*}-\sum_{l} f_{il} l^{jl}_{b} -\sum_{l} f_{jl} l^{li}_{b}\end{equation*}}  \\ \\
	
	\scalebox{0.5}{
		\begin{fmffile}{l2ipfv}
			\begin{fmfgraph*}(100,60)
				\fmfstraight
				\fmfbottom{tle,tl,t1,tr,tl2,t2,tr2,tre}     
				\fmftop{vle,vl,v1,vr,vl2,v2,vr2,vre}  
				\fmfleft{l1}
				\fmf{phantom}{tl,m1,v1}
				\fmffreeze
				\fmf{fermion}{tl,m1}
				\fmf{fermion}{v1,tr}
				\fmf{fermion}{v2,t2}  
				\fmf{fermion, label= $d$, label.side=left}{m1,v1}
				\fmf{dashes}{m1,l1}
				\fmf{plain,width=3}{vl,vr2}
				\fmfv{label=$b$,label.angle=-90}{tl}
				\fmfv{label=$j$,label.angle=-90}{tr}
				\fmfv{label=$i$,label.angle=-90}{t2}
			\end{fmfgraph*}
	\end{fmffile}} 
	& \parbox{170pt}{\begin{equation*}\sum_{d} f_{db} l^{ji}_d\end{equation*}}  \\ \\
\end{longtable}

\begin{longtable}{c  c}
	\multicolumn{2}{c}{$\bra{\Phi_{j}^{ba}} \comm{ \hL^\IPEA }{ \wtH_{\ldr} } \ket{\Phi_0}$} \\ \hline
	\\

	\scalebox{0.5}{
		\begin{fmffile}{l2eafo}
			\begin{fmfgraph*}(100,60)
				\fmfstraight
				\fmftop{tle,tl,t1,tr,tl2,t2,tr2,tre}     
				\fmfbottom{vle,vl,v1,vr,vl2,v2,vr2,vre}  
				\fmfright{r1}
				\fmf{phantom}{t2,m2,v2}
				\fmffreeze
				\fmf{fermion}{v1,tl}
				\fmf{fermion}{tr,v1}
				\fmf{fermion}{m2,t2}  
				\fmf{fermion, label= $d$, label.side=left}{v2,m2}
				\fmf{dashes}{m2,r1}
				\fmf{plain,width=3}{vl,vr2}
				\fmfv{label=$b$,label.angle=90}{tl}
				\fmfv{label=$j$,label.angle=90}{tr}
				\fmfv{label=$a$,label.angle=90}{t2}
			\end{fmfgraph*}
	\end{fmffile}} 
	& \parbox{170pt}{\begin{equation*}-\sum_{d} f_{ad} l^{bd}_{j}-\sum_{d} f_{bd} l^{da}_{j}\end{equation*}}  \\ \\
	
	\scalebox{0.5}{
		\begin{fmffile}{l2eafv}
			\begin{fmfgraph*}(100,60)
				\fmfstraight
				\fmftop{tle,tl,t1,tr,tl2,t2,tr2,tre}     
				\fmfbottom{vle,vl,v1,vr,vl2,v2,vr2,vre}  
				\fmfleft{l1}
				\fmf{phantom}{tl,m1,v1}
				\fmffreeze
				\fmf{fermion}{tl,m1}
				\fmf{fermion}{v1,tr}
				\fmf{fermion}{v2,t2}  
				\fmf{fermion, label= $l$, label.side=right}{m1,v1}
				\fmf{dashes}{m1,l1}
				\fmf{plain,width=3}{vl,vr2}
				\fmfv{label=$j$,label.angle=90}{tl}
				\fmfv{label=$b$,label.angle=90}{tr}
				\fmfv{label=$a$,label.angle=90}{t2}
			\end{fmfgraph*}
	\end{fmffile}} 
	& \parbox{170pt}{\begin{equation*}\sum_{l} f_{lj} l^{ba}_l\end{equation*}}  \\ \\
\end{longtable}

Contributions in red for the right- and left-EOM eigenvalue problem reproduce a projection of the drCCD Hamiltonian onto doubly excited determinants $\bra{\Phi_{ij}^{ab}} \wtH_\dr \ket{\Phi_0}$. 
Since these projections are the CC amplitudes equations, they add up to zero and can be ignored in the implementation.

\subsection{$\Xi$ working equations}

\begin{longtable}{c  c}
	
	\multicolumn{2}{c}{$\bra{\Phi_{ij}^{ab}} \comm{ \wtH_\dr }{ \hXi_2 } \ket{\Phi_0}$} \\ \hline
	\\
	
	\scalebox{0.5}{
		\begin{fmffile}{KsiFT2v}
			\begin{fmfgraph*}(100,60)
				\fmfstraight
				\fmftop{t1,ta,t2,ti,tj,t3,tb,t4}     
				\fmfbottom{v1,v2,bcl,v4,v5,bcr,v7,v8}
				\fmfright{r}
				\fmf{fermion}{ti,bcl}  
				\fmf{fermion}{bcl,ta} 
				\fmf{fermion}{bcr,mc,tb} 
				\fmf{fermion}{tj,bcr} 
				\fmfset{zigzag_width}{0.5mm}
				\fmf{zigzag}{v2,v7}
				\fmfv{label=$i$,label.angle=90,label.dist=3}{ti}
				\fmfv{label=$j$,label.angle=90,label.dist=2}{tj}
				\fmfv{label=$a$,label.angle=90,label.dist=3}{ta}
				\fmfv{label=$b$,label.angle=90,label.dist=3}{tb}
				\fmffreeze
				\fmf{phantom,label.side=left,label=$c$}{mc,bcr}
				\fmf{dashes}{mc,r}
			\end{fmfgraph*}
	\end{fmffile}} & 
	\parbox{170pt}{\begin{equation*} \sum_c f_{bc} \xi^{ac}_{ij}+\sum_c f_{ac} \xi^{cb}_{ij}\end{equation*}}   \\ \\
	
	\scalebox{0.5}{
		\begin{fmffile}{KsiFT2o}
			\begin{fmfgraph*}(100,60)
				\fmfstraight
				\fmftop{t1,ta,t2,ti,tb,t3,tj,t4}     
				\fmfbottom{v1,v2,bcl,v4,v5,bcr,v7,v8}
				\fmfright{r}
				\fmf{fermion}{ti,bcl}  
				\fmf{fermion}{bcl,ta} 
				\fmf{fermion}{tb,bcr} 
				\fmf{fermion}{tj,mc,bcr} 
				\fmfset{zigzag_width}{0.5mm}
				\fmf{zigzag}{v2,v7}
				\fmfv{label=$i$,label.angle=90,label.dist=3}{ti}
				\fmfv{label=$j$,label.angle=90,label.dist=2}{tj}
				\fmfv{label=$a$,label.angle=90,label.dist=3}{ta}
				\fmfv{label=$b$,label.angle=90,label.dist=3}{tb}
				\fmffreeze
				\fmf{phantom,label.side=left,label=$k$}{mc,bcr}
				\fmf{dashes}{mc,r}
			\end{fmfgraph*}
	\end{fmffile}} & 
	\parbox{170pt}{\begin{equation*}-\sum_k f_{kj} \xi^{ab}_{ik}-\sum_k f_{ki} \xi^{ab}_{kj}\end{equation*}}  \\ \\
\end{longtable}

\begin{longtable}{c  c}
	\multicolumn{2}{c}{$\bra{\Phi_0} \comm{ \hL^\IPEA }{ \comm{ \pdv{\wtH_{\ldr}}{\lambda^{ij}_{ab}} }{ \hR^\IPEA } }_+\ket{\Phi_0}$} \\ \hline
	\\
	
	\scalebox{0.5}{
		\begin{fmffile}{Ksileft_iabj}
			\begin{fmfgraph*}(100,60)
				\fmfstraight
				\fmftop{tle,tl1,t1,tr1,tl2,t2,tr2,tl3,t3,tr3,tre}     
				\fmfbottom{vle,vl1,v1,vr1,vl2,v2,vr2,vl3,v3,vr3,vre}  
				\fmf{phantom}{t2,m2,v2}
				\fmf{phantom}{t1,m1,v1}
				
				\fmffreeze
				\fmf{fermion}{tr1,m1}
				\fmf{fermion}{m1,tl1}
				\fmf{fermion}{tr3,v3}
				\fmf{fermion}{v3,tl3}
				\fmf{fermion, label= $c$, label.side=left}{vr1,m2}
				\fmf{fermion, label= $k$, label.side=left}{m2,vl3}
				\fmf{dashes}{m1,m2}
				\fmf{plain,width=3}{vr2,vr3}
				\fmf{plain,width=3}{v1,vl2}
				\fmfv{label=$a$,label.angle=90}{tl1}
				\fmfv{label=$i$,label.angle=90}{tr1}
				\fmfv{label=$b$,label.angle=90}{tl3}
				\fmfv{label=$j$,label.angle=90}{tr3}
			\end{fmfgraph*}
	\end{fmffile}} 
	& \parbox{170pt}{\begin{equation*}-P(ia|jb)  \sum_{ck} \braket{ak|ic} l^c r_{jk}^b  \end{equation*}}  \\ \\
\end{longtable}

Contributions in red in the $\bra{\Phi_0} \comm{ \hL^\IPEA }{ \comm{ \pdv{\wtH_{\ldr}}{\lambda^{ij}_{ab}} }{ \hR^\IPEA } }_+\ket{\Phi_0}$ term reproduce a projection of the drCCD Hamiltonian onto doubly excited determinants $\bra{\Phi_{ij}^{ab}} \wtH_\dr \ket{\Phi_0}$. 
Since these projections are the CC amplitudes equations, they add up to zero and can be ignored in the implementation. The complete $\bra{\Phi_0} \comm{ \hL^\IPEA }{ \comm{ \pdv{\wtH_{\ldr}}{\lambda^{ij}_{ab}} }{ \hR^\IPEA } }_+\ket{\Phi_0}$ contributions also need to be multiplied with the corresponding $\pm 1$ prefactor.

\subsection{$Z$ working equations}

\begin{longtable}{c  c}
	
	\multicolumn{2}{c}{$\bra{\Phi_0} \qty(1 + \hZ_2 ) \pdv{\wtH_\dr}{t^{ab}_{ij}} \ket{\Phi_0}$} \\ \hline
	\\
	\scalebox{0.5}{
		\begin{fmffile}{Zijab}
			\begin{fmfgraph*}(100,60)
				\fmfstraight
				\fmfbottom{t1,ta,t2,ti,tb,t3,tj,t4}     
				\fmftop{v1,v2,v3,v4,v5,v6,v7,v8}      
				\fmf{fermion}{ti,v3}  
				\fmf{fermion}{v3,ta} 
				\fmf{fermion}{tj,v6}  
				\fmf{fermion}{v6,tb} 
				\fmf{dashes}{v3,v6}
				\fmfv{label=$a$,label.angle=-90,label.dist=3}{ti}
				\fmfv{label=$b$,label.angle=-90,label.dist=2}{tj}
				\fmfv{label=$i$,label.angle=-90,label.dist=3}{ta}
				\fmfv{label=$j$,label.angle=-90,label.dist=3}{tb}
			\end{fmfgraph*}
	\end{fmffile}} 
	& \parbox{170pt}{\begin{equation*} \braket{ij|ab}\end{equation*}}  \\ \\

	\scalebox{0.5}{
		\begin{fmffile}{ZFL2v}
			\begin{fmfgraph*}(100,60)
				\fmfstraight
				\fmfbottom{t1,ta,t2,ti,tb,t3,tj,t4}     
				\fmftop{v1,v2,bcl,v4,v5,bcr,v7,v8}
				\fmfright{r}
				\fmf{fermion}{ti,bcl}  
				\fmf{fermion}{bcl,ta} 
				\fmf{fermion}{tb,bcr} 
				\fmf{fermion}{tj,mc,bcr}
				\fmfset{zigzag_width}{0.5mm}
				\fmf{zigzag}{v2,v7}
				\fmfv{label=$a$,label.angle=-90,label.dist=3}{ti}
				\fmfv{label=$b$,label.angle=-90,label.dist=2}{tj}
				\fmfv{label=$i$,label.angle=-90,label.dist=3}{ta}
				\fmfv{label=$j$,label.angle=-90,label.dist=3}{tb}
				\fmffreeze
				\fmf{phantom,label.side=right,label=$c$}{mc,bcr}
				\fmf{dashes}{mc,r}
			\end{fmfgraph*}
	\end{fmffile}}
	& 
	\parbox{170pt}{\begin{equation*}\sum_c f_{cb} \zeta_{ac}^{ij}+\sum_c f_{ca} \zeta_{cb}^{ij}\end{equation*}}   \\ \\
\end{longtable}

The above contributions to the $Z$ working equations have exactly the same form as the $\Lambda$-amplitude equations. 

\begin{longtable}{c  c}
	
	\multicolumn{2}{c}{$\bra{\Phi_0} \qty(1 + \hLam_2 ) \comm{ \pdv{\wtH_\dr}{t_{ij}^{ab}} }{ \hXi_2 } \ket{\Phi_0}$} \\ \hline
	\\
	
	\scalebox{0.5}{
		\begin{fmffile}{ZL2Ksi2Viabj}
			\begin{fmfgraph*}(100,60)
				\fmfstraight
				\fmfbottom{t1,ta,t2,ti,tm1,tm2,tm3,tmgap,tm4,tm5,tm6,tj,t3,tb,t4}     
				\fmftop{v1,v2,bcl1,v4,v5,bcr1,v7,v8,bm1,bcl2,bm3,bm4,bcr2,bm5,bm6}
				\fmfset{zigzag_width}{0.5mm}
				\fmf{plain}{v2,v7}
				\fmf{dashes}{bcl2,bcr2}
				\fmf{fermion}{ti,bcl1}  
				\fmf{fermion}{bcl1,ta}
				\fmf{fermion}{tj,bcr2}  
				\fmf{fermion}{bcr2,tb}
				\fmf{phantom}{bcr1,mc1,tm2}
				\fmf{phantom}{bcl2,mc2,tm5}
				\fmf{phantom}{tj,bm1}  
				\fmf{phantom}{bm1,tb}

				\fmffreeze
				\fmf{fermion, left=0.15,label=$k$,label.side=left,label.dist=4}{bcr1,tm2} 
				\fmf{fermion, left=0.15,label=$c$,label.side=left,label.dist=4}{tm2,bcr1} 
				\fmf{fermion, left=0.15,label=$l$,label.side=left,label.dist=4}{bcl2,tm5} 
				\fmf{fermion, left=0.15,label=$d$,label.side=left,label.dist=4}{tm5,bcl2} 
				\fmf{zigzag}{tm1,tm6}
				\fmfv{label=$a$,label.angle=-90,label.dist=3}{ti}
				\fmfv{label=$b$,label.angle=-90,label.dist=2}{tj}
				\fmfv{label=$i$,label.angle=-90,label.dist=3}{ta}
				\fmfv{label=$j$,label.angle=-90,label.dist=3}{tb}
			\end{fmfgraph*}
	\end{fmffile}} & 
	\parbox{170pt}{\begin{equation*} P(ia|jb) \sum_{klcd}\lambda^{ik}_{ac} \xi^{cd}_{kl}\braket{lj|db} \end{equation*} }
	
\end{longtable}

\begin{longtable}{c  c}
	\multicolumn{2}{c}{$\bra{\Phi_0} \comm{ \hL^\IPEA }{ \comm{ \pdv{\wtH_{\ldr}}{t_{ij}^{ab}} }{ \hR^\IPEA } }_+\ket{\Phi_0}$} \\ \hline
	\\
	
	\scalebox{0.5}{
		\begin{fmffile}{Zetaleft_iabj}
			\begin{fmfgraph*}(100,60)
				\fmfstraight
				\fmfbottom{tle,tl1,t1,tr1,tl2,t2,tr2,tl3,t3,tr3,tre}     
				\fmftop{vle,vl1,v1,vr1,vl2,v2,vr2,vl3,v3,vr3,vre}  
				\fmf{phantom}{t2,m2,v2}
				\fmf{phantom}{t1,m1,v1}
				
				\fmffreeze
				\fmf{fermion}{tr1,m1}
				\fmf{fermion}{m1,tl1}
				\fmf{fermion}{tr3,v3}
				\fmf{fermion}{v3,tl3}
				\fmf{fermion, label= $k$, label.side=right}{vr1,m2}
				\fmf{fermion, label= $c$, label.side=right}{m2,vl3}
				\fmf{dashes}{m1,m2}
				\fmf{plain,width=3}{vr2,vr3}
				\fmf{plain,width=3}{v1,vl2}
				\fmfv{label=$i$,label.angle=-90}{tl1}
				\fmfv{label=$a$,label.angle=-90}{tr1}
				\fmfv{label=$j$,label.angle=-90}{tl3}
				\fmfv{label=$b$,label.angle=-90}{tr3}
			\end{fmfgraph*}
	\end{fmffile}} 
	& \parbox{170pt}{\begin{equation*}-P(ia|jb)  \sum_{ck} \braket{ic|ak} l^k r_{bc}^j \end{equation*}}  \\ \\
\end{longtable}

The complete $\bra{\Phi_0} \comm{ \hL^\IPEA }{ \comm{ \pdv{\wtH_{\ldr}}{t_{ij}^{ab}} }{ \hR^\IPEA } }_+\ket{\Phi_0}$ contributions need to multiplied by the corresponding $ \pm 1$ prefactor.

\subsection{One-electron density matrices}

\begin{longtable}{c  c}
	\multicolumn{2}{c}{Contribution from $\bra{\Phi_0} \comm{ \hL^\IPEA }{ \comm{ \wtH_{\ldr} }{ \hR^\IPEA } }_+\ket{\Phi_0} $} \\ \hline \hline
	\multicolumn{2}{c}{\parbox{170pt}{\begin{equation*}\gamma_{ij}\end{equation*}}} \\ \hline  \\ 
	\scalebox{0.5}{
		\begin{fmffile}{eomd_ij1}
			\begin{fmfgraph*}(100,60)
				\fmfstraight
				\fmftop{t1,ta,t2,ti,t4}     
				\fmfbottom{v1,v2,v3,v4,v8}  
				\fmfright{r1}
				\fmf{phantom}{v3,m1,t2}
				\fmffreeze
				\fmf{fermion, label= $j$}{t2,m1}
				\fmf{fermion, label= $i$}{m1,v3}
				\fmf{dashes}{m1,r1}
				\fmf{plain,width=3}{v2,v4}
				\fmf{plain,width=3}{ta,ti}
				\fmfv{decor.shape=circle,decor.filled=empty,decor.size=2thick}{m1}
			\end{fmfgraph*}
	\end{fmffile}} 
	& \parbox{170pt}{\begin{equation*}  -r_{i} l^j \end{equation*}}  \\ \\
	\scalebox{0.5}{
		\begin{fmffile}{eomd_ij2}
			\begin{fmfgraph*}(100,60)
				\fmfstraight
				\fmftop{t1,ta,t2,ti,tb,t3,tj,t4}     
				\fmfbottom{v1,v2,v3,v4,v5,v6,v7,v8}  
				\fmfright{r1}
				\fmf{fermion,left=0.15, label= $k$}{t2,v3}  
				\fmf{fermion,left=0.15, label= $c$}{v3,t2} 
				\fmf{phantom}{t3,m1,v6}
				\fmffreeze
				\fmf{fermion, label= $j$}{t3,m1}
				\fmf{fermion, label= $i$}{m1,v6}
				\fmf{dashes}{m1,r1}
				\fmf{plain,width=3}{v2,v7}
				\fmf{plain,width=3}{ta,tj}
				\fmfv{decor.shape=circle,decor.filled=empty,decor.size=2thick}{m1}
			\end{fmfgraph*}
	\end{fmffile}} 
	& \parbox{170pt}{\begin{equation*} -\sum_{ck} r_{ki}^c l^{kj}_c  + \sum_{ck} r_{ik}^c l^{jk}_c \end{equation*}}  \\ \\
\end{longtable}

Contributions from $\bra{\Phi_0} \comm{ \hL^\IPEA }{ \comm{ \wtH_{\ldr} }{ \hR^\IPEA } }_+\ket{\Phi_0}$ need to be multiplied with the corresponding $\pm1$ prefactor.

\begin{longtable}{c  c}

	\multicolumn{2}{c}{Contribution from $\bra{\Phi_0} \qty(1 + \hLam_2 ) \comm{ \wtH_\dr }{ \hXi_2 } \ket{\Phi_0}$} \\ \hline \hline
	\multicolumn{2}{c}{\parbox{170pt}{\begin{equation*}\gamma_{ij}\end{equation*}}} \\ \hline  \\ 
	\scalebox{0.5}{
		\begin{fmffile}{eomksid_ij}
			\begin{fmfgraph*}(100,60)
				\fmfstraight
				\fmftop{t1,ta,t2,ti,tb,t3,tj,t4}     
				\fmfbottom{v1,v2,v3,v4,v5,v6,v7,v8}  
				\fmfright{r1}
				\fmf{fermion,left=0.1, label= $k$}{t2,v3} 
				\fmf{fermion,left=0.1, label= $c$}{v3,t2} 
				\fmf{fermion,left=0.15, label= $d$}{v6,t3}
				\fmf{phantom}{t3,m1,v6}
				\fmffreeze
				\fmfshift{(6.0,0)}{m1}
				\fmf{fermion,left=0.09, label= $j$}{t3,m1}
				\fmf{fermion,left=0.09, label= $i$}{m1,v6}
				\fmf{dashes}{m1,r1}
				\fmfset{zigzag_width}{0.5mm}
				\fmf{zigzag}{v2,v7}
				\fmf{plain}{ta,tj}
				\fmfv{decor.shape=circle,decor.filled=empty,decor.size=2thick}{m1}
			\end{fmfgraph*}
	\end{fmffile}} 
	& \parbox{170pt}{\begin{equation*}   -  \color{black}\sum_{cdk}  \lambda_{cd}^{kj} \xi_{ki}^{cd} \end{equation*}}  \\ \\
	
	\multicolumn{2}{c}{\parbox{170pt}{\begin{equation*}\gamma_{ab}\end{equation*}}} \\ \hline  \\
\end{longtable}

\subsection{Two-electron density matrices}

\begin{longtable}{c  c}
	\multicolumn{2}{c}{Contribution from $\bra{\Phi_0} \comm{ \hL^\IPEA }{ \comm{ \wtH_{\ldr} }{ \hR^\IPEA } }_+\ket{\Phi_0}$} \\ \hline \hline
	\multicolumn{2}{c}{\parbox{170pt}{\begin{equation*}\Gamma_{abij}\end{equation*}}} \\ \hline  \\

	\scalebox{0.5}{
		\begin{fmffile}{eomd_abij1}
			\begin{fmfgraph*}(100,60)
				\fmfstraight
				\fmftop{tle,tl1,t1,tr1,tl2,t2,tr2,t,tl3,t3,tr3,tre}     
				\fmfbottom{vle,vl1,v1,vr1,vl2,v2,vr2,v,vl3,v3,vr3,vre}
				\fmf{phantom}{v1,m1,t1}
				\fmf{phantom}{v,m2,t}
				\fmffreeze
				\fmf{fermion,left=0.4, label= $a$}{m1,t1}
				\fmf{fermion,left=0.4, label= $i$}{t1,m1}
				\fmf{fermion, label= $b$}{m2,t2}
				\fmf{fermion, label= $j$, label.side=right}{t3,m2}
				\fmf{dashes}{m1,m2}
				\fmf{plain,width=3}{tl1,tr2}
				\fmf{plain,width=3}{tl3,tr3}
				\fmfv{decor.shape=circle,decor.filled=empty,decor.size=2thick}{m1,m2}
			\end{fmfgraph*}
	\end{fmffile}} 
	& \parbox{170pt}{\begin{equation*}  -P(ia|jb)r_{ab}^i l^j \end{equation*}}  \\ \\
	
	\scalebox{0.5}{
		\begin{fmffile}{eomd_abij2}
			\begin{fmfgraph*}(100,60)
				\fmfstraight
				\fmftop{tle,tl1,t1,tr1,tl2,t2,tr2,t,tl3,t3,tr3,tre}     
				\fmfbottom{vle,vl1,v1,vr1,vl2,v2,vr2,v,vl3,v3,vr3,vre}
				\fmf{phantom}{v1,m1,t1}
				\fmf{phantom}{v,m2,t}
				\fmffreeze
				\fmf{fermion,left=0.4, label= $a$}{m1,t1}
				\fmf{fermion,left=0.4, label= $i$}{t1,m1}
				\fmf{fermion, label= $j$, label.side=right}{t2,m2}
				\fmf{fermion, label= $b$, label.side=left}{m2,t3}
				\fmf{dashes}{m1,m2}
				\fmf{plain,width=3}{tl1,tr2}
				\fmf{plain,width=3}{tl3,tr3}
				\fmfv{decor.shape=circle,decor.filled=empty,decor.size=2thick}{m1,m2}
			\end{fmfgraph*}
	\end{fmffile}} 
	& \parbox{170pt}{\begin{equation*}  -P(ia|jb)l^{ij}_a r_b \end{equation*}}  \\ \\
\end{longtable}

Contributions from $\bra{\Phi_0} \comm{ \hL^\IPEA }{ \comm{ \wtH_{\ldr} }{ \hR^\IPEA } }_+\ket{\Phi_0}$ need to be multiplied with the corresponding $\pm1$ prefactor.

\begin{longtable}{c  c}
	
	\multicolumn{2}{c}{Contribution from $\bra{\Phi_0} \qty(1 + \hLam_2 ) \comm{ \wtH_\dr }{ \hXi_2 }\ket{\Phi_0}$} \\ \hline \hline
	
	\multicolumn{2}{c}{\parbox{170pt}{\begin{equation*}\Gamma_{ijab}\end{equation*}}} \\ \hline
	\\
	\scalebox{0.5}{
		\begin{fmffile}{D_ksiijab1}
			\begin{fmfgraph*}(100,60)
				\fmfstraight
				\fmfbottom{t1,ta,t2,ti,tb,t3,tj,t4}     
				\fmftop{v1,v2,v3,v4,v5,v6,v7,v8}  
				\fmf{fermion,right=0.15, label= $a$}{t2,v3}  
				\fmf{fermion,right=0.15, label= $i$}{v3,t2}  
				\fmf{fermion,right=0.15, label= $b$}{t3,v6}
				\fmf{fermion,right=0.15, label= $j$}{v6,t3}
				\fmf{dashes}{v3,v6}
				\fmfset{zigzag_width}{0.5mm}
				\fmf{zigzag}{ta,tj}
				\fmfv{decor.shape=circle,decor.filled=empty,decor.size=2thick}{v3,v6}
			\end{fmfgraph*}
	\end{fmffile}} 
	& \parbox{170pt}{\begin{equation*} \xi^{ab}_{ij}\end{equation*}}  \\ \\
\end{longtable}

The one- and two-electron contributions need additional contributions from the $\bra{\Phi_0} \qty(1 + \hZ_2 ) \pdv{\wtH_\dr}{t^{ab}_{ij}} \ket{\Phi_0}$ term. 
These are not repeated here, as they have exactly the same form as the ground-state drCCD contribution, where the $\hLam_2$ operator has been substituted by $\hZ_2$. 

\section{Connecting the $G_0W_0$ non-linear equation to the IP/EA-EOM-$\ldr$CCD parameters \label{app:connection}}

In this appendix, a connection between the quantities derived from a conventional $G_0W_0$ implementation is related to the IP/EA-EOM-$\ldr$CCD quantities. 
This may aid in the implementation of the reduced density matrices within an existing $G_0W_0$ implementation. 

\subsection{Effective two-electron integrals}

The effective integrals [see \eqrref{eq:scr_2int}] can be formally rewritten using the $\ldr$CCD amplitudes and the elements of the matrices $\bX$ and $\bar{\bX}$ (or alternatively the EOM-S-drCCD eigenvectors). 
The transformed fluctuation potential operator within the direct approximation is employed 
\begin{equation}
	e^{\hT_2}\hat{V}_\mathrm{N}e^{\hT_2} + \left[\hLam_2, e^{\hT_2}\hat{V}_\mathrm{N}e^{\hT_2} \right] \approx \wtV_\ldr.
\end{equation}  
The following cases can be distinguished 
\begin{subequations} \label{eq:two_scren_trans}
\begin{align}
	\sERI{ai}{\mu} & = \bra{\Phi_{il}^{ad}}\wtV_\ldr\ket{\Phi_{0}^{}} x_{l,\mu}^d, 
	\\ 
	\sERI{ia}{\mu} & = \bra{\Phi_{l}^{d}}\wtV_\ldr\ket{\Phi_{i}^{a}} x_{l,\mu}^d, 
	\\
	\sERI{ab}{\mu} & = \bra{\Phi_{l}^{da}}\wtV_\ldr\ket{\Phi_{}^{b}} x_{l,\mu}^d, 
	\\
	\sERI{ij}{\mu} & = \bra{\Phi_{lj}^{d}}\wtV_\ldr\ket{\Phi_{i}^{}} x_{l,\mu}^d, 
	\\
	\sERI{\mu}{ai} & = \bar{x}_{l,\mu}^{d*} \bra{\Phi_{0}^{}}\wtV_\ldr\ket{\Phi_{il}^{ad}}, 
	\\ 
	\sERI{\mu}{ia} & = \bar{x}_{l,\mu}^{d*} \bra{\Phi_{i}^{a}}\wtV_\ldr\ket{\Phi_{l}^{d}}, 
	\\
	\sERI{\mu}{ab} & = \bar{x}_{l,\mu}^{d*} \bra{\Phi_{}^{b}}\wtV_\ldr\ket{\Phi_{l}^{da}}, 
	\\
	\sERI{\mu}{ij} & = \bar{x}_{l,\mu}^{d*} \bra{\Phi_{i}^{}}\wtV_\ldr\ket{\Phi_{lj}^{d}}.
\end{align} 
\end{subequations}
The corresponding contributions can be assigned to diagrams presented in Appendix \ref{app:ipea_eom_lrCCD}.

\subsection{The diagonal approximation}

Within the diagonal approximation, as presented in \eqrref{eq:diagonal_r}, the right eigenvalue singles blocks take the form
\begin{equation}
	\begin{split}
		r_i \epsilon^\GOWO_i 
		& = -r_i \epsilon^\HF_i 
		\\
		& + \sum_{dlm} \bra{\Phi_i}\wtV_\ldr \ket{\Phi_{lm}^d} r_{lm}^d 
		\\
		& - \sum_{edl}  \bra{\Phi_{il}^{ed}}\wtV_\ldr \ket{\Phi_0} r_{de}^l
	\end{split}
\end{equation}
for the IP solutions and 
\begin{equation}
	\begin{split}
		r_a \epsilon^\GOWO_a 
		& = -r_a \epsilon^\HF_a 
		\\
		&+ \sum_{dlm} \bra{\Phi_0}\wtV_\ldr \ket{\Phi_{lm}^{da}} r_{lm}^d 
		\\
		&- \sum_{edl}  \bra{\Phi_{l}^{de}}\wtV_\ldr \ket{\Phi^a} r_{de}^l
	\end{split}
\end{equation}
for the EA solutions. 
Dividing both equations by $r_{i}$ or $r_{a}$ results in 
\begin{equation} \label{eq:egw_i}
	\begin{split}
		\epsilon^\GOWO_i 
		= \epsilon^\HF_i 
		& - \sum_{dlm} \bra{\Phi_i}\wtV_\ldr \ket{\Phi_{lm}^d} r_{lm}^d/r_i 
		\\
		& + \sum_{edl}  \bra{\Phi_{il}^{ed}}\wtV_\ldr \ket{\Phi_0} r_{de}^l/r_i
	\end{split}
\end{equation}	
for the IP solutions and 
\begin{equation} \label{eq:egw_a}
	\begin{split}
		\epsilon^\GOWO_a 
		= \epsilon^\HF_a 
		& - \sum_{dlm} \bra{\Phi_0}\wtV_\ldr \ket{\Phi_{lm}^{da}} r_{lm}^d/r_a 
		\\
		& + \sum_{edl}  \bra{\Phi_{l}^{de}}\wtV_\ldr \ket{\Phi^a} r_{de}^l/r_a
	\end{split}
\end{equation}
for the EA solutions. 

Similar manipulations can be performed for the left eigenvalue problem within the diagonal approximation.
The singles take the form
\begin{equation}
	\begin{split}
		\epsilon^\GOWO_i l^i 
		& = - \epsilon^\HF_i l^i 
		\\
		& + \sum_{dlm} \bra{\Phi_{lm}^d}\wtV_\ldr \ket{\Phi_i}  l^{lm}_d 
		\\
		& - \sum_{edl}  \bra{\Phi_0} \wtV_\ldr \ket{\Phi_{il}^{ed}}  l^{de}_l
	\end{split}
\end{equation}
for the IP solutions and 
\begin{equation}
	\begin{split}
		\epsilon^\GOWO_a l^a
		& = - \epsilon^\HF_a l^a 
		\\
		&+ \sum_{dlm} \bra{\Phi_{lm}^{da}}\wtV_\ldr \ket{ \Phi_0} l^{lm}_d 
		\\
		&- \sum_{edl}  \bra{\Phi^a}\wtV_\ldr \ket{ \Phi_{l}^{de} } l^{de}_l
	\end{split}
\end{equation}
for the EA solutions. 
Dividing both sides by $l^{i/a}$ results in 
\begin{equation} \label{eq:egw_li}
	\begin{split}
		\epsilon^{G_0W_0}_i 
		& =  \epsilon^\HF_i 
		\\
		& - \sum_{dlm} \bra{\Phi_{lm}^d}\wtV_\ldr \ket{\Phi_i}  l^{lm}_d /l^i 
		\\
		& + \sum_{edl}  \bra{\Phi_0} \wtV_\ldr \ket{\Phi_{il}^{ed}}  l^{de}_l /l^i 
	\end{split}
\end{equation}
for the electron-detachment energies and 
\begin{equation} \label{eq:egw_la}
	\begin{split}
		\epsilon^{G_0W_0}_a
		& =  \epsilon^\HF_a 
		\\
		&- \sum_{dlm} \bra{\Phi_{lm}^{da}}\wtV_\ldr \ket{ \Phi_0} l^{lm}_d /l^a
		\\
		& +\sum_{edl}  \bra{\Phi^a}\wtV_\ldr \ket{ \Phi_{l}^{de} } l^{de}_l/l^a.
	\end{split}
\end{equation}
for the electron-attached energies.

Equations \eqref{eq:egw_i}, \eqref{eq:egw_a}, \eqref{eq:egw_li} and \eqref{eq:egw_la} yield two important observations. 
First, these expressions eliminate the explicit dependence on the singles (1h and 1p configurations) amplitudes and allow the eigenvalue problem to be formulated entirely in terms of the doubles space (2h1p and 2p1h configurations). 
This transforms the original linear problem into a non-linear one involving only doubles contributions, as demonstrated in Ref.~\onlinecite{Quintero_2022}. 
Second, a direct comparison with Eq.~\eqref{eq:GW_std} reveals a one-to-one correspondence between the self-energy $\Sigma_{pp}(\omega = \epsilon^\GOWO_p)$ and the doubles contributions.

By exploiting Eq.~\eqref{eq:two_scren_trans}, the EOM amplitudes can be related to contributions to the self-energy. 
For the IP case, one finds
\begin{subequations} \label{eq:IP-rdlm}
	\begin{align}
		\begin{split}
			\frac{r^d_{lm}}{r_i} &= -\sum_{\mu} \frac{\sum_{{d'}{l'}}\bra{\Phi_{l'm}^{d'}}\wtV_\ldr\ket{\Phi_{i}^{}} x_{l',\mu}^{d'} \bar{x}_{l,\mu}^{d*}} {\epsilon^{\GOWO}_i-\epsilon^\HF_m+ \Omega_\mu}  
			\\
			&=-\sum_{\mu} \frac{(im|\mu) \bar{x}_{l,\mu}^{d*}} {\epsilon^{\GOWO}_i-\epsilon^\HF_m+ \Omega_\mu},  
		\end{split}
		\\
		\begin{split}
			\frac{r^l_{de}}{r_i} &=+\sum_{\mu} \frac{ x_{l,\mu}^{d}\sum_{{d'}{l'}} \bar{x}_{l',\mu}^{d'*} \bra{\Phi_{0}^{}}\wtV_\ldr\ket{\Phi_{il'}^{ed'}}} {\epsilon^{\GOWO}_i-\epsilon^\HF_e+ \Omega_\mu}  
			\\ 
			&=+\sum_{\mu} \frac{ x_{l,\mu}^{d}(\mu|ei)} {\epsilon^{\GOWO}_i-\epsilon^\HF_e+ \Omega_\mu}.
		\end{split}
	\end{align}
\end{subequations}
For the EA case, the corresponding expressions are
\begin{subequations} \label{eq:EA-rdlm}
	\begin{align}
		\begin{split}
			\frac{r^d_{lm}}{r_a} &= -\sum_{\mu} \frac{\sum_{{d'}{l'}}\bra{\Phi_{ml'}^{ad'}}\wtV_\ldr\ket{\Phi_{0}^{}} x_{l',\mu}^{d'} \bar{x}_{l,\mu}^{d*}} {\epsilon^{\GOWO}_a-\epsilon^\HF_m+ \Omega_\mu} 
			\\
			&= -\sum_{\mu} \frac{(am|\mu) \bar{x}_{l,\mu}^{d*}} {\epsilon^{\GOWO}_a-\epsilon^\HF_m+ \Omega_\mu},
		\end{split}
		\\
		\begin{split}
			\frac{r^l_{de}}{r_a} &=+\sum_{\mu} \frac{ x_{l,\mu}^{d} \sum_{{d'}{l'}} \bar{x}_{l',\mu}^{d'*} \bra{\Phi_{}^{a}}\wtV_\ldr\ket{\Phi_{l'}^{d'e}}} {\epsilon^{\GOWO}_a-\epsilon^\HF_e+ \Omega_\mu}  
			\\
			&=+\sum_{\mu} \frac{ x_{l,\mu}^{d} (\mu|ea)} {\epsilon^{\GOWO}_a-\epsilon^\HF_e+ \Omega_\mu}.
		\end{split}
	\end{align}
\end{subequations}

Following a similar procedure to the right-EOM amplitudes, the relationship of the left-EOM amplitudes to the frequency-dependent terms reads
\begin{subequations}
	\begin{align}
		\begin{split}
			\frac{l_d^{lm}}{l^i} &= -\sum_{\mu} \frac{ x_{l,\mu}^{d} \sum_{{d'}{l'}} \bar{x}_{l',\mu}^{d'*}\bra{\Phi_{i}^{}}\wtV_\ldr\ket{\Phi_{l'm}^{d'}}} {\epsilon^{\GOWO}_i-\epsilon^\HF_m+ \Omega_\mu}  
			\\
			&=-\sum_{\mu} \frac{x_{l,\mu}^{d}(\mu|im)} {\epsilon^{\GOWO}_i-\epsilon^\HF_m+ \Omega_\mu},  
		\end{split}
		\\
		\begin{split}
			\frac{l_l^{de}}{l^i} &=+\sum_{\mu} \frac{\sum_{{d'}{l'}} \bra{\Phi_{il'}^{ed'}}\wtV_\ldr\ket{\Phi_{0}^{}} x_{l',\mu}^{d'} \bar{x}_{l,\mu}^{d*} } {\epsilon^{\GOWO}_i-\epsilon^\HF_e+ \Omega_\mu}  
			\\ 
			&=+\sum_{\mu} \frac{ (ei|\mu) \bar{x}_{l,\mu}^{d*}} {\epsilon^{\GOWO}_i-\epsilon^\HF_e+ \Omega_\mu}.
		\end{split}
	\end{align}
\end{subequations}
for the IP case, and
\begin{subequations}
	\begin{align}
		\begin{split}
			\frac{l_d^{lm}}{l^a} &= -\sum_{\mu} \frac{ x_{l,\mu}^{d} \sum_{{d'}{l'}} \bar{x}_{l',\mu}^{d'*}
				\bra{\Phi_{0}^{}}\wtV_\ldr\ket{\Phi_{ml'}^{ad'}}} {\epsilon^{\GOWO}_a-\epsilon^\HF_m+ \Omega_\mu} 
			\\
			&= -\sum_{\mu} \frac{ x_{l,\mu}^{d} (\mu|am) } {\epsilon^{\GOWO}_a-\epsilon^\HF_m+ \Omega_\mu},
		\end{split}
		\\
		\begin{split}
			\frac{l_l^{de}}{l^a} &=+\sum_{\mu} \frac{\sum_{{d'}{l'}} \bra{\Phi_{l'}^{d'e}}\wtV_\ldr\ket{\Phi_{}^{a}}  x_{l',\mu}^{d'}  \bar{x}_{l,\mu}^{d*} } {\epsilon^{\GOWO}_a-\epsilon^\HF_e+ \Omega_\mu}  
			\\
			&=+\sum_{\mu} \frac{  (ea|\mu) \bar{x}_{l,\mu}^{d*}} {\epsilon^{\GOWO}_a-\epsilon^\HF_e+ \Omega_\mu}.
		\end{split}
	\end{align}
\end{subequations}
for the EA case.